\documentclass[letterpaper,twocolumn,10pt]{article}
\usepackage{usenix2019_v3}

\usepackage{tikz}
\usepackage{amsmath}

\usepackage{filecontents}

\usepackage{caption}

\usepackage{subfigure}
\usepackage{booktabs}
\usepackage{multirow,enumitem}
\usepackage{diagbox}

\usepackage{pifont}
\newcommand{\xmark}{\ding{55}}%

\usepackage[flushleft]{threeparttable}
\usepackage[utf8]{inputenc}
\usepackage{hyperref}
\usepackage{cleveref}
\crefname{section}{§}{§§}
\Crefname{section}{§}{§§}

\usepackage{amsmath,balance}
\usepackage{amssymb}

\usepackage[ruled,linesnumbered,vlined,noend]{algorithm2e}
\makeatletter
\newcommand{\removelatexerror}{\let\@latex@error\@gobble}
\makeatother
\usepackage[noend]{algpseudocode}

\usepackage{makecell}

\usepackage{xspace}

\newcommand{\toolname}{\textsc{QFA2SR}\xspace}

\newcommand{\Softmax}{{\tt Softmax}}
\newcommand{\rank}{{\tt rnk}}

\renewcommand{\figurename}{Fig.}

\usepackage{longtable}

\usepackage{colortbl}

\definecolor{goldenrod}{rgb}{0.85, 0.65, 0.13}

\usepackage{circledsteps}

\usepackage{tcolorbox}
\tcbuselibrary{breakable}

\usepackage[normalem]{ulem}

\newcommand{\OSITARGET}{{$\mathcal{A}_{\tt OSI}^{\tt T}$}\xspace}
\newcommand{\OSIUNTARGET}{{$\mathcal{A}_{\tt OSI}^{\tt UT}$}\xspace}
\newcommand{\TDSVTARGET}{{$\mathcal{A}_{\tt TD\text{-}SV}^{\tt T}$}\xspace}

\usepackage{url}

\newcommand{\SUMGLOBAL}{Sum-Global\xspace}
\newcommand{\VOTEGLOBAL}{Vote-Global\xspace}

\usepackage{fancyhdr}
\usepackage{bbding}
\begin{document}

\date{}

\title{\Large \bf \toolname: Query-Free Adversarial Transfer Attacks to Speaker Recognition Systems}

\author{
  Guangke Chen\textsuperscript{1}, Yedi Zhang\textsuperscript{1}, Zhe Zhao\textsuperscript{1}, Fu Song\textsuperscript{1,2,3 (\Envelope)}  \\
  \textsuperscript{1} ShanghaiTech University \qquad
  \textsuperscript{2} Automotive Software Innovation Center \\
  \textsuperscript{3} Institute of Software, Chinese Academy of Sciences \& University of Chinese Academy of Sciences
}

\maketitle

\thispagestyle{fancy} 

\begin{abstract}
    Current adversarial attacks against speaker recognition systems (SRSs)
    require either white-box access or heavy black-box queries to the target SRS,
    thus still falling behind practical attacks against proprietary commercial APIs and voice-controlled devices.
    To fill this gap, we propose \toolname, an effective and imperceptible query-free black-box attack,
    by leveraging the transferability of adversarial voices.
    To improve transferability, we present three novel methods, tailored loss functions, SRS ensemble, and time-freq corrosion.
    The first one tailors loss functions to different attack scenarios.
    The latter two augment surrogate SRSs
    in two different ways.
    SRS ensemble combines diverse surrogate SRSs with new strategies,
   amenable to the unique scoring characteristics of SRSs.
    Time-freq corrosion augments surrogate SRSs by incorporating
     well-designed time-/frequency-domain modification functions,
    which simulate and approximate the decision boundary of the target SRS and distortions
introduced during over-the-air attacks.
    \toolname boosts the targeted transferability by {20.9\%-70.7\%} on four popular commercial APIs (Microsoft Azure, iFlytek, Jingdong, and TalentedSoft),
    significantly outperforming existing attacks in query-free setting, with negligible effect on the imperceptibility.
    \toolname is also highly effective when launched over the air against three wide-spread voice assistants (Google Assistant, Apple Siri, and TMall Genie)
    {with 60\%, 46\%, and 70\% targeted transferability, respectively}.
\end{abstract}

\section{Introduction}
Speaker recognition (SR) is
an automatic process recognizing the identity of a person with her voice.
SR has versatile applications,
such as authentication for financial transactions~\cite{TD-Bank},
access control for voice-controlled devices~\cite{smart-home-device-control},
and service personalization in voice assistants~\cite{Google-Voice-Match}.
However, the popularity of SR has brought new security concerns.
Recent studies have shown that SRSs
are vulnerable to adversarial attacks
as summarized in \tablename~\ref{tab:comp}.
{Such attacks aims to craft an adversarial voice from a given voice uttered by a source speaker,
so that it is misrecognized as another speaker by the target SRS,
but does not sound like the misrecognized speaker from the perception of ordinary users.}
White-box attacks
assume complete knowledge of the target SRS,
which are powerful yet remarkably unpractical
as it is {\it impossible}
to acquire any internal information about protected proprietary systems.
Black-box attacks 
do not rely on such information,
but usually require a large number of queries to the target SRS
to achieve considerable attack capabilities.
Such black-box attacks suffer from two serious drawbacks:
(1) they are cost-consuming because voice-controlled devices do not expose APIs thus voices have to be played over the air while
commercial APIs require query-charges. Furthermore, both devices and APIs often pose limitations on the query frequency;
(2) they are not very stealthy because a large bulk of queries to the target SRS leads to detectable abnormal traffics and behaviors.

\begin{table*}[htbp]
    \centering\setlength{\tabcolsep}{1.5pt}
    \caption{An overview of the state-of-the-art adversarial attacks to SRSs.}
    \label{tab:comp}%
  \resizebox{1.0\textwidth}{!}{ \begin{threeparttable}
      \begin{tabular}{|c|c|c|c|c|c|c|c|c|c|}
      \toprule
        {\bf Method}  & {\bf Threat Scenario} & {\bf Knowledge} & {\bf \#Queries} & {\bf Enrollment}  & {\bf Commercial} &{\bf Attack Type} & {\bf Attack Media} & {\bf ASR (Digital)} & {\bf ASR (Physical)} \\
      \midrule
      {\bf \cite{fool-end-2-end,li2020adversarial}} & White-box  & Gradient & N/I     & Same    & \xmark     & Untargeted     & Digital & 41\%-69\% & N/A \\
      \hline
      {\bf \cite{jati2021adversarial,Paralinguistics}} & White-box  & Gradient & N/I     & N/A      & \xmark     & Untargeted & Digital & 40\%-100\% & N/A \\
      \hline
      {\bf AS2T~\cite{AS2T}} & \begin{tabular}[c]{@{}c@{}}White-/Black-box \end{tabular} & Gradient/Scores & $\sim$ 5000 & Same & \xmark & Targeted, Untargeted & Digital, Physical & \begin{tabular}[c]{@{}c@{}} $\sim$ 100\%\end{tabular} & 89.4\%-100\%\\
      \hline
      {\bf FakeBob~\cite{chen2019real}} & Black-box & Scores & $\sim$ 2500  & Same  & TalentedSoft~\cite{Talentedsoft}, Azure~\cite{microsoft-azure-vpr} & Targeted, Untargeted & Digital, Physical & 100\% & 70\% \\
      \hline
      {\bf SirenAttack~\cite{du2020sirenattack}} & Black-box & Scores & $\sim$ 7500  & N/A   & \xmark     & Targeted, Untargeted & Digital & $\sim$ 100\% & N/A \\
      \hline
      {\bf Kenansville~\cite{Kenansville}} & Black-box & Decision & $\sim$ {15} & Same  &  Azure & Untargeted & Digital & 5\%-37\% & N/A \\
      \hline
      {\bf Occam~\cite{Occam}} & Black-box & Decision & $\sim$ 10000 & Same  & Azure, Jingdong~\cite{JingdongSRS} & {Targeted} & Digital & 100\% & N/A \\
      \hline
      {\bf \begin{tabular}[c]{@{}c@{}}\toolname \\ (Ours) \end{tabular}}  & Black-box & None  & 0     & \begin{tabular}[c]{@{}c@{}}Different, \\ Same \end{tabular}
      & \begin{tabular}[c]{@{}c@{}}TalentedSoft, Azure, iFlytek~\cite{iFlytek}, \\ Jingdong, Google Assistant~\cite{Google-Voice-Match}, \\ Apple Siri~\cite{Hey-siri}, TMall Genie~\cite{TMall-Genie} \end{tabular}
      & Targeted, Untargeted & Digital, Physical & \begin{tabular}[c]{@{}c@{}} 27.4\%-99.5\% \end{tabular} & 46\%-70\% \\
      \hline
      \end{tabular}

      \begin{tablenotes}
     \item Note:
(i)``While-box''/``Black-box'': the adversary has complete/no knowledge of the target SRS.
(ii) ``Gradient''/``Scores''/``Decisions'': the adversary requires the gradient information/the similarity scores/the identified speaker;
``None'': the adversary has no access to the target SRS when crafting adversarial voices.
(iii) \#Queries: the number of queries used for creating adversarial voices, which is not important (N/I) for white-box attacks,
and ``$\sim$'' denotes approximation.
(iv) Same (Different):  the enrollment voices used by the adversary are the same as (different from)
the ones used for enrolling the target SRS, while N/A denotes that there is no enrollment voices.
(v) ``\xmark": commercial SRSs are not considered as a target SRS.
(vi) ``Untargeted" (``Targeted"): untargeted (resp. targeted) attack where the attack succeeds
if the adversarial voice is misclassified as one of the enrolled
speakers (resp. the target speaker).
(vii) ``Digital": adversarial voices are directly fed to SRSs
in the form of audio file via exposed API;  ``Physical":
adversarial voices are played and recorded by hardware and transmitted in the air.
(viii) ``ASR'': attack success rate on commercial SRSs (if considered otherwise open-source SRSs).
    \end{tablenotes}
    \end{threeparttable}}
  \end{table*}
Our work is motivated by the following research question:
``{\it how to launch effective, stealthy, and practical
adversarial attacks against black-box commercial APIs and voice-controlled devices
without any queries to the target SRS when constructing adversarial voices (i.e., query-free)?}''.
A straightforward idea is to exploit the transferability of adversarial examples, i.e.,
crafting adversarial examples on a surrogate SRS
{(a local white-box SRS owned by the adversary)}
and then transferring them to the target SRS.
However, until now, adversarial attacks in SR
suffer from limited transferability
since adversarial voices are easy to overfit the surrogate SRS and consequently become ineffective on the target SRS.
This is because there are various aspects that the target SRS may differ in
with the surrogate SRS
{(e.g., acoustic feature~\cite{acoustic-feature-li} and scoring method~\cite{scoring-review})}
and a large number of updatable values of a seed voice due to a high audio sample rate.
Indeed, we find that the transfer attack success rate (ASR) to most target SRSs is less than 6\%
even the surrogate SRS shares the same architecture, training dataset, acoustic feature,
and scoring method with the target SRSs (cf. Appendix~\ref{sec:srs-diverse-exper}).
Thus, the main problem is how to improve the transferability of adversarial voices without
reducing imperceptibility.

In this work, we address the above problem by proposing
an attack called
{\bf Q}uery-{\bf F}ree Adversarial {\bf A}ttack
{\bf to} {\bf S}peaker {\bf R}ecognition (\toolname).
\toolname features three
novel methods: Tailored Loss Functions, SRS Ensemble, and
Time-Freq Corrosion, to improve the transferability of adversarial voices without
reducing imperceptibility.
{The first one is proposed to find optimal loss functions,
for which we design and empirically study various loss functions.
Remarkably, we find that the commonly-adopted Cross Entropy Loss~\cite{goodfellow2014explaining}
and Margin Loss~\cite{carlini2017towards} for crafting adversarial images lead to less transferable adversarial voices
than ours.
The second one combines multiple surrogate SRSs via two novel strategies,
so that adversarial voices crafted on the ensemble of surrogate SRSs
can deceive as many surrogate SRSs as possible. 
The last one incorporates various well-designed time-/frequency-domain modification functions into surrogate SRSs
to simulate and approximate unknown distribution of the target SRS and
distortions introduced during over-the-air attacks~\cite{AS2T}.}

We implement our approach in a tool and thoroughly evaluate
the performance of \toolname on various open-source SRSs, commercial APIs,
and voice assistants. The results confirm the effectiveness of our three novel methods and \toolname.
For instance, \toolname on four commercial APIs,
i.e., (Microsoft) Azure, Jingdong, iFlytek, and TalentedSoft,
improves the targeted transfer ASR by {20.9\%-70.7\%},
significantly outperforming the state-of-the-art attacks in the query-free setting,
with negligible effect on the imperceptibility in terms of both perceptual objective metric 
and subjective human study. 
In particular, \toolname achieves 89.6\%/99.6\% targeted/untargeted transfer ASR to Azure,
and 96\% targeted transfer ASR to Jingdong
(within 4 queries when launching \toolname).
\toolname on three voice assistants,
i.e., Google Assistant, Apple Siri, and Alibaba TMall Genie,
achieves 46\%-70\% targeted transfer ASR when launched over the air.

In summary, the main contribution of our work includes:
\begin{itemize}[leftmargin=*]\setlength{\itemsep}{1pt}
  \item We study various  loss functions  and find better loss functions for transferability.
  We showcase that the promising Cross Entropy loss and Margin loss in the image domain
  are sub-optimal for the transfer attack in SR.
\item  We propose two novel strategies for the ensemble of the surrogate SRSs
  which outperforms the model ensemble for crafting adversarial images~\cite{LCLS17}.
  \item We propose time-freq corrosion to enhance transferability,
accompanied with diverse  modification functions for
  simulating and approximating decision boundary of the target SRS and distortions introduced during
over-the-air attacks.
 \item We propose \toolname, a query-free black-box adversarial attack against SRSs, by leveraging the transferability of adversarial voices,
 and aided by novel methods and strategies to boost the transferability, towards a truly usable transfer attack in the physical world. 
  \item We extensively evaluate \toolname on 9 open-source SRSs, 4 commercial APIs, and 3 voice assistants,
  covering 3 attack scenarios, 2 recognition tasks, 2 attack types, 2 attack medias,
  and 3 settings of available voices to the adversary,
  with more than 144,800 adversarial voices in total.
  We find that \toolname can boost the transferability by a large margin
  with negligible effect on imperceptibility.
\end{itemize}

Voices and demo videos are available at our website~\cite{QFA2SR}.

\noindent {{\bf Abbreviations and Acronyms.}
For convenient reference, we summarize the abbreviations and acronyms in \tablename~\ref{tab:acronyms}.}

{\section{Ethical Considerations}
We make the following ethical considerations: }

\noindent {{\bf Strictly controlled experiments.}
For commercial APIs,
the target speakers in experiments are enrolled by us,
so they do not associate with any real-world financial or social accounts in the applications that exploit the APIs.
For voice assistants,
where target speakers associate with accounts, when launching our attack against them,
we stopped once the attack bypassed the authentication of the target speaker.
We did not take any further malicious actions, e.g., accessing the service exclusive to the target speaker.
Additionally, all the used voice-controlled devices are our own facilities.
}

\noindent {{\bf Responsible disclosure.}
We contacted the vendor TalentedSoft by email and other six vendors (Microsoft, iFlytek, Jingdong, Apple, Google, and Alibaba)
with their official security vulnerability report websites, to report the vulnerabilities we found.
We submitted reports with attack details, reproducibility of our attack using attached code, demonstration audios and videos,
security risks brought by our attack, reason for the vulnerabilities, and suggested countermeasures.
All vendors express their gratitude to our research and disclosure to keep their services, systems, and users secure.
For instance, iFlytek has identified our reported vulnerability as a moderate risk level
and awarded us a bounty of 1,000 RMB as recognition for our vulnerability report,
and TalentedSoft replied that they will develop a plan to fix the vulnerability.
}

\begin{table}[t]
  \centering\setlength{\tabcolsep}{1pt}
  \caption{{Abbreviation and Acronym}}
    \scalebox{0.6}{
  \begin{tabular}{c|c|c|c}
    \hline
   {\bf Acronym} & {\bf Meaning} & {\bf Acronym} & {\bf Meaning} \\
    \hline
    SR & Speaker recognition & SRS(s) & Speaker recognition system(s) \\
    \cline{1-4}
     SV & Speaker verification & OSI & \makecell[c]{Open-set speaker identification} \\
    \cline{1-4}
     \OSITARGET & \makecell[c]{Targeted attack on OSI} & TD-SV & Text-dependent speaker verification \\
    \hline
    \OSIUNTARGET & \makecell[c]{Untargeted attack on OSI} &      \TDSVTARGET & \makecell[c]{Targeted attack on TD-SV} \\
    \cline{1-4}
 Same-enroll & \makecell[c]{Surrogate \& target SRSs have\\  the same enrollment voices}  & Differ-enroll & \makecell[c]{Surrogate \& target SRSs have\\  different enrollment voices}  \\
    \cline{1-4}
   ASR$_t$ & Targeted attack success rate  & ASR$_t$-s/ASR$_t$-d & ASR$_t$ under enroll-/differ-enroll   \\
    \cline{1-4}
   ASR$_u$ & Untargeted attack success rate  &ASR$_u$-s/ASR$_t$-d & ASR$_u$ under enroll-/differ-enroll    \\
    \hline
     RD & Reverberation-distortion&   \SUMGLOBAL & \makecell[c]{Summation-based  global score ranking} \\
    \cline{1-4}
     NF & Noise-flooding &  \VOTEGLOBAL & \makecell[c]{Voting-based  global score ranking}\\
    \cline{1-4}
      SA & Speed-alteration & CD & Chunk-dropping \\
    \cline{1-4}
      FD & Frequency-dropping & TW & Time-warping \\
    \cline{1-4}
      TM & Time-masking & FM & Frequency-masking \\
    \hline
  \end{tabular}
    }\vspace{-3mm}
    \label{tab:acronyms}
\end{table}

\section{Background \& Related Work}\label{sec:background}
\subsection{Speaker Recognition System (SRS)}
\begin{figure}[t]
    \centering
    \includegraphics[width=0.46\textwidth]{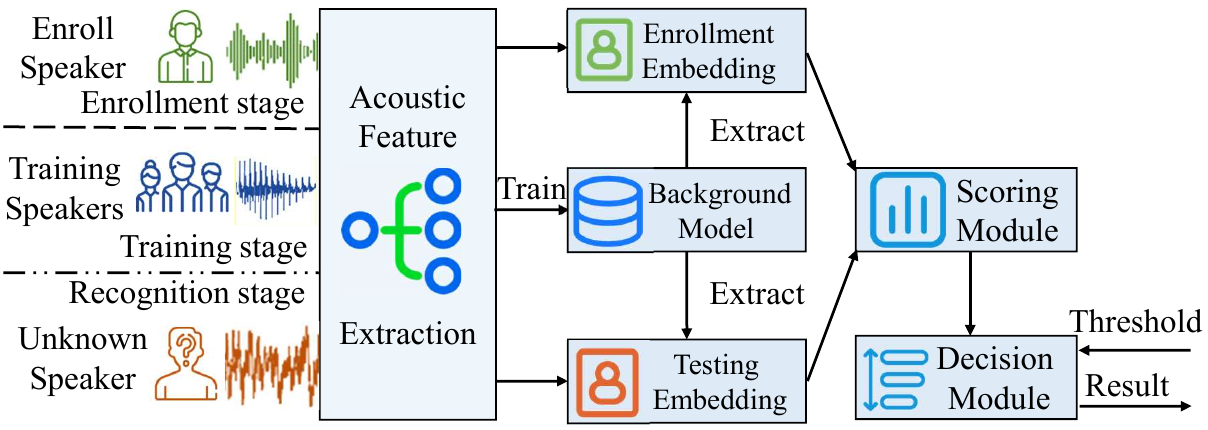}
    \caption{Framework of SRSs.}
    \label{fig:typical-SRSs}\vspace{-3mm}
\end{figure}

{\bf Speaker recognition}.
Speaker recognition (SR) is the task of automatically recognizing individual speakers
from their voices, typically representing acoustic characteristics as fixed-dimensional vectors via speaker embedding~\cite{wang2020simulation}.
An architecture of generic speaker recognition systems (SRSs) is shown in \figurename~\ref{fig:typical-SRSs},
comprising three stages: \emph{training}, \emph{enrollment}, and \emph{recognition}.
All of them extract acoustic features from raw speech signals
via an acoustic feature extraction module, yielding the acoustic characteristics.
Common acoustic features include speech spectrogram~\cite{hannun2014deep},
fBank~\cite{FilterBanks}, and MFCC~\cite{muda2010voice}.
The training stage trains a background model 
which learns a mapping from training voices to embeddings. 
Classic background model utilizes Gaussian Mixture Model (GMM)~\cite{reynolds2000speaker},
to produce identity-vector (ivector) embeddings~\cite{DehakDKBOD09}.
Recent promising background model
utilizes deep neural networks (DNNs) to produce deep embeddings, e.g., xvector~\cite{snyder2018x}.
The enrollment stage maps a voice uttered by an enrolling speaker to an \emph{enrollment embedding} using the background model.
The recognition stage first retrieves the \emph{testing embedding} of a given voice $x$ from the background model
and then  measures the similarity between the enrollment
and testing embeddings via the scoring module.
The scoring module produces a score vector $S(x)$
 based on which the decision module produces the result.
Probabilistic Linear Discriminant Analysis (PLDA)~\cite{nandwana2019analysis}
and COSine Similarity (COSS)~\cite{dehak2010cosine} are two widely-adopted scoring methods.

\noindent{\bf SR task}.
The SR can be classified into two major tasks: speaker identification and speaker verification (SV),
where the former can be further classified into
open-set identification (OSI) and close-set identification (CSI) both allowing multiple speakers to be enrolled forming a speaker group $G$.
OSI determines if a given voice is uttered by either one of the enrolled speakers
or imposter (i.e., an unenrolled speaker),  according to the scores of all the enrolled speakers  and a predefined score threshold $\theta$.
Formally, assuming $G=\{1,\cdots,n\}$,
given a voice $x$, the decision module outputs $D(x)$:
\begin{center}
    $D(x)=\left\{
    \begin{array}{ll}
    \arg\max\limits_{i\in G} \ [S(x)]_i, &\mbox{if } \max\limits_{i\in G} \ [S(x)]_i\geq \theta; \\
    {\tt imposter}, &\mbox{otherwise}.
    \end{array}\right.$
    \end{center}
where $[S(x)]_i$ denotes the $i$-th entry of the score vector $S(x)$,
namely, the score of the voice $x$ that is likely uttered by the enrolled speaker $i$.
Intuitively, the speaker $i$ that gives the maximal score is assigned as the speaker of the voice $x$,
if  $[S(x)]_i$ is no less than the threshold $\theta$.
Otherwise, the voice $x$ is rejected, regarding it as being uttered by an imposter.
In contrast, CSI always identifies the speaker that gives the maximal score as the speaker of the voice $x$,
i.e.,  the decision module outputs $D(x)=\arg\max_{i\in G} \ [S(x)]_i$,
and SV is a restricted case of OSI, which has exactly \emph{one} enrolled speaker.

\noindent{\bf Text dependency}.
SR can be text-dependent  (TD) and
text-independent (TID).
TD requires speakers to utter some predefined phrases or words
during both the enrollment and recognition stages
while TID does not pose any such constraints.
TD can achieve good performance on short voices,
but needs a large number of training voices with the same phrases or words,
thus it is only used in the SV task, called TD-SV.
TID needs longer voices to achieve good performance,
but is more convenient and can be used in all tasks.

\subsection{Attacks on SRS}
{\bf Adversarial attack}.
An adversarial attack on SRS aims to craft an adversarial voice from a given voice uttered by a \emph{source} speaker,
so that the SRS under attack misclassifies it as one of the enrolled speakers (untargeted attack)
or the target speaker (targeted attack),
but ordinary users do not determine it as the recognized speakers by the SRS (imperceptibility).

The problem of finding such an adversarial voice $x'$ from a voice $x$ can be formalized as the  optimization problem:
\begin{center}
$\text{argmin}_{x'} f(x') \mbox{ subject to $d(x',x)\leq \varepsilon$ and $x'\in [-1,1]$}$
\end{center}
where $f$ is a loss function measuring the effectiveness of the attack,
$d(x',x)$ is a distance metric quantifying the similarity between $x'$ and $x$ (imperceptibility),
and $\varepsilon$ is the budget of added adversarial perturbation to ensure imperceptibility.
The most widely adopted distance metric is $L_p$ norm~\cite{carlini2017towards}, i.e., $d(x',x)=\sqrt[p]{\sum_{i}|x'_i-x_i|^p}$.
Under the white-box setting where the adversary has full knowledge of the target SRS,
the optimization problem can be solved by gradient descent
using the exact gradient obtained by backpropagation~\cite{fool-end-2-end,li2020adversarial,jati2021adversarial,Paralinguistics, AS2T, LiW00020}.
Under the black-box setting where the exact gradient is not available,
the attack either estimates the gradient
(e.g., FakeBob~\cite{chen2019real} and AS2T~\cite{AS2T})
or utilizes gradient-free optimization approaches
(e.g., SirenAttack~\cite{du2020sirenattack}, Kenansville~\cite{Kenansville},
and Occam~\cite{Occam}). All these black-box attacks access the target SRS
as an oracle, i.e., providing a series of carefully crafted inputs
to the model and observing its outputs (either scores~\cite{chen2019real,
du2020sirenattack,AS2T} or decisions~\cite{Kenansville,Occam}).

\noindent{\bf Hidden voice and spoofing attacks}.
Hidden voice attack~\cite{AbdullahGPTBW19} perturbs given a voice uttered by a target speaker
so that the resulting voice is treated as mere noise by humans,
but still correctly recognized as the target speaker by the SRS.
The spoofing attack~\cite{wu2015spoofing} (e.g., replay attack~\cite{shirvanian2019quantifying}
and voice cloning attack~\cite{wenger2021hello})
aims to obtain a voice that is correctly classified as the target speaker by the SRS
and also sound like the {target speaker} listened to by ordinary users.
Specifically, a replay attack aims to bypass the SRS
using pre-recorded voices surreptitiously captured from the target speaker,
and is usually used for attacking \emph{text-independent} SV,
as the collected voices usually do not contain the required text by \emph{text-dependent} SV.
In contrast, given a few voices of a speaker and the desired text,
voice cloning attack creates a voice that sounds like the speaker and
contains the specified speech content,
thus can be exploited to attack text-dependent SV.

Hidden voice and spoofing attacks have different attack purposes and scenarios
from adversarial attacks~\cite{chen2019real,wenger2021hello,SpeakerGuard}.
The perception of human listeners is inconsistent with that of the SRS
under adversarial and hidden voice attacks,
while it is consistent under the spoofing attack (cf. \cref{sec:commercial-APIs}).
Furthermore,
we will show that our adversarial attack \toolname
achieves a higher attack success rate than hidden voice and spoofing attacks
in the query-free setting (cf. \cref{sec:commercial-APIs}).

\section{Methodology of \toolname}\label{sec:methodology}

\subsection{Threat model}
We consider so far the most practical threat model
concerning the knowledge of the target SRS and attack capability
in the adversarial speaker recognition domain.

\noindent {{\bf Target SRS.}}
Regarding the target SRS, we assume that the adversary
neither has white-box access to any
of its internal information (e.g., architecture, parameters, training algorithm, and dataset),
nor perform queries to the SRS during the generation of adversarial voices, so-called \emph{query-free block-box} setting.
First, it is almost \emph{impossible} for the adversary
to acquire internal information of a strictly protected proprietary SRS in the real life, e.g.,
commercial service APIs and voice controlled devices, thus preventing from white-box attacks~\cite{fool-end-2-end,li2020adversarial,jati2021adversarial,Paralinguistics,AS2T}.
{Second, query-free is necessary and significant for achieving truly practical attacks in the real world considering that:
(1) Voice assistants can \emph{only}
be interacted via the air channel, while the generation of adversarial voices via air channels would be difficult and time-consuming as the generation is an iterative process,
and at each iteration, intermediate voices have to be played by loudspeakers.
(2) Commercial APIs usually pose a limit on the query frequency, e.g., Jingdong SRS resticts 2 queries per second with a maximum of 500 queries per day.
The limit can be solved by using time slots between queries, but making attacks time-consuming.
(3) Commercial APIs may charge on the query, e.g., JingDong SRS charges 500 RMB for 1,000 queries, making attacks expensive.
(4) Voice assistants and some commercial APIs only return  final decision without any scores, thus stopping all the  score-based black-box attacks~\cite{chen2019real,du2020sirenattack,AS2T}.
Query-free attacks overcome all the above limitations.}

\noindent {{\bf Voice resources.}}
{Regarding voice resources, we assume that the adversary: (1)
has a large number of voices for training the background model of \emph{surrogate} SRSs but
could be different from those used for training the background model of the \emph{target} SRS
and (2) knows all the enrolled speakers of the \emph{target} SRS
and has some voices for each of them which are used to enroll \emph{surrogate} SRSs
but also could be different from those used for enrolling the \emph{target} SRS.
The first assumption is reasonable thanks to many large-scale open-source speech corpora, e.g, Librispeech~\cite{panayotov2015librispeech} and
VoxCeleb1~\cite{nagrani2017voxceleb}. 
The second assumption is also reasonable as the adversary can either use enrolled speakers' public videos on social media or record their speeches
via social engineering.
In \cref{sec:impact-knowledge-enrolled-speakers}, we will relax the second assumption
by considering that the adversary only has the target speaker's voice instead of all the enrolled speakers of the \emph{target} SRS.
In contrast, prior works~\cite{fool-end-2-end,li2020adversarial,jati2021adversarial,
Paralinguistics,AS2T,chen2019real,du2020sirenattack,Kenansville,Occam}
are either white-box or query-based black-box attacks, thus require neither voice datasets to train
 \emph{surrogate} SRSs
nor voices of enrolled speakers of the \emph{target} SRS
to enroll \emph{surrogate} SRSs, but used the same enrollment speakers and
the same voices between the surrogate and target SRSs when launching transfer attacks.}

\noindent {{\bf Attack scenarios and risks.}}
Regarding attack scenarios, different combinations of source/target speaker, and {recognition} task enables the adversary to achieve different goals,
e.g., unauthorized access, denial-of-service, anonymous access, evasion, and privacy protection~\cite{AS2T}.
we consider three combinations, denoted by \OSITARGET, \OSIUNTARGET, and \TDSVTARGET,
all of which attempt to craft an adversarial voice from a given benign voice uttered by an \emph{imposter}
such that the adversarial voice is accepted by the target SRS.
Both  \OSITARGET and \OSIUNTARGET focus on the OSI task, but
\OSITARGET is a targeted attack that specifies an enrolled speaker as the target speaker,
while \OSIUNTARGET is an untargeted attack that succeeds when the adversarial voice is accepted as any enrolled speaker.
{\TDSVTARGET focuses on the text-dependent SV task (i.e., TD-SV),
where the adversary has target speakers' voices \emph{not} containing the desired text but knows the text in advance.
It is practical as systems should inform customers of the text, e.g., ``Hey Siri''.
Adversary can use voices of imposters with such text to craft adversarial voices.
We found our attack rarely alters the text as it focuses on identity instead of speech content.}
Since SV is a binary classification problem with only one enrolled speaker,
the target speaker of \TDSVTARGET is the unique enrolled speaker.
We do not consider the CSI task since the OSI task is more difficult to attack than the CSI task~\cite{AS2T},
and to the best of our knowledge, no commercial SRSs use the CSI task.

{Our attack exposes the following risks.
(1) Speaker recognition has been used for access control in smart home~\cite{smart-home-device-control},
smartphones~\cite{phoen-unlock-VPR}, and mobile applications~\cite{VPR-log-in}.
Our attack may enable unauthorized access, e.g., controlling over critical appliances,
unlocking and logging target speakers' smartphones and applications.
(2) Speaker recognition has been used for identity verification
in banks' telephone-communication~\cite{TD-Bank, Citi-Bank}
and password-free payment~\cite{TMall-Genie-payment-VPR},
so our attack may lead to property damage.
(3) Speaker recognition has been used in key-word detection of voice assistants~\cite{Google-Voice-Match},
so our attack can activate assistants and then issue malicious instructions
(e.g., reading messages, deleting reminders, circumventing the confidentiality and integrity of data),
or launch follow-up attacks targeting speech-to-text, e.g., Dolphin-attack~\cite{dolphin-attack} and CommanderSong~\cite{yuan2018commandersong}.
Readers are recommended to watch recorded videos on our} {website~\cite{QFA2SR}}.
{However, our attack cannot achieve certain objectives, e.g.,
(i) denial-of-service to the target speaker,
or (ii) actively hiding the identity of the target speaker  to achieve anonymous access to illegal services,
protect personal privacy, or evade being detected~\cite{AS2T}.
Realizing these purposes requires crafting adversarial voices from the \emph{target} speaker's benign voices
such that they are rejected or recognized as other speakers by the target SRS,
which is beyond the scope of \OSITARGET, \OSIUNTARGET, and \TDSVTARGET.
}

\subsection{Technical Challenges}
Under the query-free black-box setting, all the prior attacks cannot
be directly mounted, as they are either white-box or query-based black-box attacks.
To tackle this issue, one has to exploit the intriguing property of adversarial examples, i.e., transferability --
an adversarial example crafted with respect to one model is often found effective against other models as well.
Thus, the adversary can first craft an adversarial voice on a local surrogate SRS
and then transfer it to the target SRS.
While advanced transfer attacks against computer vision systems have been extensively studied in the literature (e.g., \cite{LCLS17,Momentum-attack,Nesterov-attack,MaoFWJ0LZLB022}),
current transfer attacks on SRSs are considerably limited  (e.g., the targeted/untargeted transfer attack success rate
to most target SRSs is less than 6\% (cf. Appendix~\ref{sec:srs-diverse-exper}))
due to the following technical challenges.

\noindent{\bf Challenge CH-I}.
The target SRS may be different from the surrogate SRS in various aspects,
such as dataset and hyper-parameters for training background model,
architecture (e.g., GMM and DNN),
acoustic feature (e.g., fBank and MFCC),
scoring method (e.g., PLDA and COSS), and
input pre-processing, all of which can largely affect the transferability~\cite{AS2T,chen2019real}.
More specifically, different datasets may obey different voice distributions
due to different recording environments, hardware, and subjects,
while different voice pre-processing can change the voice distributions in different ways.
Thus, SRSs trained with different datasets and input pre-processing
may learn different voice distributions.
As a result, an adversarial voice crafted from a surrogate SRS would be highly sensitive
to the voice distribution of the surrogate SRS, leading to low transferability. 
A piece of evidence is that adversarial voices are more likely to be destroyed by some input transformation~\cite{SpeakerGuard,SEC4SR}.
Similarly, SRSs with different training hyper-parameters, architectures, acoustic features
and scoring methods may learn different voice distributions and decision boundaries.
For instance, removing an MFCC acoustic feature extraction module
from the surrogate system improves the transferability in the speech recognition
domain~\cite{abdullah2021demystifying}.
We highlight that in the audio domain,
the surrogate system may differ from the target one in more aspects than in the image domain
because audio systems are usually more complicated and own several unique components and pipelines,
e.g., acoustic feature extraction module and scoring method,
making the transfer attacks more challenging~\cite{AWBPT20,Occam,du2020sirenattack}.

\noindent{\bf Challenge CH-II}. The iterative generation process of adversarial examples can be seen as
the ``training" of the input data with a fixed model,
in contrast to the standard training where the model is trained with a fixed input dataset.
Due to the high audio sampling rate (e.g., 16 khz), an audio has a large number of trainable variables,
leading to the curse of dimensionality.
For instance, a 1-second audio with 16k Hz sampling rate
has totally 16,000 updatable variables,
much larger than 784 ($28\times 28$) and 3072 ($32\times 32\times 3$)
variables of an image from MNIST and CIFAR-10, respectively.
As a result, similar to significant overfitting and poor generalization of
training DNNs with a larger number of parameters~\cite{SrivastavaHKSS14,LawrenceGT97},
the crafted adversarial voices are easy to over-fit to the surrogate SRS,
resulting in ineffective transfer attacks~\cite{Momentum-attack,WYGWei21}.
This phenomenon has been reported in the image domain~\cite{LCLS17},
where targeted transfer attacks that are effective on low-resolution images (e.g., MNIST or CIFAR-10) become significantly less effective
on high-resolution images (e.g., ImageNet).

\noindent{\bf Challenge CH-III}.
To attack voice-controlled devices,
adversarial voices should be played by loudspeakers, transmitted over the air, and recorded by microphones,
during which loudspeakers and microphones will induce distortions to voices
due to their non-uniform frequency selectivity~\cite{LiW00020}.
Even worse, different loudspeakers and microphones may exhibit distinct frequency responses, thus incurring distinct
distortions~\cite{LiW00020}. Moreover,
both ambient noise and reverberation could distort  adversarial voices and undermine the attack as well,
and their impacts depend on the specific attack environments~\cite{AS2T}.
Therefore, over-the-air transfer attacks
undergo additional challenges,
compared to pure API transfer attacks.

\subsection{Overview of \toolname}
A straightforward idea to improve the transferability
is to enlarge the perturbation budget or increase the confidence of adversarial voices~\cite{AS2T,chen2019real}.
However, it not only makes the adversarial voices less imperceptible,
thus much easier to increase the awareness of human, but only is almost ineffective when there is a large gap between the surrogate and the target SRSs,
as SRS-specific factors (e.g., architecture) are dominant factors
over attack-specific ones (e.g., perturbation budget and confidence)~\cite{AS2T}.
We also note that confidence is not a good
tool to increase the transferability in commercial computer vision platforms (cf.~\cite[Observation 9]{MaoFWJ0LZLB022}).

\begin{figure}[t]
    \centering
    \includegraphics[width=0.46\textwidth]{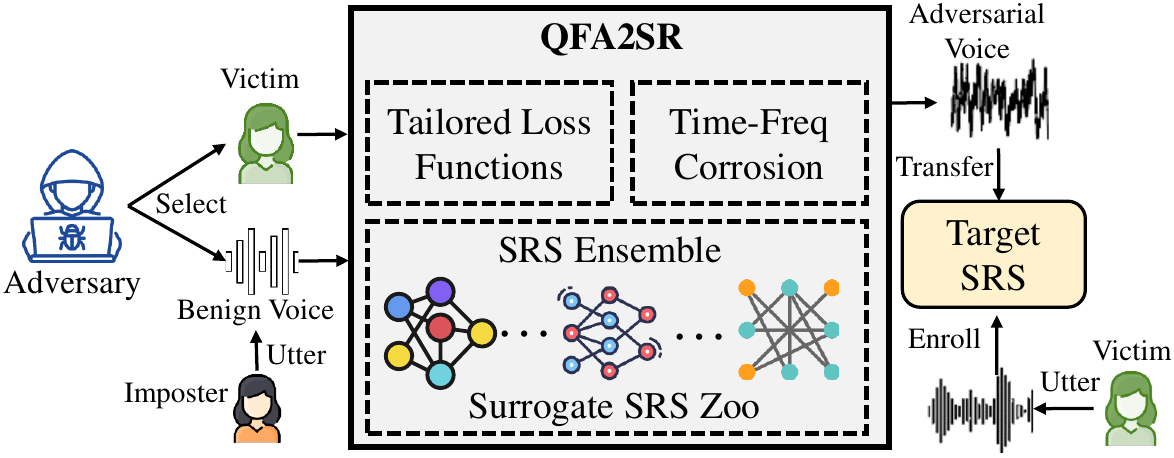}
    \caption{The overview of our attack.}
    \label{fig:attack-overview}\vspace{-3mm}
\end{figure}

In this work, we propose an effective and imperceptible adversarial transfer attack
on SRSs, named \toolname, addressing all the above three challenges.
The overview of  \toolname is depicted in \figurename~\ref{fig:attack-overview},
which consists of three key components: tailored loss functions,
time-frequency (time-freq) corrosion and SRS ensemble,
designed to increase the transferability without sacrificing imperceptibility,
where the latter two  are proposed to address the above three challenges.

\noindent
{\bf Tailored loss functions}.
We study and evaluate various loss functions for achieving the optimal transferability
for each attack scenario
(i.e., \OSITARGET, \OSIUNTARGET and \TDSVTARGET) (cf.~\cref{sec:loss}).
It is essential to explore different loss functions for improving the transferability,
as their effectiveness may vary in different attack scenarios (cf. Appendix~\ref{sec:loss-tailor-exper}).
The evaluation leads to the best tailored loss function for each attack scenario.

\noindent
{\bf SRS ensemble}.
Inspired by the ensemble-based approach for improving the transferability in the image domain~\cite{LCLS17},
we propose SRS ensemble (cf.~\cref{sec:srs-diversify}),
which builds a surrogate SRS zoo with multiple surrogate SRSs.
To alleviate the overfitting problem of adversarial voices
to a single surrogate SRS, adversarial voices are crafted to fool as many as surrogate SRSs simultaneously,
so they will be more transferable to an unknown target SRS.
We emphasize that our SRS ensemble differs from the one in the image domain~\cite{LCLS17} (cf.~\cref{sec:srs-diversify}).

\noindent
{\bf Time-freq corrosion}.
We propose time-frequency corrosion (cf.~\cref{sec:time-freq-corr}),
which randomly manipulates voice signals in the time domain
and acoustic features in the frequency domain using well-designed modification functions.
These  functions are inserted into proper positions of the surrogate SRSs
(before the acoustic feature extraction for time-domain modification functions
and after the acoustic feature extraction for frequency-domain modification functions).
During the generation of adversarial voices, intermediate voices are randomly modified
in both the time and frequency domains. 
Each modification function is intentionally designed to be random
and changes the distribution of the surrogate SRS in a different way,
thus we can simulate and approximate as many distributions as possible.
The adversarial voices crafted in this way will be more robust
against different distributions (e.g., the unknown distribution of the target SRS) and
the distortions introduced during over-the-air attacks,
thus  more transferable to an unknown target SRS even being played over the air.

\section{Design of \toolname}\label{sec:our-attack}
In this section, we present the details of our attack \toolname.

\subsection{Tailored Loss Functions}\label{sec:loss}
We study various loss functions for the attack scenarios \OSITARGET, \OSIUNTARGET, and \TDSVTARGET
whose effectiveness will be thoroughly evaluated to choose the best one for better transferability.

\noindent{\bf Attack scenario \OSITARGET}.
Given a benign voice uttered by an imposter $s$,
the adversarial attack in  \OSITARGET aims to craft a voice such that the OSI SRS recognizes it being uttered
by a given target speaker $t\in G$. We define the following loss functions:  
\begin{align}
   f_{\text{CE}}(x) &= -\log[{\Softmax}(S(x))]_t \qquad f_{1}(x) = -[S(x)]_t \nonumber \\
    f_{\text{M}}(x) &= {\tt max}_{i\in G,i\neq t}[S(x)]_i - [S(x)]_t  \nonumber  \\
    f_{2}(x) &= {\tt max}\{\theta, {\tt max}_{i\in G,i\neq t}[S(x)]_i\} - [S(x)]_t \nonumber
\end{align}
where $\theta$ is a preset score threshold,
 $f_{\text{CE}}$ and $f_{\text{M}}$ are respectively the Cross Entropy Loss~\cite{goodfellow2014explaining}
and the Margin Loss~\cite{carlini2017towards} that have been widely used to craft adversarial images.
$f_1$ is designed to increase the score of the target speaker $t$ \emph{only},
in contrast to the loss function $f_\text{M}$ which is designed to simultaneously increase the score of the target speaker $t$
and reduce the scores of the other enrolled speakers.
$f_2$ is designed such that $f_2(x)\leq 0 \Leftrightarrow D(x)=t$,
when minimized, the score $[S(x)]_t$ of the target speaker $t$ is maximized to exceed  $\theta$ and the scores of all the other enrolled speakers.
Note that $\theta$ is the threshold of the surrogate SRS, which is known to the adversary.

\noindent{\bf Attack scenario \OSIUNTARGET}.
Given a benign voice $x_0$ uttered by an imposter $s'$,
the adversarial attack in  \OSIUNTARGET
aims to craft a voice
such that it is accepted as an arbitrary enrolled speaker $t\in G$ by the OSI SRS.
We define the following loss functions:
\begin{align}
f_{\text{CE}}^{s}(x) &= -\log[\Softmax(S(x))]_s \nonumber \\
f_{2}^{s}(x) &= {\tt max}\{\theta, {\tt max}_{i\in G,i\neq s}[S(x)]_i\} - [S(x)]_s \nonumber \\
f_{\text{M}}^{s}(x) &= {\tt max}_{i\in G,i\neq s}[S(x)]_i - [S(x)]_s  \nonumber \\
f_{1}^{s}(x) &= -[S(x)]_s \qquad f_{3}(x) = \theta-{\tt max}_{i\in G}[S(x)]_i \nonumber
\end{align}
where $s=\arg\max_{i}[S(x_0)]_i$ and $x_0$ is the input voice.
The loss functions $f_{\text{CE}}^{s}(x)$,
$f_{1}^{s}(x)$, $f_{\text{M}}^{s}(x)$ and $f_{2}^{s}(x)$
are defined the same as  $f_{\text{CE}}(x)$,
$f_{1}(x)$, $f_{\text{M}}(x)$ and $f_{2}(x)$ except that
the enrolled speaker $s$ giving the maximal score on
the \emph{input} voice $x_0$ is used as the target speaker.
$f_{3}$ is designed such that $f_{3}(x)\leq 0\Leftrightarrow D(x)=\text{any enrolled speaker}$.
Minimizing $f_{3}$ makes the maximal score among all the enrolled speakers exceed the threshold $\theta$,
thus the adversarial voice is accepted. 
Unlike the others which always optimize towards the speaker $s$ that gives the maximal score on the {\it input} voice $x_0$
throughout the optimization,
$f_{3}$ dynamically adjusts the target speaker based on the scores of each {\it intermediate} voice.

\noindent{\bf Attack scenario \TDSVTARGET}.
Given a benign voice uttered by an imposter,
the adversarial attack in \TDSVTARGET aims to
craft a voice that is recognized as the enrolled speaker by the SV SRS.
We consider the following two loss functions for this goal:
\begin{align}
  f_{\text{BCE}}(x) = -\log (\varphi(S(x))) \qquad f_{3B}(x) = \theta - S(x) \nonumber
\end{align}
where $\varphi$ denotes the sigmoid function. Intuitively, $f_{\text{BCE}}$ is the binary Cross Entropy Loss function,
and $f_{3B}(x)$ is the special case of $f_{3}(x)$ for the binary classification task SV.
We note that $f_{3B}(x)$ is also equivalent to the loss functions
$f_1$, $f_{M}$ and $f_2$ when only one speaker is enrolled.

\subsection{SRS Ensemble}\label{sec:srs-diversify}
\noindent{\bf Ensemble of multiple SRSs}.
To combine multiple SRSs,
a straightforward idea is to adopt {\it loss-level fusion}~\cite{LCLS17},
originally proposed for the ensemble of image classification models.
The loss-level fusion computes the loss of the ensemble model
using the weighted sum of losses of multiple SRSs.
Formally, the loss function of the ensemble model is defined as $f_{\tt ens}=\sum_{k=1}^{K}w_k\times f(x;R_k)$,
where $K$ is the number of surrogate models, $f(x;R_k)$ is the loss function of the $k$-th surrogate model $R_k$
with the ensemble weight $w_k$.

We tried uniform weights, i.e., $w_k=\frac{1}{K}$ for $k=1,\cdots,K$,
which works well for the ensemble of  multiple image classification models~\cite{LCLS17}.
However, it has limited effectiveness and sometimes even reduces the transferability
compared to the attack using a single surrogate SRS (cf. Appendix~\ref{sec:srs-diverse-exper}),
probably because  different SRSs produce scores with different ranges and scales.
For example, the scoring method PLDA produces unconstrained scores,
while COSS outputs scores within the range $[-1, 1]$.
The loss function also varies with SRSs in the range and scale,
due to its dependency on the scores.
Thus, uniform weights cause the optimization to
concentrate more on the SRSs with large losses than the SRSs with small losses,
 definitely reducing the effect of SRS ensemble.
An intuitive way to address this issue is to treat the weights as hyper-parameters
and manually tune them. But, searching for (approximately) optimal
weights is prohibitively expensive and  difficult with the increase of surrogate SRSs~\cite{multi-task-learning}.
Moreover, the weights obtained via  tuning
depend on both the surrogate and subjunctive target SRSs,
thus may have to be re-tuned when
either the surrogate SRSs or target SRSs change.

We propose to craft adversarial voices using multiple surrogate SRSs
as multi-task learning~\cite{multi-task-learning} and use the following method
to automatically and adaptively choose appropriate weights {(called dynamic weighting)} for balancing different loss terms.
During each iteration of crafting adversarial voices, we normalize the loss of the $k$-th SRS $f_k=f(x;R_k)$
by its mean ${\mu_k}$ and standard derivative ${\sigma_k}$,
i.e., $f'_k=\frac{f_k-\mu_k}{\sqrt{\sigma_k}}$.
Remark that both ${\mu_k}$ and ${\sigma_k}$ are SRS-specific and are iteratively updated via
$\mu_k=\mu_k+\frac{f_k-\mu_k}{n}$ and $\sigma_k=\sigma_k+\frac{1}{n}((f_k-\mu_k)^2-\sigma_k)$,
where $n$ is the current iteration.
Finally, the loss function of the ensemble model is defined as  $f_{\tt ens} = \sum_{k=1}^{K}f'_k$.

\noindent{\bf Global ranking for untargeted attack}.
We now  face another problem when combining the surrogate SRSs
for untargeted attack (i.e., scenario \OSIUNTARGET).
Recall that the loss functions for \OSIUNTARGET (i.e., $f_{1}^{s}$, $f_{\text{M}}^{s}$, $f_{2}^{s}$ and $f_{3}$)
depend on the maximal score among the enrolled speakers.
Due to the diversity, the ranking of the enrolled speakers according to their scores
on each surrogate SRS (called \emph{local rank}) may differ from that of the others.
If we solely use the maximal scores based on the local ranks in the loss functions,
the optimization directions on the surrogate SRSs may differ,
definitely reducing the effect of SRS ensemble.
This is in contrast to the targeted attack (i.e., \OSITARGET and \TDSVTARGET)
where the target speaker is the same in all the surrogate SRSs.
To solve this problem, instead of using local ranks,
we utilize the global rank which aggregates the local ranks of all the surrogate SRSs.
We define the following two different global ranks, i.e., summation and voting.

Consider the surrogate SRS zoo $\{R_1,\cdots,R_K\}$
that has the same group $G$ of enrolled speakers.
Let $\rank_{k,x}$ be the \emph{local rank} of the SRS $R_k$ on a voice $x$ that maps enrolled speakers
to ranks according to their scores, i.e.,
speaker $i\in G$ has the $\rank_{k,x}(i)$-th maximal score
in the score vector $S(x)$ of the SRS $R_k$.
We define the \emph{summation-based global ranking} (\SUMGLOBAL) as
$\rank_{x}(i)=\sum_{k=1}^K \rank_{k,x}(i)$, where the global rank of the speaker $i$ is $\rank_{x}(i)$.
We define the \emph{voting-based global ranking} (\VOTEGLOBAL) as
$\rank_{x}(i)=\arg\max_{j\in G\setminus\{\rank_{x}(1),\cdots,\rank_{x}(i-1)\}}
\sum_{1\leq k\leq K}{\mathbb{I}(\rank_{k,x}(j)\leq i)}$,
where $\mathbb{I}(\rank_{k,x}(j)\leq i)$ is $1$ if $\rank_{k,x}(j)\leq i$ otherwise $0$.

The loss functions $f_{\text{CE}}^{s}(x)$,
$f_{1}^{s}(x)$, $f_{\text{M}}^{s}(x)$, $f_{2}^{s}(x)$ and $f_{3}$ for untargeted attacks against
the ensemble of the surrogate SRSs
are now generalized as follows:\\
\begin{align}
  f_{\text{CE}}^{s}(x) &= -\log[\Softmax(S(x))]_s \qquad    f_{1}^{s}(x) = -[S(x)]_s \nonumber \\
f_{2}^{s}(x) &= {\tt max}\{\theta, {\tt max}_{i\in G,i\neq s}[S(x)]_i\} - [S(x)]_s \nonumber \\
f_{\text{M}}^{s}(x) &= {\tt max}_{i\in G,i\neq s}[S(x)]_i - [S(x)]_s \quad f_{3}(x) = \theta- [S(x)]_{s'} \nonumber
\end{align}
where $x_0$ is the input voice,
$s=\arg\min_{i}[\rank_{x_0}(i)]$ and $s'=\arg\min_{i}[\rank_{x}(i)]$ for \SUMGLOBAL,
and $s=\rank_{x_0}(1)$ and $s'=\rank_{x}(1)$ for \VOTEGLOBAL.
Finally, the loss function of the ensemble model
$f_{\tt ens} = \sum_{k=1}^{K}f'_k$ is defined the same as above.
{Remark that the above loss functions, adapted by replacing
the local rank with a global rank,
only differ from the original loss functions for \OSIUNTARGET in the enrolled speaker (i.e., $s$ or $s'$).
}

\noindent
{\bf Attack with multiple SRSs}.
An  adversarial attack with SRS ensemble is shown in Alg.~\ref{al:SRS-diversify}.
It first initializes the means $\mu$ and derivatives $\sigma$ for each surrogate SRS (Line~\ref{alg-srs-diver-nor-init}).
The second for-loop (Lines \ref{alg-srs-diver-nor-2ndforstart}--\ref{alg-srs-diver-nor-2ndforend}) iteratively searches for an adversarial voice
starting from a seed voice $x_0$.
In each iteration,
the third for-loop (Lines \ref{alg-srs-diver-nor-3rdforstart}--\ref{alg-srs-diver-nor-3rdforend}) iteratively computes the loss of each surrogate SRS,
normalizes them by their individual means and standard derivatives,
based on which we compute SRS ensemble loss $f_{\tt ens}$ as the sum of these normalized losses (Line~\ref{alg-srs-diver-ens}).
Since the surrogate SRSs may introduce some randomness (e.g., the randomized pre-processing),
we independently draw a randomness $r$ for each surrogate SRS $\beta$ times (Lines \ref{alg-srs-diver-nor-4thforstart}--\ref{alg-srs-diver-nor-4thforend})
and obtain the loss using the average of the $\beta$ losses (Line~\ref{alg-srs-diver-avg}).
We found that this leads to better transferability.
Next, the intermediate voice $x_{n-1}$ is updated according to
the gradient sign of $f_{\tt ens}$ w.r.t. $x_{n-1}$ and the step size $\alpha$ (Line~\ref{alg-srs-diver-update}),
which is further clipped into the $L_\infty$ $\varepsilon$-neighbourhood of the seed
$x_0$ and the valid range of voices $[-1,1]$ (Line~\ref{alg-srs-diver-clip}).

\begin{figure}[t]\footnotesize\removelatexerror
\begin{algorithm*}[H]
      \caption{SRS Ensemble}
      \label{al:SRS-diversify}
      \KwIn{seed voice $x_0$; $L_\infty$ perturbation budget $\varepsilon$; number of steps $N$;  step size $\alpha$;
      surrogate SRS zoo $\{R_1,\cdots,R_K\}$; sampling size $\beta$;
      loss function $f(\cdot)$}
      \KwOut{adversarial voice}
      \lFor{$k$ from $1$ to $K$} {$\mu_k\gets 0$; $\sigma_k\gets 1$} \label{alg-srs-diver-nor-init}
      \For{$n$ from $1$ to $N$}{ \label{alg-srs-diver-nor-2ndforstart}
          $f_{\tt ens}\gets 0$\;
          \For{$k$ from $1$ to $K$}{ \label{alg-srs-diver-nor-3rdforstart}
              $f_k\gets 0$\;
              \For(\Comment{\textcolor{goldenrod}{$R_k^r$ denotes the SRS $R_k$ with}}){$r$ from $1$ to $\beta$}{  \label{alg-srs-diver-nor-4thforstart}
                $f_k\gets f_k+f(x_{n-1};R_k^r)$; \Comment{\textcolor{goldenrod}{the sampled randomness $r$}}
              }\label{alg-srs-diver-nor-4thforend}
              $f_k\gets\frac{f_k}{\beta}$\; \label{alg-srs-diver-avg}
              $\mu_k\gets \mu_k+\frac{f_k-\mu_k}{n}$; \label{alg-srs-diver-nor-1}
              $\sigma_k\gets \sigma_k+\frac{1}{n}((f_k-\mu_k)^2-\sigma_k)$\; \label{alg-srs-diver-nor-2}
              $f_{\tt ens} \gets f_{\tt ens} + \frac{f_k-\mu_k}{\sqrt{\sigma_k}}$; \label{alg-srs-diver-ens} \textcolor{goldenrod}{\Comment{SRS ensemble loss}} 
          } \label{alg-srs-diver-nor-3rdforend}
          $x_n \gets x_{n-1} -\alpha\times {\tt sign}(\nabla_{x_{n-1}}f_{\tt ens})$\; \label{alg-srs-diver-update}
          $x_n \gets \max\{\min\{x_n,1,x_0+\varepsilon\},-1,x_0-\varepsilon\}$\; \label{alg-srs-diver-clip}
      } \label{alg-srs-diver-nor-2ndforend}
      \Return{$x_{N}$}\;
  \end{algorithm*}\vspace{-3mm}
\end{figure}

\subsection{Time-Freq Corrosion}\label{sec:time-freq-corr}
Due to the time-varying non-stationary property,
voices are not resilient enough to noises and other variations,
and waveform signals themselves cannot effectively represent speaker characteristics~\cite{voice-feature-review}.
Thus, to achieve better
performance~\cite{xiao2016speech},
a raw input voice is often transformed into a two-dimensional time-frequency representation via an acoustic feature extraction (cf. \figurename~\ref{fig:typical-SRSs}).
This motivates us to design functions for manipulating voices in both the time and frequency domains.

\subsubsection{Time Domain Modification Functions}
We consider the following five modification functions for manipulating
voice signals in the time domain.

 \noindent {\bf Reverberation-distortion (RD)}~\cite{reverber}.
Reverberation occurs when a signal propagates through multiple paths (direct and reflected paths) in a room,
where the direct sound and reflections overlap with each other.
Room Impulse Response (RIR), denoted by $r$, can characterize the acoustic properties
of a room regarding sound transmission and reflection.
Given an input voice $x$, the reverberant voice is created by convolving $r$ with $x$.
Given a list of RIRs ($\mathcal{R}$),
each of which models a room configuration, RD randomly applies one RIR each step.

 \noindent {\bf Noise-flooding (NF)}~\cite{DeepSpeech}.
NF modifies a voice by superimposing it with a random white Gaussian noise.
The magnitude of the noise is controlled via the
signal-to-noise ratio (SNR) $10\log_{10}\frac{P_v}{P_n}$, where $P_v$ and $P_n$
are the power of the input voice and the noise, respectively.
The SNR is randomly chosen from $[\text{SNR}_l,\text{SNR}_h]$ during each step,
where SNR$_l$ and SNR$_h$ denote the lower bound and upper bound of the SNR.

 \noindent {\bf Speed-alteration (SA)}~\cite{speed-aug}.
Given a voice $x(t)$ and the speed ratio $\alpha$ between the new and original speeds,
SA produces the time-scaled voice $x(\alpha t)$,
which sounds faster (resp. slower) when $\alpha>1$ (resp. $\alpha<1$).
SA changes the duration of the utterance,
thus affects the number of frames of the voice.
The randomized version of SA selects a one-speed ratio
from a candidate list of speed ratios ($\mathcal{A}$) each step.

 \noindent {\bf Chunk-dropping (CD)}~\cite{SpeechBrain}.
Given a voice with $T$ sample points,
CD sets the magnitudes of the sample points within $[t_0,t_0+t)$ to zero,
where $t$ and $t_0$ are randomly chosen from $[T_l,T_u]$ and $[0, T-t]$, respectively.
$T_l$ and $T_u$ are the lower and upper bounds of the chunk lengths to be dropped.
In addition, given the lower $C_l$ and upper $C_u$ bounds of the number of the chunks to be dropped,
the process is independently repeated $c$ times
where $c$ is randomly selected from $[C_l, C_u]$.

 \noindent {\bf Frequency-dropping (FD)}~\cite{SpeechBrain}.
A voice signal can be decomposed into multiple pure tones with different frequencies.
Given the lower $F_l$ and upper $F_u$ bounds of the frequencies to be dropped,
FD applies a notch filter to remove the pure tone with  frequency $f$ which is randomly chosen from $[F_l,F_u]$.
This process can also be independently repeated $c$ times
for a randomly chosen number $c$ between the lower $C_l$ and upper  $C_u$ bounds of the number of the frequencies to be dropped.

\subsubsection{Frequency Domain Modification Functions}
We consider three modification functions for manipulating
voice signals in the frequency domain which was used for voice data augmentation in~\cite{SpecAugment}.
We denote by $M\in \mathbb{R}^{T\times F}$ the acoustic feature matrix,
where $T$ and $F$ are the number of time frames and frequency channels, respectively.

 \noindent {\bf Time-warping (TW)}.
TW introduces deformations in the time frame dimension of $M$.
First, an entry $p$ of $M$ is selected such that its horizontal coordinate is the center and the vertical coordinate $P$ is randomly chosen from $[W, T-W]$ where $W$ is the time warping parameter.
Then, the sub-region above the horizontal line passing $p$ with the size $P\times F$ is scaled to the size $w\times F$,
while the sub-region below the horizontal line passing $p$ with the size $(T-P)\times F$ is scaled to the size $(T-w)\times F$,
where $w$ is randomly chosen from $[P-W, P+W]$.

 \noindent {\bf Time-masking (TM)}.
TM introduces deformations in the time frame dimension of $M$ by applying zero masking to
$t$ consecutive time frames $[t_0,t_0+t)$ where $t$ is randomly chosen from $[0,t']$ for a given TM parameter $t'$ and $t_0$ is randomly chosen from $[0, T-t]$.

 \noindent {\bf Frequency-masking (FM)}.
FM introduces deformations in the frequency channel dimension of $M$
by replacing the coefficients of $f$ consecutive frequency channels $[f_0,f_0+f)$ with $0$
where $f$ is randomly chosen from $[0,f']$ for a given FM parameter $f'$,
and $f_0$ is randomly chosen from $[0, F - f]$.

\begin{table*}
	\centering\setlength{\tabcolsep}{1mm}
	\caption{{Details of Datasets.}}
	\resizebox{1\textwidth}{!}{\begin{threeparttable}
	\begin{tabular}{|c|c|c|c|}
		\hline
	{\bf Name} 	 &  {\bf \#Voices} & \makecell[c]{{\bf Voice Source \& Attack Scenario}} & {\bf Description} \\
		\hline
		\textbf{Spk$_{10}$-enroll-P$_1$} & 10 $\times$ 5 & \multirow{4}{*}{\makecell[c]{Provided by  SEC4SR~\cite{SEC4SR} that are \\   derived  from  LibriSpeech~\cite{panayotov2015librispeech}, \\  for \OSITARGET and \OSIUNTARGET}}
		& \makecell[c]{Used to enroll OSI surrogate and target SRSs for same-enroll} \\
		\cline{1-2}\cline{4-4}
		\textbf{Spk$_{10}$-enroll-P$_2$} & 10 $\times$ 5 & & \makecell[c]{Same speakers but different voices as Spk$_{10}$-enroll-P$_1$, used to enroll OSI target SRSs for differ-enroll} \\
		\cline{1-2}\cline{4-4}
		\textbf{Spk$_{10}$-test} & 10 $\times$ 100 & & \makecell[c]{Same speakers but different voices as Spk$_{10}$-enroll-P$_1$\&P$_2$, used to test the performance of SRSs} \\
		\cline{1-2}\cline{4-4}
		\textbf{Spk$_{10}$-imposter} & 10 $\times$ 100 & & \makecell[c]{Speakers different from Spk$_{10}$-test, used to test the performance of SRSs and as seeds for crafting adversarial voices} \\
		\hline
		\textbf{Spk$_5$-TD-P$_1$} & 5 $\times$ 10 $\times$ 4 & \multirow{2}{*}{\makecell[c]{Recruiting  volunteers \\ to record voices for \TDSVTARGET}} & \makecell[c]{Ten different sentences in Appendix~\ref{sec:TD-phrase}, used to enroll TD-SV target SRSs and  as seeds for crafting adversarial voices} \\
		\cline{1-2}\cline{4-4}
		\textbf{Spk$_5$-TD-P$_2$} & 5 $\times$ 10 $\times$ 1 &  & \makecell[c]{Ten sentences from \cite{wenger2021hello}, same speakers but different texts as Spk$_5$-TD-P$_1$, used to enroll surrogate SRSs} \\
		\hline
		\textbf{Spk$_{1000}$-enroll-P$_1$} & 1000 $\times$ 5 & \multirow{2}{*}{\makecell[c]{Derived from LibriSpeech, \\ for \OSITARGET and \OSIUNTARGET}}
		& \makecell[c]{Expand Spk$_{10}$-enroll-P$_1$ with 990 speakers, used to enroll OSI surrogate and target SRSs for same-enroll} \\
		\cline{1-2}\cline{4-4}
		\textbf{Spk$_{1000}$-enroll-P$_2$} & 1000 $\times$ 5 & & \makecell[c]{Same speakers but different voices as Spk$_{1000}$-enroll-P$_1$, used to enroll OSI target SRSs for differ-enroll} \\ 
		\hline
	\end{tabular}
	\begin{tablenotes}
		\item Note: In the \#Voices column, x $\times$ y denotes x speakers and y voices per speaker,
		and x $\times$ y $\times$ z denotes x speakers, y different texts, and z voices per text and per speaker.
	\end{tablenotes}
\end{threeparttable}
	}
	\label{tab:dataset-detail}
\end{table*}

\subsubsection{Combination of Modification Functions}
We explore the combinations of the above modification functions to improve transferability.
Denoting the individual modification functions by $m_1,\cdots,m_K$,
we consider two combination strategies: {\it serial} and {\it parallel}.
Consider the indices $i_1,\cdots,i_K\in \{1,\cdots,K\}$
such that if $m_{i_j}$ is a time domain modification function,
then $m_{i_{1}},\cdots, m_{i_{j-1}}$ are all time domain modification functions.
The \emph{serial composite modification function} $M_s(\cdot)=m_{i_K}(m_{i_{K-1}}(\cdots m_{i_1}(\cdot)))$
 sequentially applies the functions $m_{i_{1}},\cdots, m_{i_{K}}$
either at signal-level or feature-level depending on the function.
{$M_s$ is achieved by building the \emph{simulated} SRS $R_{M_s}$ from a given surrogate SRS $R$
where all the modification functions $m_i$ of $M_s$ are inserted at proper positions,
i.e., before (resp. after) the acoustic extraction module if $m_i$ is a time (resp. frequency)
domain function.}
The \emph{parallel composite modification function} $M_p(\cdot)=M_1||\cdots ||M_K$
modifies an input voice by applying the functions $M_1,\cdots,M_K$ in parallel,
leading to $K$ different modified voices and $K$ loss values $v_1,\cdots,v_K$.
Note that $M_i$ in $M_p$ could be a serial composite modification function.
{$M_p$ is achieved by building $K$ \emph{simulated} SRSs $\{R_{M_1},\cdots, R_{M_K}\}$ from a given surrogate SRS $R$
where $R_{M_i}$ is the simulated SRS of the surrogate SRS $R$ for the modification functions $M_i$.
The $K$ \emph{simulated} SRSs $\{R_{M_1},\cdots, R_{M_K}\}$ can be combined using our ensemble method (cf. \cref{sec:srs-diversify}).}
In this work,
we consider three serial composite modification functions:
RD+NF,
SA+FD+CD,
and TW+TM+FM.
 For {\it parallel} combination,
{we consider the combination of these three serial composite functions.}
We leave other composite functions as future work.

\begin{figure}[t]\footnotesize\removelatexerror
\begin{algorithm*}[H]
  \caption{\toolname}
  \label{al:final-attack}
  \KwIn{seed voice $x_0$;  modification functions $\mathcal{M}=\{\cdots,M_j,\cdots\}$;
  sampling size $\beta$; surrogate SRS zoo $\mathcal{R}=\{\cdots,R_i,\cdots\}$;
  number of steps $N$; the step size $\alpha$; $L_\infty$ perturbation budget $\varepsilon$;
  the optimal loss function for the attack scenario $f_{\tt opt}(\cdot)$}
  \KwOut{adversarial voice $x^{adv}$}
$\mathcal{Z}\gets \{ R_M\mid R\in \mathcal{R}, M\in \mathcal{M}\}$ if $\mathcal{M}\neq \emptyset$ else $\mathcal{R}$; \\
  $x^{adv}\gets$ invoke Alg.~\ref{al:SRS-diversify} with the surrogate SRS zoo $\mathcal{Z}$
  and parameters ($x_0$, $\beta$, $N$, $\alpha$, $\varepsilon$, and $f_{\tt opt}(\cdot)$)\; \label{alg-QFA2SR-invoke}
  \Return{$x^{adv}$}\;
\end{algorithm*}
\end{figure}

\subsection{\toolname: Our Final Attack}\label{sec:attack-final}

\toolname for one seed voice is shown in Alg.~\ref{al:final-attack}, {where $\mathcal{M}$
is a set of (serial composite) modification functions $\{M_1,\cdots, M_K\}$ and used as
a parallel composite modification function $M_p(\cdot)=M_1||\cdots ||M_K$ if $k\geq 2$.
Alg.~\ref{al:final-attack} first builds a \emph{simulated} surrogate SRS zoo $\mathcal{Z}$ by combining each surrogate SRS $R\in \mathcal{R}$ with each modification function $M\in \mathcal{M}$,
to get rid of modification functions.
Then, it invokes Alg.~\ref{al:SRS-diversify} with $\mathcal{Z}$ as the surrogate SRS zoo
and other necessary input parameters to craft an adversarial voice.
We note that the surrogate SRSs $\{ R_M\mid R\in \mathcal{R}, M\in \mathcal{M}\}$
are combined using our ensemble method (cf. \cref{sec:srs-diversify}) when crafting
adversarial voices.}

In practice, the adversary can generate many adversarial voices
but can only query the target SRS few times during transfer attack.
Thus, we propose a method to select the adversarial voices
which are the most likely transferable to the target SRS,
thus largely improves the success rate of QFA2SR with few allowed queries.
Details refer to Appendix~\ref{sec:generalized-attack}.

\begin{table*}[htbp]
  \centering\setlength{\tabcolsep}{.6pt}
 \begin{minipage}{0.508\textwidth}\centering
  \caption{Results on commercial APIs in \OSITARGET.}
 \resizebox{1.0\textwidth}{!}{
    \begin{tabular}{|c|c|c|c|c|c|c|c|c|c|c|c|c|}
    \hline
    \multirow{3}[6]{*}{} & \multicolumn{4}{c|}{\textbf{Microsoft Azure}} & \multicolumn{4}{c|}{\textbf{TalentedSoft}} & \multicolumn{4}{c|}{\textbf{IFlytek}} \\
\cline{2-13}
 & \textbf{ASR$_t$-s} & \textbf{ASR$_t$-d} & \textbf{SNR}&   \textbf{PESQ}    & \textbf{ASR$_t$-s} & \textbf{ASR$_t$-d}  & \textbf{SNR} &    \textbf{PESQ}   & \textbf{ASR$_t$-s} & \textbf{ASR$_t$-d}  & \textbf{SNR} & \textbf{PESQ} \\
    \hline
    \textbf{SirenAttack} & 1     & 2.1   & 8.02  & 1.12  & 1.4   & 1.3   & 10.07 & 1.18  & 0     & 0     & 8     & 1.12 \\
    \hline
    \textbf{Kenansville} & 0     & 0     & {\bf 16.23} & {\bf 1.75}  & 0     & 0     & {\bf 16.23} & {\bf 1.75}  & 0     & 0     & {\bf 16.23} & {\bf 1.75} \\
    \hline
     \textbf{FakeBob} & 4.2   & 3.1   & 12.23 & 1.22  & 5.0   & 2.4   & 12.50 & 1.23  & 0   & 0   & 12.16 & 1.24 \\
     \hline
     \textbf{FakeBob} + {\small {\Circled[outer color=green]{1}}} & 6.2   & 4.1   & 12.23 & 1.23  & 5.6   & 2.7   & 12.51 & 1.24  & 1.9   & 1.9   & 12.16 & 1.23 \\
     \hline
     \textbf{FakeBob} + {\small {\Circled[outer color=green]{1}}}  {\small {\Circled[outer color=blue]{2}}} & 17.5 & 17.2 & 12.22 & 1.24  & 9.3 & 4.7 & 12.22 & 1.24  &  9.1 & 8.8 & 12.22 & 1.24 \\
     \hline
     \makecell[c]{\textbf{FakeBob} + {\small {\Circled[outer color=green]{1}}}  {\small {\Circled[outer color=blue]{2}}}  {\small {\Circled[outer color=red]{3}}}} & 3.8 & 2.7 & 12.71 & 1.28  & 4.0 & 2.5 & 12.71 & 1.28  &  0.6 & 0.6 & 12.71 & 1.28 \\
     \hline
    \textbf{BIM} & 18.9  & 12.7  & 11.49  & 1.18  & 8.9   & 6.5   & 11.28  & 1.19  & 16    & 15.5  & 11.50  & 1.18  \\
    \hline
    \makecell[c]{\textbf{BIM} +  {\small {\Circled[outer color=green]{1}}}} & 27.2  & 21.8  & 11.50  & 1.18  & 9.3   & 6.6   & 11.28  & 1.19  & 24    & 17.5  & 11.52  & 1.19  \\
    \hline
    \makecell[c]{\textbf{BIM} + {\small {\Circled[outer color=green]{1}}}  {\small {\Circled[outer color=blue]{2}}}} & 42.8 & 34.2 & 11.29 & 1.18 & 16.9 & 12.5 & 11.29 & 1.18 & 25.9 & 21.6 & 11.29 & 1.18 \\
    \hline
    \makecell[c]{\textbf{BIM} + {\small {\Circled[outer color=green]{1}}}  {\small {\Circled[outer color=blue]{2}}}  {\small {\Circled[outer color=red]{3}}} \\ (\textbf{\toolname})}
    & \makecell[c]{\textbf{89.6} \\ $\uparrow$ 70.7} & \makecell[c]{\textbf{82.8} \\ $\uparrow$ 70.1} & 10.85  & 1.18
    & \makecell[c]{\textbf{40.1} \\ $\uparrow$ 31.2} & \makecell[c]{\textbf{27.4} \\ $\uparrow$ 20.9} & 10.85  & 1.18
    & \makecell[c]{\textbf{46.1} \\ $\uparrow$ 30.1} & \makecell[c]{\textbf{39.5} \\ $\uparrow$ 24} & 10.85  & 1.18  \\
    \hline
    \end{tabular}%
  }
  \label{tab:commercial-B1}
  \end{minipage}
  \begin{minipage}{0.485\textwidth}\centering
   \caption{Results of \toolname on commercial APIs  in \OSIUNTARGET.}
 \resizebox{1.0\textwidth}{!}{
    \begin{tabular}{|c|c|c|c|c|c|c|c|c|c|c|c|c|}
    \hline
  \multirow{3}[6]{*}{} & \multicolumn{4}{c|}{\textbf{Microsoft Azure}} & \multicolumn{4}{c|}{\textbf{TalentedSoft}} & \multicolumn{4}{c|}{\textbf{IFlytek}} \\
\cline{2-13}
 & \textbf{ASR$_u$-s} & \textbf{ASR$_u$-d} & \textbf{SNR}&   \textbf{PESQ}    & \textbf{ASR$_u$-s} & \textbf{ASR$_u$-d}  & \textbf{SNR} &    \textbf{PESQ}   & \textbf{ASR$_u$-s} & \textbf{ASR$_u$-d}  & \textbf{SNR} & \textbf{PESQ} \\
    \hline
    \textbf{SirenAttack} & 16.67 & 8.25  & 8.16  & 1.12  & 23.9  & 18.7  & 10.07 & 1.18  & 0     & 0     & 8.07  & 1.12 \\
    \hline
    \textbf{Kenansville} & 0     & 0     & {\bf 16.97} & {\bf 1.8}   & 7     & 4     & {\bf 17.58} & {\bf 1.84}  & 0     & 0     & {\bf 16.66} & {\bf 1.77} \\
    \hline
    \textbf{Hidden} & 21.4  & 23    & -2.84 & 1.14  & 22.9  & 21.9  & -2.9  & 1.18  & 0     & 0     & -2.95 & 1.15 \\
    \hline
    \textbf{FakeBob} & 33.33 & 15.46 & 12.24 & 1.23  & 26.8  & 24    & 12.41 & 1.24  & 11.5 & 5.8 & 12.12 & 1.23 \\
    \hline
    \textbf{FakeBob} + {\small {\Circled[outer color=green]{1}}} & 33.33 & 15.46 & 12.24 & 1.23  & 26.8  & 24    & 12.41 & 1.24  & 11.5  & 5.8 & 12.12 & 1.23 \\
    \hline
    \textbf{FakeBob} + {\small {\Circled[outer color=green]{1}}}  {\small {\Circled[outer color=blue]{2}}} & 47.92 & 37.11 & 12.22 & 1.22  & 31 & 26.7  & 12.22 & 1.22  & 19.2 & 13.5 & 12.22 & 1.22 \\
    \hline
    \textbf{FakeBob} + {\small {\Circled[outer color=green]{1}}}  {\small {\Circled[outer color=blue]{2}}}  {\small {\Circled[outer color=red]{3}}} & 15.42 & 6.41 & 12.55 & 1.27  & 11.7 & 7.2  & 12.55 & 1.27  & 5.0 & 2.7 & 12.55 & 1.27 \\
    \hline
    \textbf{BIM} & 61.22  & 47.21  & {11.55} & 1.18  & 17.8  & 16.2  & 11.37  & {1.18} & 60    & 58  & {11.53} & 1.17  \\
    \hline
    \textbf{BIM} + {\small {\Circled[outer color=green]{1}}} & 68.4  & 50.8  & {11.54} & 1.18  & 22.7  & 19.9  & 11.37  & {1.19} & 64    & 61.9  & {11.54} & 1.18  \\
    \hline
    \textbf{BIM} + {\small {\Circled[outer color=green]{1}}} {\small {\Circled[outer color=blue]{2}}} & 80.62  & 66.53  & 11.37  & {1.19} & 30.1  & 23.5  & 11.37  & {1.19} & 69    & 62.9  & 11.37  & {1.19}\\
    \hline
    \makecell[c]{\textbf{BIM} + {\small {\Circled[outer color=green]{1}}}  {\small {\Circled[outer color=blue]{2}}}  {\small {\Circled[outer color=red]{3}}} \\ (\textbf{\toolname})}
    & \makecell[c]{\textbf{99.49} \\ $\uparrow$ 38.27} & \makecell[c]{\textbf{92.39} \\ $\uparrow$ 45.18} & 11.01  & {1.19}
    & \makecell[c]{\textbf{55} \\ $\uparrow$ 28.2} & \makecell[c]{\textbf{39.6} \\ $\uparrow$ 15.6} & 11.01  & {1.19}
    & \makecell[c]{\textbf{70} \\ $\uparrow$ 10} & \makecell[c]{\textbf{68} \\ $\uparrow$ 10} & 11.01  & {1.19}\\
    \hline
    \end{tabular}%
  }
  \label{tab:commercial-B2}
  \end{minipage}
   \begin{tablenotes}\scriptsize
       \item Note: {\tiny {\Circled[outer color=green]{1}}}, {\tiny {\Circled[outer color=blue]{2}}}, {\tiny {\Circled[outer color=red]{3}}}
       denote Tailored Loss Functions, SRS Ensemble, and Time-Freq Corrosion, respectively.
       $\uparrow$ is the improvement  of \toolname over the most effective baseline.
    \end{tablenotes}
\end{table*}

\section{Evaluation of Attack}\label{sec:evaluation}

\subsection{Experimental Setting and Design}\label{sec:setting}

\noindent {\bf Enrollment settings.}
The enrollment voices used in the target SRS may be the same as (resp. different from)
that used by the adversary to enroll surrogate SRSs,
called {\emph{same-enroll}} (resp. {\emph{differ-enroll}}).
Note that all the prior works consider same-enroll \emph{only} which is less realistic than differ-enroll.
Here we consider both same-enroll and differ-enroll except for {text-dependent verification in the scenario \TDSVTARGET
for which we consider differ-enroll \emph{only} where the target speaker's voices available to the adversary do \emph{not} contain the desired text.}

\noindent {\bf Datasets.}
{Our evaluation is mainly based on eight datasets, the details of which are shown in \tablename~\ref{tab:dataset-detail}.}

\noindent{\bf SRSs}.
We use 9 open-source SRSs:
Ivector-PLDA (IV)~\cite{kaldi-ivector-plda},
\mbox{ECAPA-TDNN (ECAPA)~\cite{ECAPA-TDNN}},  Xvector-PLDA (XV-P)~\cite{kaldi-xvector-plda},
Xvector-COSS (XV-C)~\cite{SpeechBrain},
Resnet18 trained for OSI (Res18-I) and SV (Res18-V)~\cite{resnet18},
Resnet34 trained for OSI (Res34-I) and SV (Res34-V)~\cite{resnet34},
and AutoSpeech (Auto)~\cite{auto-speech}.
We also include four commercial APIs: (Microsoft) Azure~\cite{microsoft-azure-vpr}, TalentedSoft~\cite{Talentedsoft},
iFlytek~\cite{iFlytek}, and Jingdong~\cite{JingdongSRS},
and three voice assistants: Google Assistants~\cite{Google-Voice-Match},
Apple Siri~\cite{Hey-siri}, and TMall Genie~\cite{TMall-Genie}.
Details of these SRSs, and their threshold $\theta$ and performance are given
in Appendix~\ref{sec:srs-performance}.

\noindent{\bf Metrics}.
We use transfer attack success rate (ASR) to measure attack effectiveness,
and denote by ASR$_u$ (resp. ASR$_t$) the untargeted (resp. targeted) ASR,
which refers to the proportion of adversarial voices
that are misrecognized as any enrolled speakers (resp. target speaker) by the target SRS.
{Let ASR$_u$-s (resp.  ASR$_u$-d) denote ASR$_u$ under the same-enroll (resp. differ-enroll) setting.
ASR$_t$-s and ASR$_t$-d are defined similarly.
The ASR improvement $x\%$ by our attack compared over a baseline is calculated as $x=b-a$,
where b\% (resp. a\%) is the ASR of our attack (resp. baseline).}
To quantify imperceptibility, we use Signal-to-Noise Ratio (SNR)~\cite{SpeakerGuard} and
Perceptual Evaluation of Speech Quality (PESQ)~\cite{rix2001perceptual}.
SNR is defined as $10\log_{10}\frac{P_x}{P_\delta}$,
where $P_x$ and $P_\delta$ are the power of the original voice and the perturbation.
PESQ is an objective perceptual measure that simulates the human auditory system~\cite{xiang2017digital}.
Higher SNR/PESQ indicates better imperceptibility.

\noindent {\bf Experimental design.}
{We first summarize the results of tuning parameters of \toolname (\cref{sec:tuning-parameter-attack}).
We then evaluate \toolname on commercial APIs
where adversarial voices are directly fed to the exposed APIs as audio files (\cref{sec:commercial-APIs})
and voice assistants where adversarial voices are played over the air to attack voice assistants (\cref{sec:exper-voice-assistant}).
We finally study 
the effect of adversarial knowledge on the enrolled speakers of target SRSs (\cref{sec:impact-knowledge-enrolled-speakers}),
and  the attack scalability of \toolname (\cref{sec:scalability}).}

{\subsection{Tuning Parameters of \toolname}\label{sec:tuning-parameter-attack}
We tune the parameters of \toolname on open-source SRSs,
simulating a real-world adversary who tunes parameters within the surrogate SRS zoo,
and attacks commercial SRSs
in \cref{sec:commercial-APIs} and \cref{sec:exper-voice-assistant} using the resulting parameters.
Due to space limit, here we only summarize the results of parameter tuning.
Details are given in
Appendixes~\ref{sec:loss-tailor-exper},  \ref{sec:time-freq-curr-exper}, and \ref{sec:srs-diverse-exper},  respectviely.}

\noindent {\bf Tailored loss functions.}
The losses $f_1$ and $f_2$ are comparable and outperform the others for \OSITARGET,
$f_3$ in general performs better than the others for \OSIUNTARGET,
and $f_\text{BCE}$ and $f_{3B}$ have the same performance for \TDSVTARGET.
The comparison results among loss functions
keep consistent across different surrogate and target SRSs, and between a single surrogate SRS
and the ensemble of multiple SRSs with adapted losses (cf.~\cref{sec:srs-diversify}).
Thus, we will  use $f_1$ for \OSITARGET,
$f_3$ for \OSIUNTARGET, and $f_{3B}$ for \TDSVTARGET for all systems and settings,
rather than performing the comparison and selection repeatedly.
Note that $f_1$ is preferable than $f_2$
since $f_1$ depends \emph{only} on the score of the targeted speaker.

\noindent 
{\bf SRS ensemble.}
Our dynamic weighting outperforms 
 the uniform weighting used in the image domain~\cite{LCLS17}.
Summation-based global ranking (\SUMGLOBAL) and voting-based global ranking (\VOTEGLOBAL)
are comparable, and both of them
perform better than the local ranking.
Hence, we will use our dynamic weighting and \SUMGLOBAL for SRS ensemble.

\noindent 
{\bf Time-freq corrosion.}
All individual modification functions can improve transferability.
The  serial composite functions (RD+NF, SA+FD+CD, and TW+TM+FM)
achieve higher transferability than individual functions.
Their parallel combination yields the best transferability,
hence will be utilized as the default modification function for time-freq corrosion.

\subsection{\toolname against Commercial APIs}\label{sec:commercial-APIs}

\noindent {\bf Setting.}
For OSI task (i.e., \OSITARGET and \OSIUNTARGET),
we attack 3 commercial SRSs: Azure, TalentedSoft, and iFlytek.
For TD-SV task (i.e., \TDSVTARGET), we attack 2 commercial SRSs: Azure and Jingdong.
Note that Jingdong does not support OSI while TalentedSoft and iFlytek
do not support TD-SV.
For surrogate SRSs,
we only consider IV, ECAPA, XV-P, and XV-C
since they yield the best transferability in general
according to the results in Appendix~\ref{sec:srs-diverse-exper}.
We mainly compare \toolname with baselines: Basic-Iterative-Method (BIM)~\cite{BIM}, FakeBob, SirenAttack,
and Kenansville. Occam is not available and AS2T is based on BIM and FakeBob, thus are not compared.
In \OSIUNTARGET, we also compare with the hidden voice attack~\cite{AbdullahGPTBW19},
where 100 voices are randomly selected from Spk$_{10}$-test as the seed voices.
Note that the hidden voice attack can neither launch targeted attack
(cf.~\cref{sec:background})
nor change the speech content, so is not applicable to \OSITARGET and \TDSVTARGET.
In \TDSVTARGET, we also compare with the voice cloning attack using the few-shot voice cloning toolkit
Real-Time-Voice-Cloning~\cite{SV2TTS,SV2TTS-github}.
It produces a voice with the desired speech content given a set of the target speaker's voice samples
and a  speech content.
We use the voices in Spk$_{5}$-TD-P2 as voice samples
and the ten sentences from Azure (cf. Appendix~\ref{sec:TD-phrase}) as the desired contents.

We set $L_\infty$ perturbation budget $\varepsilon=0.02$, step size $\alpha=\frac{\varepsilon}{5}=0.004$,
number of steps $N=300$, and sampling size $\beta=5$ for \toolname,
and detailed setups of the baselines refer to Appendix~\ref{sec:detail-compared-attack}.
As we focus on query-free attacks (i.e., no query to target SRSs during adversarial voice generation),
all the baselines are used to launch transfer attacks.
We report the best {transferability} among different surrogate SRSs for them.

\noindent {\bf Results of scenario \OSITARGET.}
are shown in \tablename~\ref{tab:commercial-B1}.
\toolname achieves {20.9\%-70.7\%} higher ASR$_t$ than BIM which is generally the most effective one
among the baselines.
\toolname can achieve more than 82\% ASR$_t$ on Azure.

\noindent {\bf Results of scenario \OSIUNTARGET.}
The results for \OSIUNTARGET
are shown in \tablename~\ref{tab:commercial-B2}.
Compared with the most effective baseline,
\toolname improves the ASR$_u$ by {10\%-45.1\%},
achieving more than 92\% ASR$_u$
on Azure.
It also achieves much higher ASR$_u$ than the hidden voice attack,
probably because the least incomprehensible hidden voices
crafted with respect to the source SRS are difficult for the target to correctly recognize.

\noindent {\bf Results of scenario \TDSVTARGET.}
The results for \TDSVTARGET\
are shown in \tablename~\ref{tab:commercial-B3}.
Compared to the best baseline,
\toolname improves the ASR$_t$ by 48.85\% and 54\% on Azure
and Jingdong, respectively.
\toolname also achieves 26\%-51.86\% higher ASR$_t$ than the voice cloning attack.
It is because the voice cloning attack generates artificially fake voices,
which usually contain artifacts and suffer from low quality,
e.g., the characteristic prosody is lost~\cite{SV2TTS}.
As a result, the cloned voice does not have sufficient acoustic similarity
with the genuine enrollment voice of the target speaker and thus fails to bypass the SRS.
In contrast, \toolname starts from genuine voices of an imposter
and only add to them imperceptible perturbations that sound like background noise
to improve the score of the target speaker,
thus the adversarial voices crafted by \toolname have sufficient acoustic similarity to bypass the target SRS.

\noindent {\bf Imperceptibility.}
{In \OSITARGET, \OSIUNTARGET, and \TDSVTARGET,} \toolname has higher SNR and PESQ than SirenAttack and hidden voice attack.
Kenansville and FakeBob have better imperceptibility than \toolname,
but their transferability is too low to effectively
mislead the target SRS and thus far from being practical.
The SNR of \toolname is slightly lower than BIM, but PESQ is the same or even larger in most cases.
Note that PESQ is an objective perceptual measure that simulates human auditory system,
but SNR is not, we believe PESQ can better characterize the imperceptibility.

As SNR and PESQ may not fully measure human imperceptibility,
we conduct a human study on MTurk~\cite{amazon_mturk} with approval from the Institutional Review Board (IRB) of our institute.
The participants are presented with a pair of voices 
and asked to tell after listening whether they are uttered by the same speaker, 
provided with three options: {\it same}, {\it different}, and {\it not sure}. 
We compare the imperceptibility of \toolname with BIM 
and voice cloning attack, while other attacks are excluded since their transfer success rates
are too low to be practical. Furthermore, we conduct the human study in \TDSVTARGET
because voice cloning attack is text-dependent. 
Specifically, we build 24 pairs:
4 normal pairs (two clean voices from distinct speakers),
10 adversarial pairs (one adversarial voice from an imposter and one clean voice from the target speaker;
5 pairs are from \toolname and 5 pairs are from BIM),
and 10 cloning pairs (one voice generated by voice cloning and one clean voice from the target speaker).
To guarantee the quality of the answers and validity of the results,
we filter out the answers that are randomly chosen by participants.
In particular, we insert 6 special voice pairs (two clean voices from different speakers with opposite gender)
as the concentration test.
All the submitted answers from a participant will be excluded as long as
she/he does not choose the {\it different} option for any one of the special pairs.

\begin{figure}[t]
	\centering\subfigcapskip=-5pt
 \begin{minipage}[t]{0.22\textwidth}
   \centering
		\includegraphics[width=.85\textwidth]{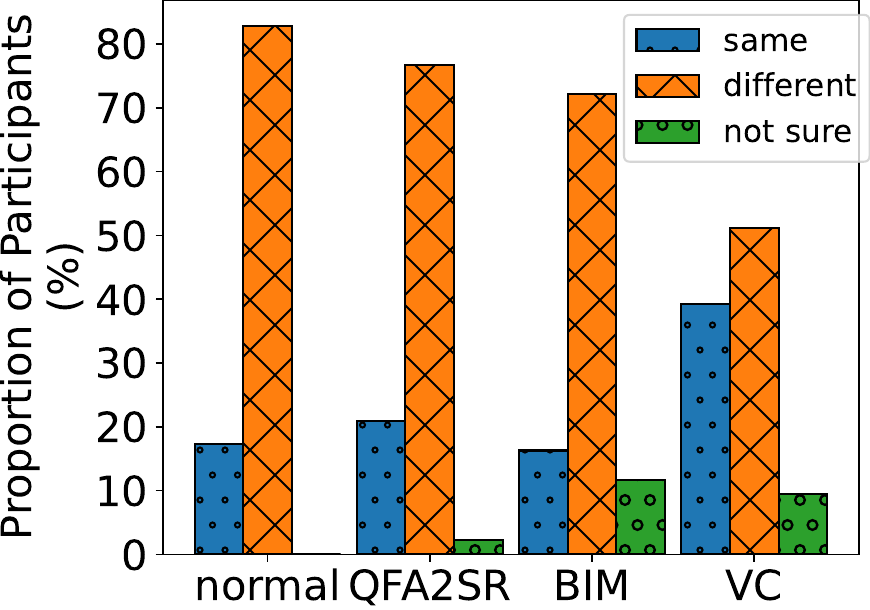}
 \caption{{Results of human study. VC=voice cloning.}}
 \label{fig:human-study}
 \end{minipage}\quad
 \begin{minipage}[t]{0.22\textwidth}
  \centering
  \includegraphics[width=.9\textwidth]{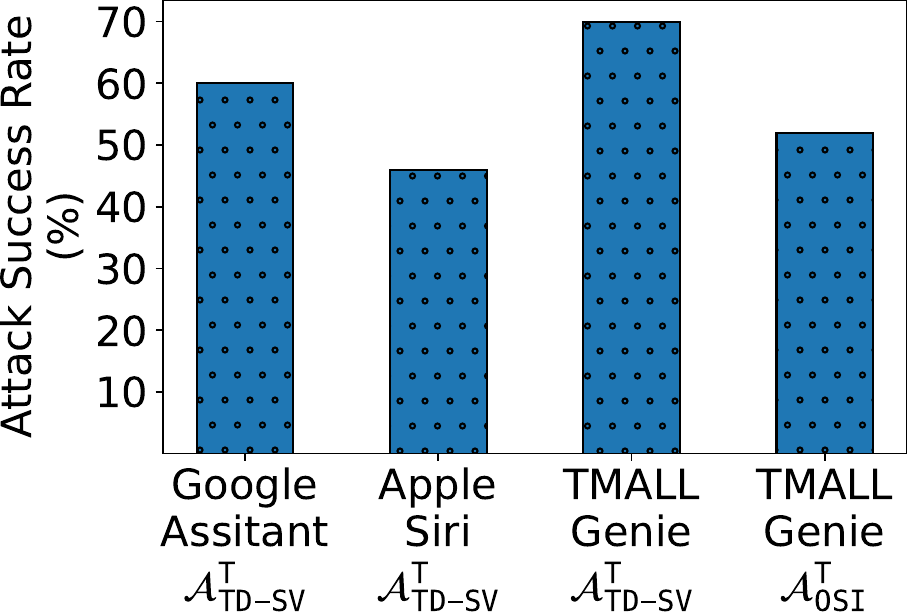}
\caption{{Results on voice assistants.}}
\label{fig:voice-assistant}
\end{minipage}
\end{figure}

After excluding 14 participants who failed to pass our concentration tests,
we finally received the answers from 126 participants.
The results of human study is shown in \figurename~\ref{fig:human-study}.
76.7\% of participants think that the adversarial voices crafted by \toolname
do not sound like the target speaker, merely 6\% lower than that of the normal pairs
and even 4.6\% higher than that of BIM.
This demonstrates that \toolname enhances the transferability without harming the human imperceptibility.
Interestingly, 39.3\% of participants choose the {\it same} option for the cloning pairs,
very close to the ASR$_t$ against Jingdong SRS in \tablename~\ref{tab:commercial-B3}
and much higher than that for adversarial pairs.
In contrast, only 20\% of participants choose the {\it same} option for \toolname,
although \toolname achieves more than 60\% ASR$_t$ against target SRSs in \tablename~\ref{tab:commercial-B3}.
This confirms the difference between adversarial and voice cloning attacks
regrading the human-machine perception consistency.

\begin{table}[t]
  \centering
  \setlength{\tabcolsep}{6pt}
   \caption{Results on commercial APIs
   in \TDSVTARGET.}
  \scalebox{0.62}{  \begin{tabular}{|c|c|c|c|c|c|c|}
     \hline
     \multirow{3}{*}{} & \multicolumn{3}{c|}{\textbf{Microsoft Azure}} & \multicolumn{3}{c|}{\textbf{Jingdong}} \\
 \cline{2-7}          & \textbf{differ-enroll} & \textbf{SNR} & \multirow{2}{*}{\textbf{PESQ}} & \textbf{differ-enroll} & \textbf{SNR} & \multirow{2}{*}{\textbf{PESQ}} \\
 \cline{2-2}\cline{5-5}          & \textbf{ASR$_t$} & \textbf{(dB)} &       & \textbf{ASR$_t$} & \textbf{(dB)} &  \\
     \hline
     \textbf{SirenAttack} & 0.49  & 8.97  & {1.15} & 0     & 10.15 & 1.18 \\
     \hline
     \textbf{Kenansville} & 0     & {\bf 20.64} & {\bf 2.11}  & 0     & {\bf 20.64}  & {\bf 2.11}  \\
     \hline
     {\bf Voice Cloning} & 10 & - & - & 40 & - & - \\
     \hline
     \textbf{FakeBob} & 0.52  & 13.16 & 1.28  & 8     & 13.32 & 1.28 \\
     \hline
     \textbf{FakeBob} +  {\small {\Circled[outer color=green]{1}}} & 0.52  & 13.16 & 1.28  & 8 & 13.32 & 1.28 \\
     \hline
     \textbf{FakeBob} +  {\small {\Circled[outer color=green]{1}}}  {\small {\Circled[outer color=blue]{2}}} & 16.67  & 13.14 & 1.28  & 11 & 13.14 & 1.28 \\
     \hline
     \textbf{FakeBob} +  {\small {\Circled[outer color=green]{1}}}  {\small {\Circled[outer color=blue]{2}}}  {\small {\Circled[outer color=red]{3}}} & 0.1  & 13.45 & 1.30  & 3 & 13.45 & 1.30 \\
     \hline
     \textbf{BIM} & 13.01 & 12.40  & {1.24} & 12    & 12.21  & 1.23  \\
     \hline
     \textbf{BIM} + {\small {\Circled[outer color=green]{1}}} & 13.01 & 12.40  & {1.24} & 12    & 12.21  & 1.23  \\
     \hline
     \textbf{BIM} + {\small {\Circled[outer color=green]{1}}}  {\small {\Circled[outer color=blue]{2}}} & 27.78 & 12.21  & 1.23  & 23.5  & 12.21  & 1.23  \\
     \hline
     \makecell[c]{\textbf{BIM} + {\small {\Circled[outer color=green]{1}}}  {\small {\Circled[outer color=blue]{2}}}  {\small {\Circled[outer color=red]{3}}} \\ (\textbf{\toolname})}
     & \makecell[c]{\textbf{61.86} \\ $\uparrow$ 48.85} & 11.84  & {1.24}
     & \makecell[c]{\textbf{66} \\ $\uparrow$ 26} & 11.84  & {1.24} \\
     \hline
     \end{tabular}%
     \label{tab:commercial-B3}%
   }
\end{table}%

  {Summary: \textit{\uline{\toolname significantly improves  transferability
  under all the three attack scenarios,
    with negligible effect on imperceptibility in terms of both perceptual objective metric and subjective human study,
    compared to the best baseline.}}}

\noindent {\bf Ablation Study.}
To understand the contributions of tailored loss functions, SRS ensemble, and time-freq corrosion,
we perform ablation study by gradually incorporating
them into BIM and FakeBob, which are in general the most effective baselines. 
Note that \toolname bases on BIM.
From \tablename{s}~\ref{tab:commercial-B1}--\ref{tab:commercial-B3},
we observe that:
{\it \uline{all the three methods improve transferability,
and in general, time-freq corrosion contributes the most,
while tailored loss functions contribute the least,
regardless of attacks, scenarios, and enrollment settings, with the following two exceptions.}}

\begin{figure}[t]
	\centering\subfigcapskip=-6pt
    \subfigure[CW]{\label{fig:ablation-study-cw}
 		\begin{minipage}[b]{0.22\textwidth}\centering
			\includegraphics[width=1\textwidth]{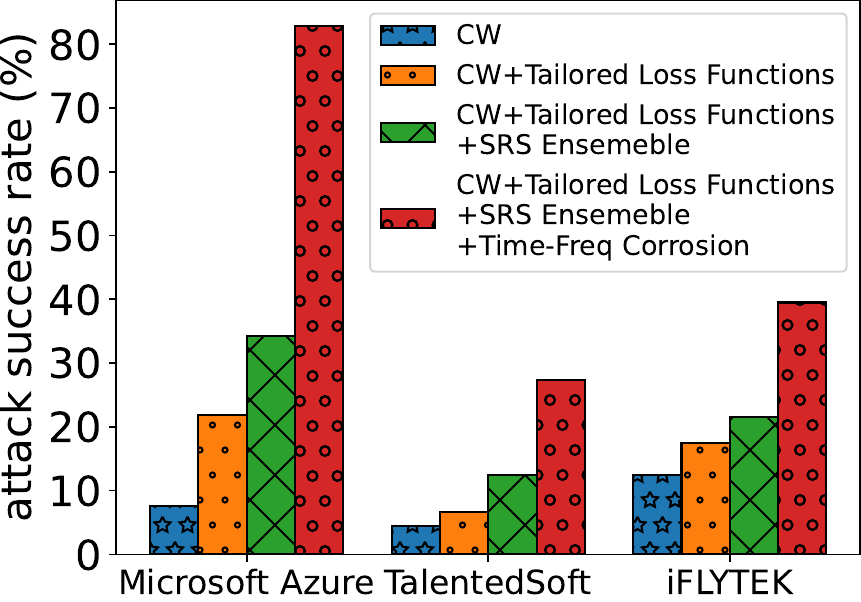}
		\end{minipage}	
    }
    \subfigure[PGD]{\label{fig:ablation-pgd}
 		\begin{minipage}[b]{0.22\textwidth}\centering
			\includegraphics[width=1\textwidth]{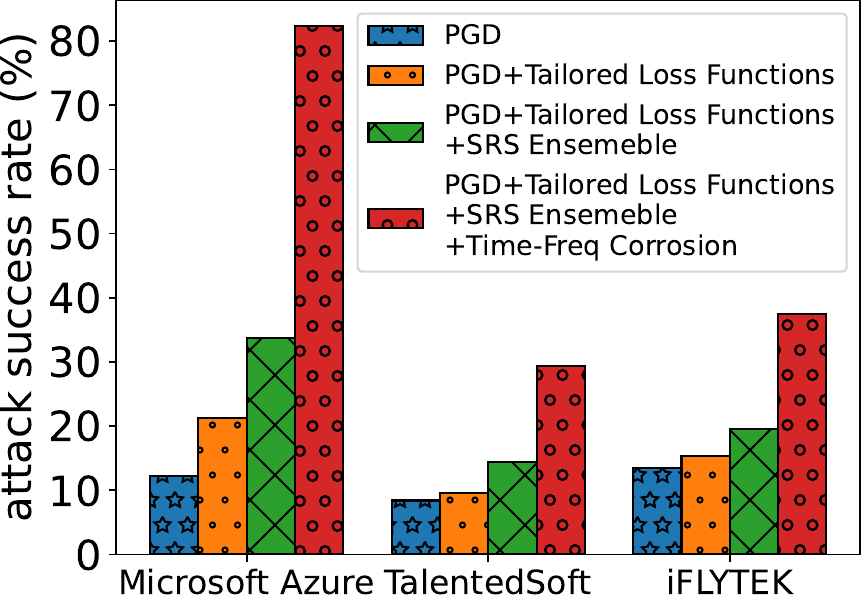}
		\end{minipage}	
    }
	\caption{{Ablation study using the CW and PGD attacks in \OSITARGET under different-enrollment setting.}}
	\label{fig:ablation-study-cw-pgd}
\end{figure}

{First, the tailored loss function $f_{3B}$ (resp. $f_{3}$) does not enhance the transferability
of BIM and FakeBob on \TDSVTARGET (resp. FakeBob on \OSIUNTARGET).
This is because FakeBob uses the same loss $f_{3B}$ and $f_{3}$ for \TDSVTARGET and \OSIUNTARGET, respectively,
and BIM uses the loss function $f_\text{BCE}$ that has the same performance as $f_{3B}$ for \TDSVTARGET\
(see \cref{sec:tuning-parameter-attack}).
Second, time-freq corrosion does not improve or even worsens the transferability of FakeBob.
It is because the black-box attack FakeBob estimates gradients instead of using exact gradients
as BIM, and the randomness introduced by time-freq corrosion makes the estimated gradients uninformative
and hence the optimization direction unreliable, consistent with the finding in \cite{SpeakerGuard}.
We try to address this by enlarging the parameter of FakeBob
that is positively correlated with the precision of estimated gradients from 50 to 1,000,
but the improvement is rather limited, and the computation cost is totally unacceptable
(1000 $\times$ 4 surrogate SRSs $\times$ 50 steps = $2e^5$ queries for a \emph{single} adversarial voice).
These suggest that time-freq corrosion is more compatible with white-box attacks that utilize exact gradients.
To confirm this, we perform additional ablation study using two white-box attacks:
Carlini and Wagner's attack (CW)~\cite{carlini2017towards}
and Projected Gradient Descent (PGD) attack~\cite{madry2017towards} (cf. Appendix~\ref{sec:detail-compared-attack}).
The results are shown in \figurename~\ref{fig:ablation-study-cw-pgd}. {\it \uline{All the three methods enhance the transferability,
demonstrating their generalizability for incorporating into white-box attacks.}}
Note that we also perform the ablation study on open-source SRSs in Appendix~\ref{sec:evaluate-qfa2sr-open},
where we can draw the same conclusion on the contributions of individual methods of \toolname.
}

\begin{figure}[t]
	\centering\subfigcapskip=-4pt
     \subfigure[ASR$_t$ of \toolname in \OSITARGET]{\label{fig:SV-TO-OSI}
           \includegraphics[width=.22\textwidth]{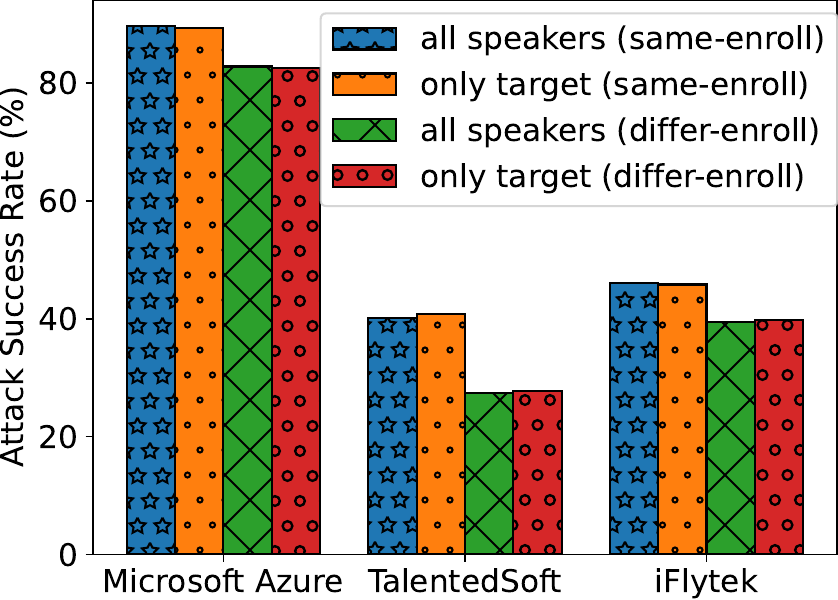}}
     \subfigure[ASR$_u$ of \toolname/BIM in \OSIUNTARGET]{\label{fig:SV-TO-OSI-B2}
    \includegraphics[width=.225\textwidth]{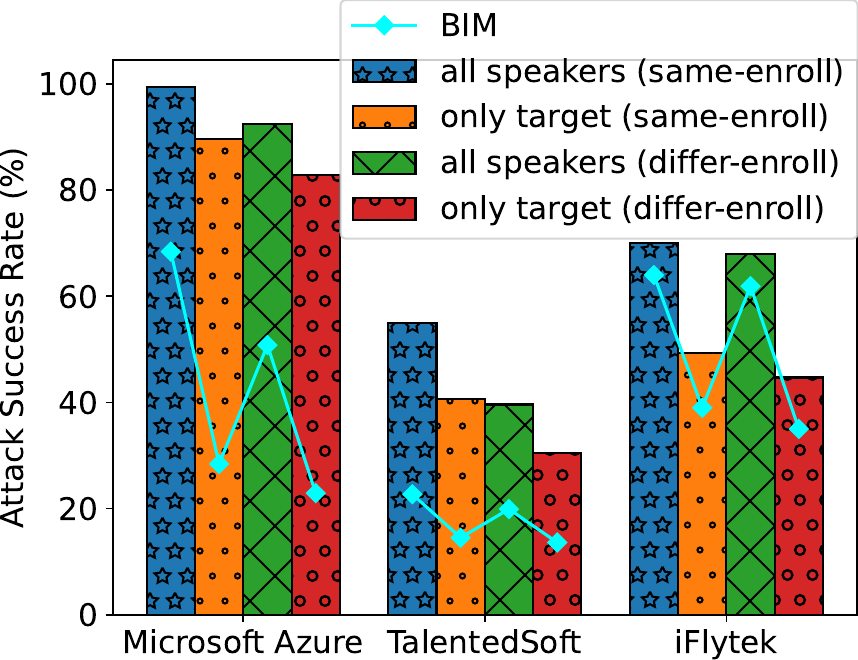} }
        \caption{{Effect of knowledge on  enrolled speakers of target SRSs.}}
       \label{fig:impact-knowledge-enrolled-speakers}
\end{figure}

\subsection{\toolname against Voice Assistants}\label{sec:exper-voice-assistant}

\noindent {\bf Settings.}
For \TDSVTARGET, we consider three voice assistants supporting speaker recognition,
i.e., Google Assistant in Google Pixel 5~\cite{Google-Voice-Match},  Siri in Apple iPad Pro 10.5~\cite{Hey-siri},
and TMall Genie in smart speaker X5~\cite{TMall-Genie}.
For \OSITARGET (when the adversary only has voices of target speakers),
we only consider  TMall Genie as the others
do not support speaker identification.
\OSIUNTARGET is omitted since it is easier than \OSITARGET and \TDSVTARGET.
To be diverse, we adopt JBL clip3 portable loudspeaker~\cite{JBL-clip3},
TMall smart speaker X5, and iPad Pro 10.5 as the loudspeaker to play adversarial voices
when attacking Google Assistant, Apple Siri, and TMall Genie, respectively.
We conduct experiments  in a meeting room with air-conditioner noise
and the distance between voice assistants and loudspeakers is set to 1.5 meters.

The enrollment and test voices of these voice assistants are text-dependent,
so we recruited six volunteers (four male and two female) to utter the desired phrases.
To cover the differ-enroll setting, we also ask them to utter the ten English sentences
used in Spk$_5$-TD-P$_2$ for Google Assistant and Apple Siri, and five Chinese quotes for TMall Genie,
which are used to enroll surrogate SRSs.
Details refer to Appendix~\ref{sec:voice-assistant-dataset}.

The results are depicted in \figurename~\ref{fig:voice-assistant}.
\toolname achieves 60\%, 46\%, and 70\% ASR$_t$
in \TDSVTARGET\ on Google Assistant, Apple Siri, and TMall Genie, respectively,
indicating that different voice assistants have  different frangibility to
adversarial attacks.
For TMall Genie, the ASR$_t$ for \OSITARGET is lower than that for \TDSVTARGET,
because without the voices of other enrolled speakers in \OSITARGET,
the crafted adversarial voices may be recognized as another enrolled speaker
whose voiceprint is similar to the target speaker.
{
  {\it \uline{These results demonstrate the effectiveness of \toolname in crafting transferable adversarial voices
  which can be played over the air against popular voice assistants.}}
}

{\subsection{Effect of Knowledge on Enrolled Speakers}\label{sec:impact-knowledge-enrolled-speakers}
We show that \toolname is still effective when
the surrogate SRS is \emph{only} enrolled with the target speaker,
which relaxes the assumption that the adversary knows and has some voices of \emph{all} the enrolled speakers of a target SRS.
\TDSVTARGET is not considered as only one speaker is enrolled for speaker verification.}

{We conduct experiments on commercial APIs
in the same settings as \cref{sec:commercial-APIs}.
The results are depicted in \figurename~\ref{fig:impact-knowledge-enrolled-speakers}.
We observe that {\it \uline{whether knowing and having voices of the other enrolled speakers
have almost no effect for \OSITARGET,}}
and the minor difference in ASR$_t$ is due to the randomness in crafting adversarial voices.
It is no surprising as the optimal loss function ($f_1(x)=-S[(x)]_t$) \emph{only} depends on the score of the target speaker
which are independent on the scores of the other enrolled speakers.
{\it \uline{For \OSIUNTARGET, the ASR$_u$ decreases moderately if the adversary only knows the target speaker.}}
It is because the optimal loss function of \OSIUNTARGET ($f_3(x)=\theta-\max_{i\in G}[S(x)]_i$)
can dynamically select the most transferable enrolled speaker as the optimization direction
when the enrolled speakers of surrogate and target SRSs are the same,
but when only the target speaker is  enrolled in the surrogate SRS, $f_3$ becomes $\theta-[S(x)]_t$ that always optimizes towards the ``target speaker''
which may not be the most transferable one.
{\it \uline{This problem also occurs in the most effective baseline attack (BIM),
and \toolname still improves its  transferability by a large margin.}}}

\begin{figure}[t]
	\centering\subfigcapskip=-4pt
    \subfigure[ASR$_t$ of \toolname in \OSITARGET]{\label{fig:scalability-B1}
       \includegraphics[width=.22\textwidth]{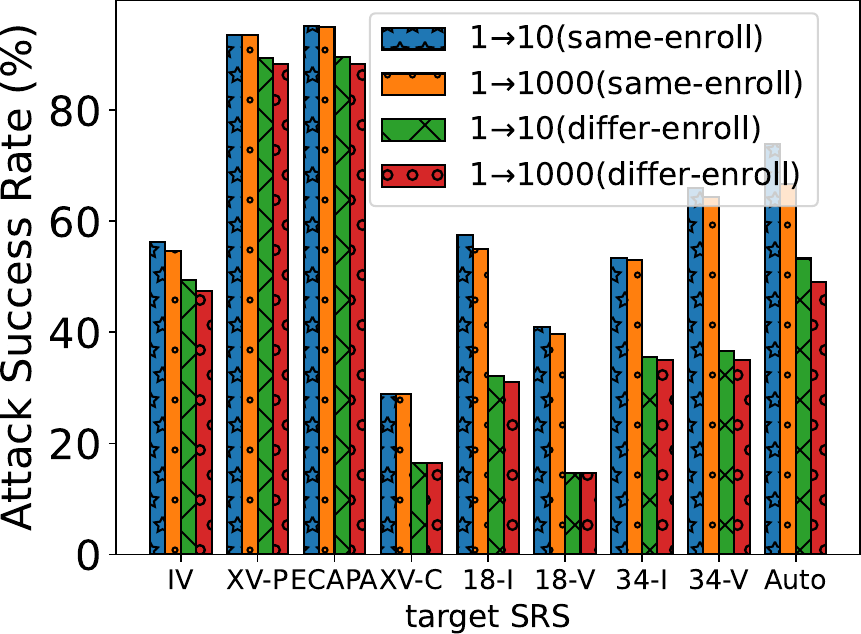}   }
   \subfigure[ASR$_u$ of \toolname/BIM in \OSIUNTARGET]{\label{fig:scalability-B2}
   \includegraphics[width=.225\textwidth]{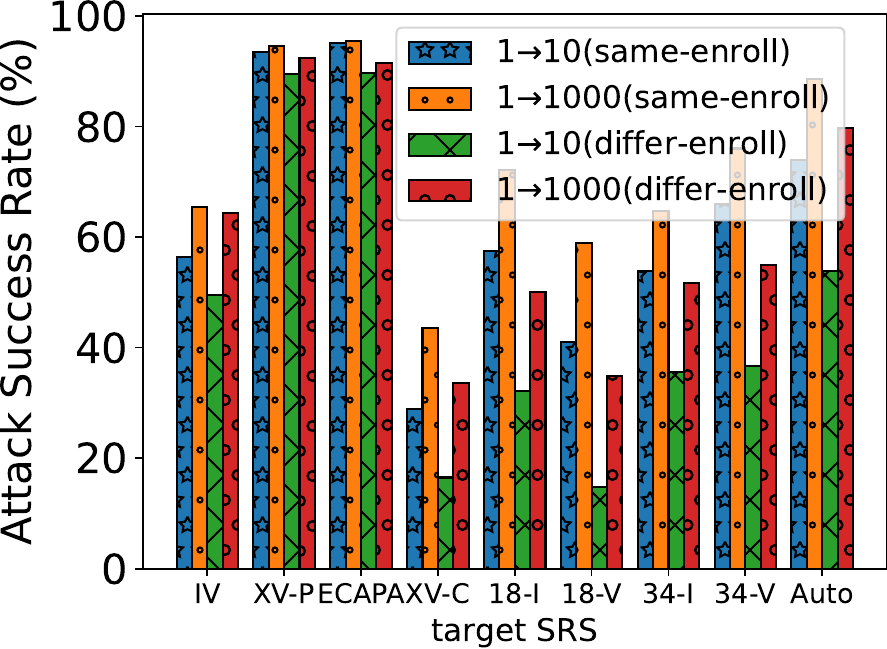}   }
    \caption{{The scalability of \toolname.}}
    \label{fig:scalability}
\end{figure}

{\subsection{Scalability of \toolname}\label{sec:scalability}
We have shown that \toolname is effective
in attacking target SRSs with no more than 10 enrolled speakers.
Now we evaluate attack scalability by increasing the enrolled speakers  to 1,000,
while the surrogate SRS is \emph{only} enrolled with the target speaker.
We use all the nine open-source SRSs as target SRSs
and Spk$_{1000}$-enroll-P1\&P2 as enrollment voices.
\figurename~\ref{fig:scalability} compares the ASR of \toolname between
10 and 1,000 enrolled speakers.
{\it \uline{With the increase of enrolled speakers,
the ASR$_t$ of \OSITARGET  decreases slightly on some target SRSs,
while the ASR$_u$ of \OSIUNTARGET increases, indicating the scalability of \toolname.}}
It is because those adversarial voices, optimized towards the target speaker and successfully transferring to the target SRS,
often have higher scores on the target speaker than on other enrolled speakers,
thus rarely get recognized as other enrolled speakers when increasing enrolled speakers.
Thus, the ASR$_t$ of \OSITARGET does not decrease too much.
On the other hand, those that fail to transfer to the target SRS
are more likely to be recognized as other enrolled speakers by the target SRS when increasing enrolled speakers,
thus, the ASR$_u$ of \OSIUNTARGET increases.}

{\section{Countermeasures}\label{sec:countermeasures}
We discuss and evaluate possible countermeasures
by considering transformation-based defenses and liveness detection.}

{\subsection{Transformation-based Defenses}
Transformation-based defenses apply some transformations
to input voices to disrupt adversarial perturbations.
We consider the seven most efficient such defenses in \cite{SpeakerGuard},
i.e., Quantization (QT), Audio Turbulence (AT), Average Smoothing (AS), Median Smoothing (MS),
Down Sampling (DS), Low Pass Filter (LPF), and Band Pass Filter (BPF),
some of which are reported promising for mitigating existing adversarial attacks.
We incorporate each of them to target SRSs and check the accuracy of normal voices of
enrolled speakers (Spk$_{10}$-test) and imposters (Spk$_{10}$-imposter),
and the accuracy of adversarial voices crafted from Spk$_{10}$-imposter.
We use XV-P and Res18-V as target SRSs which have different architectures,
and all nine open-source SRSs except for the target SRS are used as surrogate SRSs.
We conduct the evaluation in \OSITARGET and set perturbation budget $\varepsilon=0.02$,
step size $\alpha=\varepsilon/5$, number of steps $N=300$, and sampling size $\beta=5$.
The results are shown in \tablename~\ref{tab:defense}.
We find that {\it \uline{they are either effective in defending \toolname but significantly sacrificing the accuracy on the normal voices of enrolled speakers,
or ineffective, thus are not suitable for mitigating \toolname regarding the trade-off between normal and adversarial accuracy.}}
This is because they mitigate \toolname by
lowering the scores of adversarial voices 
to fall below the threshold $\theta$ of target SRSs,
which also incurs the same side-effect on normal voices.}

\begin{table}[t]
  \centering\setlength{\tabcolsep}{6pt}
  \caption{{The accuracy (\%) of normal voices and adversarial voices crafted by \toolname on target SRSs with defenses.}}
  \scalebox{0.66}{
    \begin{threeparttable}
    \begin{tabular}{|c|c|c|c|c|c|c|c|c|c|}
    \hline
    \multicolumn{2}{|c|}{} & \textbf{Baseline} & \textbf{QT} & \textbf{AT} & \textbf{AS} & \textbf{MS} & \textbf{DS} & \textbf{LPF} & \textbf{BPF} \\
    \hline
    \multirow{3}{*}{\textbf{XV-P}} & \textbf{Enrolled} & 97.2  & 75.3  & 71.4  & 91.5  & 29.6  & 44.8  & 62.8  & 70.6 \\
\cline{2-10}     & \textbf{Imposter} & 97.4  & 99.6  & 99.5  & 96.4  & 100   & 100   & 99.8  & 99.6 \\
\cline{2-10}      & \textbf{\toolname} & 10.4  & 16.5  & 26.8  & 12.1  & 84.6  & 88.6  & 47.2  & 42 \\
\hline
    \multirow{3}{*}{\textbf{Res18-V}} & \textbf{Enrolled} & 92.5  & 46    & 6.4   & 7.1   & 3.9   & 8.1   & 10.6  & 0 \\
\cline{2-10}     & \textbf{Imposter} & 92.6  & 92.2  & 100   & 99.9  & 98.4  & 100   & 100   & 100 \\
\cline{2-10}     & \textbf{\toolname} & 85.3  & 91.5  & 97.9  & 99.8  & 100   & 100   & 100   & 100 \\
\hline
    \end{tabular}%
    \begin{tablenotes}
      \item (1) Baseline:  target SRS without any defense. (2) Enrolled/imposter:  normal voices from enrolled speakers/imposters.
      (3) Only differ-enroll is considered, an easier setting for defenses.
    \end{tablenotes}
  \end{threeparttable}
  }
  \label{tab:defense}
\end{table}%

\subsection{Liveness Detection}
By exploiting the different characteristics
of the voices generated by human vocal tract and electronic loudspeakers,
liveness detection predicts whether or not input voices are uttered by humans.
Such defense can be used to defend against \toolname when launched over the air,
e.g., when attacking voice assistants deployed in voice-controlled devices.

{We use three recent liveness  detectors that are open sourced
and reported promising in the ASVspoof challenge~\cite{ASVspoof-2021,Void}:  
Void~\cite{Void}, LFCC-LCNN~\cite{LFCC-LCNN}, and LFCC-GMM~\cite{ASVspoof-2021}.
These detectors are trained using the physical access dataset of ASVspoof.
Following~\cite{ASVspoof-2021},
we compute True/False Positive/Negative Rate (i.e., TPR, TNR, FPR, and FNR) on
the adversarial and benign voices used in \cref{sec:exper-voice-assistant} (i.e., experiments on voice assistants).
To void confusing, we use \emph{Physical} to refer the adversarial voices that are played and recorded with 3 loudspeakers
(JBL clip3 portable loudspeaker, TMall Genie smart speaker X5, and DELL laptop)
and 3 microphones (Google Pixel 5 and  iPhone 6 Plus, and iPad Pro 10.5),
leading to 9 different hardware setups, and use \emph{Digital} to refer the adversarial voices
that are directly fed to the detector using the audio files. 
The average results are shown in \tablename~\ref{tab:liveness-detection-result}.
{\it \uline{These detectors can detect adversarial voices in the physical world (i.e., played over the air)
at the cost of falsely rejecting many  benign voices (more than 20\%).
Unsurprisingly, they have a remarkably high FNR (at least 75\%) on adversarial voices in the digital world,
indicating that liveness detection cannot defeat our attack
when adversarial voices are launched via APIs.}}
This is no surprising since these adversarial voices do not contain the characteristics of loudspeakers.}

\begin{table}
  \centering
  \caption{Results of liveness detection.}
  \resizebox{0.45\textwidth}{!}{%
  \begin{tabular}{|c|c|c|c|c|c|c|}
      \hline
   \multirow{3}{*}{\bf Detector} & \multicolumn{2}{c|}{\bf Benign voices} & \multicolumn{4}{c|}{\bf Adversarial voices} \\ \cline{2-7}
                    &  \multirow{2}{*}{{\bf TNR}} & \multirow{2}{*}{{\bf FPR}} & \multicolumn{2}{c|}{\bf Physical} & \multicolumn{2}{c|}{\bf Digital}   \\ \cline{4-7}
                    &                           &                   &  {\bf TPR} & {\bf FNR} &  {\bf TPR} & {\bf FNR} \\ \hline
    {\bf Void}  & {\bf 79.1\%} & {\bf 20.9\%} &  {\bf 80.2\%} & {\bf 19.8\%}  & 11.0\% & 89.0\%  \\ \hline
    {\bf LFCC-LCNN} & 59.2\% & 40.8\% & 59.3\% & 40.7\% & 15.5\%& 84.5\% \\ \hline
    {\bf LFCC-GMM}  & 61.8\% & 38.2\% & 61.6\% & 38.4\% &  {\bf25.0\%} & {\bf 75.0\%} \\ \hline
  \end{tabular}
  }
  \label{tab:liveness-detection-result}
\end{table}

\section{{Discussion}}\label{sec:open-question}
We discuss {the generalizability of our methods for improving transferability and} 
interesting future works.

\noindent {{\bf Generalizability of the three methods.}
The optimal loss functions we selected are scenario-dependent, so they may not be
optimal for other scenarios other than \OSITARGET, \OSIUNTARGET, and \TDSVTARGET. 
It is interesting  to consider other scenarios and design specific loss functions for them in future.
SRS ensemble and time-freq corrosion are
scenario-independent, but their effectiveness should still be evaluated in other scenarios.}

\noindent {\bf How to further improve \toolname?}
While \toolname significantly improves the transferability,
there is still space for improvement.
Possible directions include using more advanced optimization methods
(e.g., momentum-based gradient~\cite{Momentum-attack, Hao-Tan-speaker}
and Nesterov accelerated gradient~\cite{Nesterov-attack})
and adopting more effective loss balancing strategies for SRS ensemble,
(e.g., uncertainty-based balancing~\cite{multi-task-learning}).

\noindent {\bf How to launch effective transfer attack without voices of the target speaker?}
It is challenging to craft adversarial voices on surrogate SRSs when the adversary has no voices of the target speaker,
due to the lack of optimization guidance by the embedding of the target speaker.
One potential solution is dictionary attack~\cite{DA-attack},
which creates a master voice that matches the identity of a large population
such that it is likely to bypass the authentication of the target speaker.
However, this attack is extremely limited in the query-free black-box setting.
Future works can address this by incorporating the methods of \toolname into dictionary attack.

\section{Conclusion}\label{sec:conclusion}
We proposed \toolname, so far the most effective
query-free black-box adversarial attacks against SRSs.
It leverages the transferability of adversarial voices
and features three novel methods, i.e., tailored loss functions, SRS ensemble,
and time-freq corrosion, which significantly improves the transferability.
From the adversary perspective, our work unveils the feasibility of launching realistic and practical adversarial attacks
against strictly protected proprietary commercial SRS APIs and voice-controlled devices
in a complete black-box manner without queries them when crafting adversarial voices, thus enabling lots of follow-up attacks,
e.g., those targeting speech recognition systems.
From the perspective of SRSs maintainers and inspectors,
our attack can serve as a strong baseline for measuring  adversarial robustness
under a realistic setting.


\section*{Acknowledgments}
We thank the reviewers and our shepherd for their constructive feedbacks.
This work is supported by the National Key Research Program
(2020AAA0107800), National Natural Science Foundation of China (62072309), CAS Project for Young Scientists in Basic
Research (YSBR-040), and ISCAS New Cultivation Project (ISCAS-PYFX-202201).


\bibliographystyle{plain}
\bibliography{ref}

\appendix

\section{Generalized \toolname}\label{sec:generalized-attack}
In practice, the adversary can generate a large number of adversarial voices
but can only query the target SRS few times during transfer attack.
Thus, it is highly desired to select the adversarial voices
which are the most likely transferable to the target SRS.
To tackle this practical problem, we generalize \toolname for targeted attack in Alg.~\ref{al:strengthened-QFA2SR}.
Given $m$ seed voices, it
first crafts $m$ adversarial voices
by applying Alg.~\ref{al:final-attack} (Line~\ref{alg:generalized:init}).
Then, for each surrogate SRS $R$, it computes the scores $\mathcal{S}$ of all the adversarial voices on $R$ (Line~\ref{alg:generalized:score})
based on which the (local) ranking $\lambda_R$ of adversarial voices in $\mathcal{S}$ is computed (Line~\ref{alg:generalized:rank}),
where $x^{adv}$ has rank $i$, i.e., $\lambda_R(x^{adv})=i$, if its score $[S(x^{adv};R)]_{\tt ts}$ for the target speaker ${\tt ts}$ is $i$-th largest
in $\mathcal{S}$ (note that higher score, lower rank).
Next, a global ranking $\lambda$ is computed by summing all the local rankings $\lambda_R$'s (Line~\ref{alg:generalized:grank}).
Finally, we iteratively mount transfer attacks from the voice with least rank in $\lambda$ until
the attack succeeds,
where the smaller rank in
the global $\lambda$ indicates the higher score over all the surrogate SRSs.
Note that the selection is performed locally without querying to the target SRS.

\begin{figure}[t]\footnotesize\removelatexerror
  \begin{algorithm*}[H]
    \caption{Generalized \toolname}
    \label{al:strengthened-QFA2SR}
    \KwIn{target speaker ${\tt ts}$; set of seed voices $X=\{x_0^1,\cdots,x_0^{m}\}$;
     modification functions $\mathcal{M}=\{\cdots,M_i,\cdots\}$;
    sampling size $\beta$;
    surrogate SRS zoo $\mathcal{R}=\{\cdots,R_j,\cdots\}$;
    number of steps $N$; the step size $\alpha$; $L_\infty$ perturbation budget $\varepsilon$;
    the optimal loss function for the attack scenario $f_{\tt opt}(\cdot)$;
    maximal number $\gamma$ of queries to the target SRS $S$ s.t., $\gamma\ll m$}
    $X^{adv}\gets\{x^{adv}\mid x\in X, x^{adv} \mbox{ is crafted by applying Alg.~\ref{al:final-attack} on } x\}$\; \label{alg:generalized:init}
    \For{$R\in \mathcal{R}$}{
       $\mathcal{S}\gets\{[S(x^{adv};R)]_{\tt ts}\mid x^{adv}\in X^{adv}\}$; \textcolor{goldenrod}{\Comment{Compute all scores on $R$}} \label{alg:generalized:score} \\
      \For(\textcolor{goldenrod}{\Comment{Ranking the voices by their all scores}}){$x^{adv}\in X^{adv}$}  {
           $\lambda_R(x^{adv})=i$ if $[S(x^{adv};R)]_{\tt ts}$ is the $i$-th largest one in $\mathcal{S}$\; \label{alg:generalized:rank}
       }
    }
    \For{$x^{adv}\in X^{adv}$}{
        $\lambda(x^{adv})\gets \Sigma_{R\in \mathcal{R}}\lambda_R(x^{adv})$; \textcolor{goldenrod}{\Comment{Compute global ranking}}  \label{alg:generalized:grank}
    }
    \For{$i$ from $1$ to $\gamma$}{
       Select a voice $x^{adv}$ s.t. $\lambda(x^{adv})$ with the $i$-least rank in $\lambda$\;  \label{alg:generalized:attack}
       Transfer attack using $x^{adv}$ and abort if succeeds\;
    }
  \end{algorithm*}
  \vspace*{-5mm}
  \end{figure}

\begin{figure*}
	\centering\subfigcapskip=-2pt
    \subfigure[\OSITARGET same-enroll]{\label{fig:multi-attempt-B1-same}
 		\begin{minipage}[b]{0.25\textwidth}\centering
			\includegraphics[width=1\textwidth]{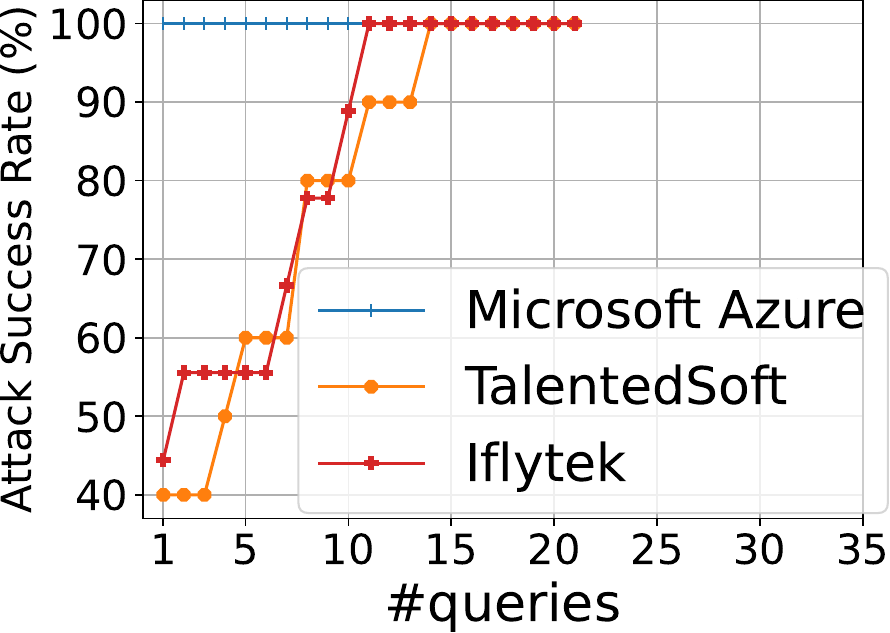}
		\end{minipage}	
    }\qquad
    \subfigure[\OSITARGET differ-enroll]{\label{fig:multi-attempt-B1-differ}
 		\begin{minipage}[b]{0.25\textwidth}\centering
			\includegraphics[width=1\textwidth]{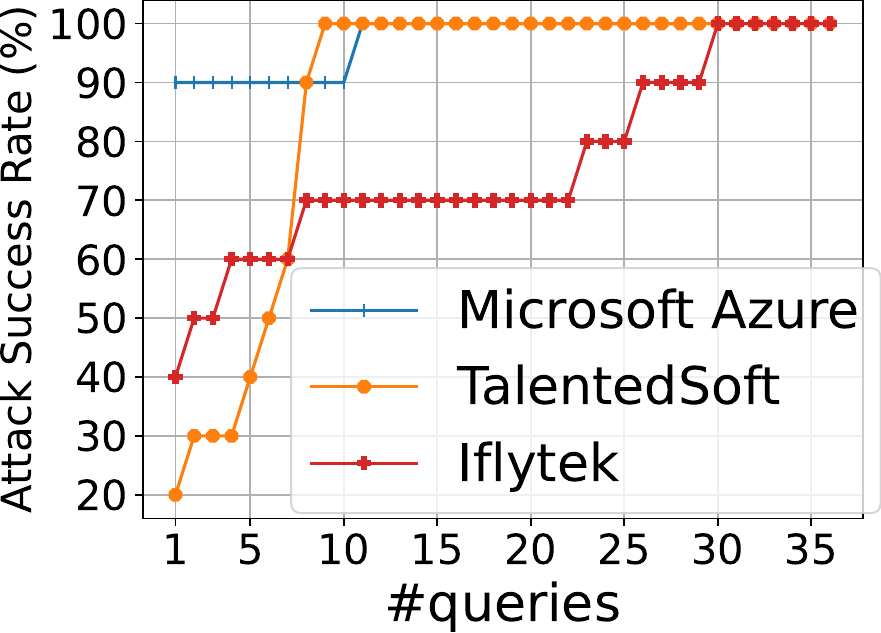}
		\end{minipage}	
    }\qquad
     \subfigure[\TDSVTARGET differ-enroll]{\label{fig:commercial-B3-multi-attempt}
 		\begin{minipage}[b]{0.25\textwidth}\centering
			\includegraphics[width=1\textwidth]{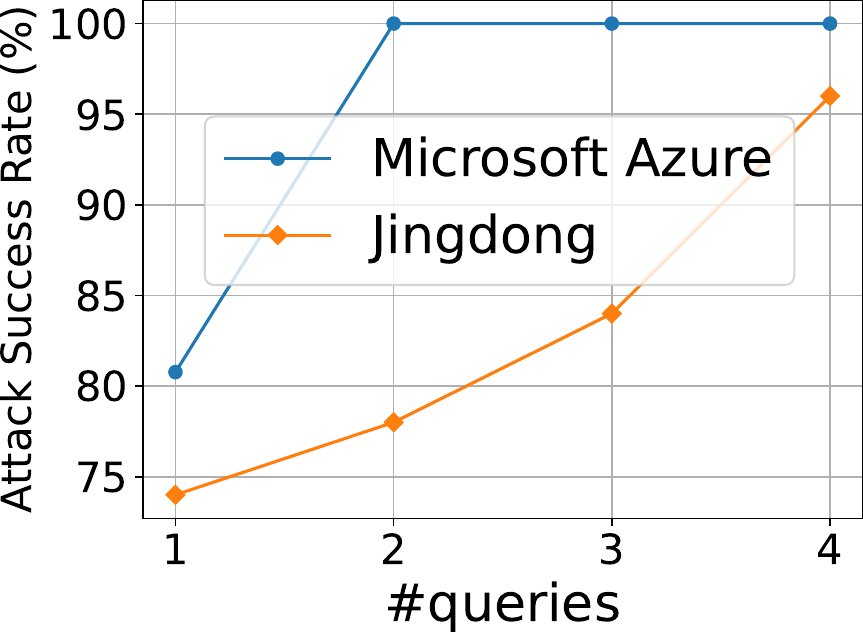}
		\end{minipage}	
    }
    \vspace{-3mm}
	\caption{ASR of \toolname w.r.t. the number of queries.}
	\label{fig:commercial-multi-attempt}\vspace{-2mm}
\end{figure*}

\begin{table*}
  \centering  \setlength\tabcolsep{8pt}
  \caption{Details of the 9 open-source SRSs where Arch denotes architecture.}\vspace{-3mm}
  \resizebox{0.75\textwidth}{!}{%
    \begin{tabular}{|c|c|c|c|c|c|}
    \hline
     {\bf Arch} & {\bf Name} & {\bf \#Params} & \makecell{\textbf{Acoustic feature}} & \makecell{{\bf Training dataset}}  & \makecell{{\bf Scoring Backend}} \\
    \hline
    {\bf GMM} & \textbf{Ivector-PLDA (IV)}~\cite{kaldi-ivector-plda} & 80.37M & MFCC &  VoxCeleb1\&2& PLDA \\
    \hline
    \multirow{3}[1]{*}{\bf TDNN} & \textbf{ECAPA-TDNN (ECAPA)}~\cite{ECAPA-TDNN} & 20.77M & fBank & VoxCeleb1 & COSS  \\
    \cline{2-6}
     & \textbf{Xvector-PLDA (XV-P)}~\cite{kaldi-xvector-plda} & 5.79M & MFCC & VoxCeleb1\&2  & PLDA  \\
     \cline{2-6}
    & \textbf{Xvector-COSS (XV-C)}~\cite{SpeechBrain} & 4.21M & fBank & VoxCeleb1  & COSS  \\
    \hline
    \multirow{5}[1]{*}{\bf CNN} &
    \textbf{Res18-Identification (Res18-I)}~\cite{resnet18} & 11.17M & spectrogram & VoxCeleb1 & COSS \\
     \cline{2-6}
    & \textbf{Res18-verification (Res18-V)}~\cite{resnet18} & 11.17M & spectrogram & VoxCeleb1 & COSS \\
    \cline{2-6}
    & \textbf{Res34-Identification (Res34-I)}~\cite{resnet34} & 21.28M & spectrogram & VoxCeleb1 & COSS \\
    \cline{2-6}
    & \textbf{Res34-Verification (Res34-V)}~\cite{resnet34} & 21.28M & spectrogram & VoxCeleb1 & COSS \\
    \cline{2-6}
    & \textbf{AutoSpeech (Auto)}~\cite{auto-speech} & 15.11M & spectrogram & VoxCeleb1 & COSS \\
    \hline
    \end{tabular}%
  }
  \label{tab:system-info}\vspace{-2mm}
\end{table*}%

  \noindent {\bf Results of Generalized \toolname with Few Queries.}
We evaluate the targeted attacks (i.e., \OSITARGET and \TDSVTARGET) of generalized \toolname by varying the number of allowed queries to the target SRS.
We consider transfer attack success rate (ASR), which
refers to the proportion of the \emph{target speakers} that can be successfully imitated on the target SRS,
where a target speaker can be successfully imitated if
the target SRS recognizes one of adversarial voices as the target speaker within
the query limit $\gamma$.
Other experimental settings are the same as \cref{sec:commercial-APIs}.
The results are depicted in \figurename~\ref{fig:multi-attempt-B1-same}, \figurename~\ref{fig:multi-attempt-B1-differ}
and \figurename~\ref{fig:commercial-B3-multi-attempt}.
Note that the ASR$_u$ of generalized \toolname for \OSIUNTARGET is 100\% regardless of the number of allowed queries,
thus is omitted here.
Generally, ASR$_t$ improves with the increase of allowed queries,
and finally becomes stable.
In detail, for \OSITARGET, under same-enroll,
\toolname achieves 100\% ASR$_t$ on Azure, iFlytek, and TalentedSoft
with 1, 11, and 14 number of allowed queries, respectively.
Similar results can be observed under differ-enroll.
For \TDSVTARGET, \toolname achieves 100\% and 96\% ASR$_t$
on  Azure and Jingdong within 4 queries, respectively,
38\% and 30\% higher than the results in \tablename~\ref{tab:commercial-B3}.
{ \it
\uline{These results demonstrate that the effectiveness and feasibility of \toolname
can be largely improved with few allowed queries to the target SRS using the adversarial voice selection method in Alg.~\ref{al:strengthened-QFA2SR}.
}
}

\section{Phrases for Text-Dependent SV}\label{sec:TD-phrase}
We use the following ten phrases supported by Microsoft Azure~\cite{microsoft-azure-vpr}:
\begin{itemize}\setlength{\itemsep}{0pt}  \vspace*{-1mm}
  \item I am going to make him an offer he cannot refuse.  \vspace*{-1mm}
  \item Houston we have had a problem.  \vspace*{-1mm}
  \item My voice is my passport verify me.  \vspace*{-1mm}
  \item Apple juice tastes funny after toothpaste.  \vspace*{-1mm}
  \item You can get in without your password.  \vspace*{-1mm}
  \item You can activate security system now.  \vspace*{-1mm}
  \item My voice is stronger than passwords.  \vspace*{-1mm}
  \item My password is not your business.  \vspace*{-1mm}
  \item My name is unknown to you.  \vspace*{-1mm}
  \item Be yourself everyone else is already taken.
\end{itemize}

\section{More Detail about SRSs}\label{sec:srs-performance}
The details of the nine adopted open-source SRSs are shown in \tablename~\ref{tab:system-info}.
They cover three architectures, i.e., the typical GMM~\cite{DehakDKBOD09}
and the state-of-the-art deep neural networks (TDNN~\cite{ECAPA-TDNN} and CNN~\cite{resnet18}).
GMM is a generative model, while the others are discriminative models.
Auto is an automatically searched architecture by~\cite{auto-speech}
while the others are manually designed by the existing works.
They also cover three most popular acoustic features~\cite{acoustic-feature-li} (i.e.,
spectrogram~\cite{hannun2014deep}, fBank~\cite{FilterBanks}, and MFCC~\cite{muda2010voice}),
and two commonly-used scoring methods (i.e., PLDA~\cite{nandwana2019analysis} and COSS~\cite{dehak2010cosine}).
They are trained using two datasets, i.e., VoxCeleb1~\cite{nagrani2017voxceleb}
and VoxCeleb2~\cite{resnet34},
which have different number of speakers, utterances, and subjects background
(e.g., ethnicities, accents, age, and profession).

We tune the threshold $\theta$ of the open-source SRSs listed in \tablename~\ref{tab:system-info}
based on the Equal Error Rate (EER) meaning the same FAR and FRR,
 where False Acceptance Rate (FAR) is
the proportion of voices that are uttered by unenrolled speakers but accepted by the SRS,
and False Rejection Rate (FRR is the proportion of voices that are uttered by
enrolled speakers but rejected. The tuned threshold and the performance of SRSs
are shown in \tablename~\ref{tab:performance} where
column (IER) denotes Identification Error Rate, i.e., the proportion of voices uttered by enrolled speakers
which should not be rejected but incorrectly classified by the SRS~\cite{chen2019real}.

\begin{table}[t]
  \centering\setlength\tabcolsep{6pt}\footnotesize
  \caption{The threshold and performance of SRSs.}\vspace{-3mm}
  \resizebox{0.48\textwidth}{!}{%
  \begin{threeparttable}
    \begin{tabular}{|c|c|c|c|c|c|}
    \hline
    \multirow{2}[1]{*}{\textbf{SRS}} & \multicolumn{2}{c|}{\textbf{SV}} & \multicolumn{3}{c|}{\textbf{OSI}} \\
\cline{2-6}          & \textbf{EER (\%)} & \textbf{$\theta$} & \textbf{EER (\%)} & \textbf{IER (\%)} & \textbf{$\theta$} \\
    \hline
    \textbf{IV} & 1.40  & 10.41  & 6.50  & 0     & 12.90  \\
    \hline
    \textbf{ECAPA} & 1.43  & 0.40  & 3.01  & 0     & 0.48  \\
    \hline
    \textbf{XV-P} & 1.12  & 12.64  & 3.02  & 0     & 16.23  \\
    \hline
    \textbf{XV-C} & 6.10  & 0.60  & 11.57  & 0     & 0.69  \\
    \hline
    \textbf{Res18-I} & 1.92  & 0.45  & 6.91  & 0     & 0.57  \\
    \hline
    \textbf{Res18-V} & 2.83  & 0.41  & 6.80  & 0     & 0.55  \\
    \hline
    \textbf{Res34-I} & 1.50  & 0.46  & 9.60  & 0     & 0.57  \\
    \hline
    \textbf{Res34-V} & 2.80  & 0.43  & 5.83  & 0     & 0.56  \\
    \hline
    \textbf{Auto} & 1.52  & 0.29  & 5.61  & 0     & 0.38  \\
    \hline
    \hline
    \textbf{Microsoft Azure} & 0.72  & 0.49  & 1.6   & 0     & 0.53 \\
    \hline
    \textbf{Talentedsoft} & -   & -  & 5.1   & 0     & 0.19 \\
    \hline
    \textbf{iFlytek} & -   & -  & 14    & 0     & 0.64 \\
    \hline
    \textbf{Jingdong} &   {0.5}$^{\sharp}$    &    {0}$^{\natural}$   &    -   &   -    & - \\
    \hline
    \hline
    \textbf{Google Assistant} & {0.8}$^{\sharp}$ & {0}$^{\natural}$ &    -   &   -    & - \\
    \hline
    \textbf{Apple Siri} & {1.2}$^{\sharp}$ & {0}$^{\natural}$ &    -   &   -    & - \\
    \hline
    \textbf{TMall Genie} & {0.4}$^{\sharp}$ & {0}$^{\natural}$ &    {0.5}$^{\sharp}$   &   0    & {0}$^{\natural}$ \\
    \hline
  \end{tabular}%

    \begin{tablenotes}
        \item Note: the number with ``$\sharp$'' and ``$\natural$'' superscript denote FAR and FRR, respectively.
        ``-'' means unsupported, i.e., Jingdong, Google Assistant, Apple Siri do not support OSI,
        and TalentedSoft and iFltek do not support TD-SV.
    \end{tablenotes}
  \end{threeparttable}
  }
  \label{tab:performance}\vspace{-5mm}%
\end{table}%

\begin{table}[!t]
  \centering\footnotesize\setlength\tabcolsep{12pt}
  \caption{ASR$_t$ and imperceptibility of time-freq corrosion with reverberation-distortion (RD) w.r.t.
  the RIR set $\mathcal{R}$.}\vspace{-1mm}
 \scalebox{1}{ 
  \begin{threeparttable}
    \begin{tabular}{|c|c|c|c|c|}
    \hline
    \textbf{$\mathcal{R}$} & \textbf{ASR$_t$-s}  &  \textbf{ASR$_t$-d}& \textbf{SNR} & \textbf{PESQ} \\
    \hline
    \textcolor{orange}{\textbf{Simulate}} & 52.2  & \textbf{36.3} & 11.24  & 1.16  \\
    \hline
    \textbf{Real} & \textbf{53.4} & 35.7  & 11.27  & 1.16  \\
    \hline
    \end{tabular}%
    \begin{tablenotes}
      \item Note: Simulate and real denote the RIR generated
      by the simulation approach Image Source Method~\cite{allen1979image}
      and collected in real-world measurement, respectively.
    \end{tablenotes}
  \end{threeparttable}
  }\vspace{-1mm}
  \label{tab:RD}%
\end{table}%

For the commercial SRSs, the responses from Microsoft Azure, TalentedSoft, and iFltek
only contain the scores given by the enrolled speakers
without the final decision results, which should be determined
by the developers to adapt to the specific applications.
Therefore, we tune the threshold $\theta$ of these commercial SRSs the same as for open-source SRSs.
In contrast, Jingdong, Google Assistant, Apple Siri, and TMall Genie
only provide the decision result without any scores,
so there is no need to tune the threshold $\theta$.

\section{Tuning Parameters of Modification Functions}\label{sec:tuning-para}
Since each modification function involves at least one adjustable parameter
that may impact the transferability,
we empirically tune them here in the attack scenario \OSITARGET,
and use the voices in Spk$_{10}$-imposter as seed voices.
We consider the transfer attack {XV-C} $\to$ {XV-P},
since it produces promising ASR$_t$ without applying our approach
(cf. \tablename~\ref{tab:B1-mutual}).
We set $L_\infty$ perturbation budget $\varepsilon=0.02$, step size $\alpha=\frac{\varepsilon}{5}=0.004$,
the number of steps $N=300$, and sampling size $\beta=5$.

\begin{table}[!t]
  \centering
  \caption{ASR$_t$ and imperceptibility of time-freq corrosion with noise-flooding (NF) w.r.t. SNR$_l$ and SNR$_h$.}\vspace{-1mm}
  \resizebox{0.36\textwidth}{!}{%
    \begin{tabular}{|c|c|c|c|c|c|}
    \hline
    \textbf{SNR$_h$} & \textbf{SNR$_l$} &  \textbf{ASR$_t$-s}   &  \textbf{ASR$_t$-d} & \textbf{SNR} & \textbf{PESQ} \\
    \hline
    \textbf{0} & \textbf{0} & 58.6  & 40.1  & 10.95  & 1.14 \\
    \hline
    \multirow{2}{*}{\textcolor{orange}{\textbf{5}}} & \textcolor{orange}{\textbf{0}} & \textbf{59.0 } & \textbf{40.6 } & 10.86  & 1.14 \\
\cline{2-6}          & \textbf{5} & 52.9  & 34.7  & 10.86  & 1.14 \\
    \hline
    \multirow{3}{*}{\textbf{10}} & \textbf{0} & 57.2  & 39.8  & 10.90  & 1.14 \\
\cline{2-6}          & \textbf{5} & 50.1  & 29.9  & 10.97  & 1.14 \\
\cline{2-6}          & \textbf{10} & 46.1  & 28.9  & 11.12  & 1.15 \\
    \hline
    \multirow{4}{*}{\textbf{15}} & \textbf{0} & 56.8  & 37.2  & 10.92  & 1.14 \\
\cline{2-6}          & \textbf{5} & 49.9  & 31.1  & 11.00  & 1.15 \\
\cline{2-6}          & \textbf{10} & 46.9  & 28.9  & 11.22  & 1.15 \\
\cline{2-6}          & \textbf{15} & 43.7  & 27.2  & 11.37  & 1.16 \\
    \hline
    \multirow{5}{*}{\textbf{20}} & \textbf{0} & 56.8  & 35.8  & 10.95  & 1.14 \\
\cline{2-6}          & \textbf{5} & 50.5  & 32.4  & 11.09  & 1.15 \\
\cline{2-6}          & \textbf{10} & 44.9  & 26.9  & 11.30  & 1.15 \\
\cline{2-6}          & \textbf{15} & 42.4  & 24.7  & 11.45  & 1.16 \\
\cline{2-6}          & \textbf{20} & 41.9  & 24.4  & 11.53  & 1.16 \\
    \hline
    \multirow{6}{*}{\textbf{25}} & \textbf{0} & 56.5  & 36.1  & 11.02  & 1.14 \\
\cline{2-6}          & \textbf{5} & 49.3  & 30.6  & 11.17  & 1.15 \\
\cline{2-6}          & \textbf{10} & 46.1  & 27.7  & 11.36  & 1.16 \\
\cline{2-6}          & \textbf{15} & 42.6  & 25.3  & 11.50  & 1.16 \\
\cline{2-6}          & \textbf{20} & 41.8  & 24.0  & 11.57  & 1.16 \\
\cline{2-6}          & \textbf{25} & 41.1  & 24.1  & 11.60  & 1.16 \\
    \hline
    \multirow{7}{*}{\textbf{30}} & \textbf{0} & 55.0  & 36.1  & 11.09  & 1.14 \\
\cline{2-6}          & \textbf{5} & 49.1  & 30.7  & 11.23  & 1.15 \\
\cline{2-6}          & \textbf{10} & 45.3  & 26.8  & 11.40  & 1.16 \\
\cline{2-6}          & \textbf{15} & 41.0  & 25.4  & 11.54  & 1.16 \\
\cline{2-6}          & \textbf{20} & 41.2  & 23.5  & 11.59  & 1.16 \\
\cline{2-6}          & \textbf{25} & 40.8  & 23.8  & 11.64  & 1.16 \\
\cline{2-6}          & \textbf{30} & 39.5  & 23.4  & 11.65  & 1.16 \\
    \hline
    \end{tabular}%
  }
  \label{tab:NF}\vspace{-1mm}%
\end{table}%

\begin{table}[!t]
  \centering
  \caption{ASR$_t$ and imperceptibility of time-freq corrosion with speed-alteration (SA)
  w.r.t. the speed ratio set $\mathcal{A}$.}\vspace{-1mm}
  \resizebox{0.38\textwidth}{!}{%
    \begin{tabular}{|c|c|c|c|c|}
    \hline
    \textbf{$\mathcal{A}$} &  \textbf{ASR$_t$-s} &  \textbf{ASR$_t$-d} & \textbf{SNR} & \textbf{PESQ} \\
    \hline
    \textcolor{orange}{\textbf{range(95,105,5)}} & \textbf{53.9 } & \textbf{38.1 } & 11.51 & 1.15 \\
    \hline
    \textbf{range(90,110,5)} & 52.4  & 33.3  & 11.41 & 1.15 \\
    \hline
    \textbf{range(85,115,5)} & 42.9  & 24.9  & 11.36 & 1.15 \\
    \hline
    \textbf{range(80,120,5)} & 35.5  & 18.9  & 11.37 & 1.15 \\
    \hline
    \textbf{range(75,125,5)} & 28.2  & 12.1  & 11.36 & 1.15 \\
    \hline
    \textbf{range(70,130,5)} & 22.9  & 10.0  & 11.36 & 1.15 \\
    \hline
    \end{tabular}%
  }
  \label{tab:SA}\vspace{-1mm}%
\end{table}%

\begin{table}[!t]
     \centering\setlength\tabcolsep{8pt}\scriptsize
  \caption{ASR$_t$ and imperceptibility of time-freq corrosion
  with chunk-dropping (CD) w.r.t. $T_l$ and $T_h$.}\vspace{-1mm}
  \resizebox{.45\textwidth}{!}{
    \begin{threeparttable}
    \begin{tabular}{|c|c|c|c|c|c|}
    \hline
    \textbf{$T_h$} & \textbf{$T_l$} &  \textbf{ASR$_t$-s}   &   \textbf{ASR$_t$-d} & \textbf{SNR} & \textbf{PESQ} \\
    \hline
    \textbf{1000} & \textbf{1000} & 42.5  & 24.5  & 11.65 & 1.16 \\
    \hline
    \multirow{2}{*}{\textbf{2000}} & \textbf{1000} & 43    & 25.7  & 11.67 & 1.16 \\
\cline{2-6}          & \textbf{2000} & 43.7  & 27.1  & 11.66 & 1.16 \\
    \hline
    \multirow{3}{*}{\textbf{3000}} & \textbf{1000} & 45.6  & 27.4  & 11.67 & 1.16 \\
\cline{2-6}          & \textbf{2000} & 46.1  & 28.4  & 11.67 & 1.16 \\
\cline{2-6}          & \textbf{3000} & 49    & 31.5  & 11.66 & 1.16 \\
    \hline
    \multirow{4}{*}{\textbf{4000}} & \textbf{1000} & 48.2  & 28.5  & 11.65 & 1.16 \\
\cline{2-6}          & \textbf{2000} & 49.5  & 30.9  & 11.64 & 1.16 \\
\cline{2-6}          & \textbf{3000} & 51.1  & 31.5  & 11.62 & 1.16 \\
\cline{2-6}          & \textbf{4000} & 53.1  & 33.8  & 11.59 & 1.16 \\
    \hline
    \multirow{5}{*}{\textcolor{orange}{\textbf{5000}}} & \textbf{1000} & 50.6  & 32.2  & 11.63 & 1.16 \\
\cline{2-6}          & \textbf{2000} & 52.4  & 31.5  & 11.61 & 1.16 \\
\cline{2-6}          & \textbf{3000} & 54.3  & 34    & 11.59 & 1.16 \\
\cline{2-6}          & \textbf{4000} & 56.2  & 35.4  & 11.58 & 1.16 \\
\cline{2-6}          & \textcolor{orange}{\textbf{5000}} & \textbf{57.7} & \textbf{36.1} & 11.58 & 1.16 \\
    \hline
    \end{tabular} %
    \begin{tablenotes}
      \item Note: (1) $C_l$ and $C_h$ are fixed to {4} and {5}, respectively.
      (2) The best parameter for \OSIUNTARGET and \TDSVTARGET is $T_l=2000,T_h=4000,C_l=0, C_h=5$.
    \end{tablenotes}
  \end{threeparttable}}
  \label{tab:CD}\vspace{-1mm}%
  \end{table}

 \begin{table}[!t]
    \centering
    \caption{ASR$_t$ and imperceptibility of time-freq corrosion
    with frequency-dropping (FD) w.r.t. $C_l$ and $C_h$.}\vspace{-1mm}
    \resizebox{.36\textwidth}{!}{
      \begin{threeparttable}
      \begin{tabular}{|c|c|c|c|c|c|}
      \hline
      \textbf{$C_h$} & \textbf{$C_l$} &  \textbf{ASR$_t$-s} &  \textbf{ASR$_t$-d} & \textbf{SNR} & \textbf{PESQ} \\
      \hline
      \multirow{2}{*}{\textbf{1}} & \textbf{0} & 42.4  & 25.7  & 11.61 & 1.16 \\
  \cline{2-6}          & \textbf{1} & 43.4  & 27.1  & 11.57 & 1.15 \\
      \hline
      \multirow{3}{*}{\textbf{2}} & \textbf{0} & 44.6  & 26.9  & 11.56 & 1.15 \\
  \cline{2-6}          & \textbf{1} & 43.5  & 26.5  & 11.53 & 1.15 \\
  \cline{2-6}          & \textbf{2} & 44.8  & 27.6  & 11.51 & 1.15 \\
      \hline
      \multirow{4}{*}{\textbf{3}} & \textbf{0} & 44.8  & 28.2  & 11.55 & 1.16 \\
  \cline{2-6}          & \textbf{1} & 44.1  & 28.9  & 11.53 & 1.15 \\
  \cline{2-6}          & \textbf{2} & 44.5  & 28.6  & 11.49 & 1.15 \\
  \cline{2-6}          & \textbf{3} & 45.7  & 30.5  & 11.47 & 1.15 \\
      \hline
      \multirow{5}{*}{\textbf{4}} & \textbf{0} & 45.1  & 28.4  & 11.51 & 1.15 \\
  \cline{2-6}          & \textbf{1} & 45.6  & 29.4  & 11.49 & 1.15 \\
  \cline{2-6}          & \textbf{2} & 45.5  & 27.7  & 11.47 & 1.15 \\
  \cline{2-6}          & \textbf{3} & 45.8  & 30.2  & 11.44 & 1.15 \\
  \cline{2-6}          & \textbf{4} & 46.2  & 30.1  & 11.45 & 1.15 \\
      \hline
      \multirow{6}{*}{\textcolor{orange}{\textbf{5}}} & \textbf{0} & 46    & 29.7  & 11.5  & 1.15 \\
  \cline{2-6}          & \textbf{1} & 44.7  & 29.2  & 11.48 & 1.15 \\
  \cline{2-6}          & \textbf{2} & 46.8  & 29.6  & 11.45 & 1.15 \\
  \cline{2-6}          & \textbf{3} & 45.2  & 30.2  & 11.45 & 1.15 \\
  \cline{2-6}          & \textcolor{orange}{\textbf{4}} & 46.8  & \textbf{31} & 11.42 & 1.15 \\
  \cline{2-6}          & \textbf{5} & \textbf{47.6} & 29.3  & 11.4  & 1.15 \\
      \hline
      \end{tabular}%
      \begin{tablenotes}
        \item Note: $F_l$ and $F_h$ are fixed to {0 Hz} and {8000 Hz}, respectively.
      \end{tablenotes}
    \end{threeparttable}
    }\vspace{-1mm}
    \label{tab:FD}%
\end{table}

\begin{table}[!t]
\centering
  \caption{ASR$_t$ and imperceptibility of time-freq corrosion with time-warping (TW) w.r.t. $W$. }\vspace{-1mm}
  \resizebox{.33\textwidth}{!}{%
    \begin{tabular}{|c|c|c|c|c|}
    \hline
    \textbf{W} &  \textbf{ASR$_t$-s} &  \textbf{ASR$_t$-d}& \textbf{SNR} & \textbf{PESQ} \\
    \hline
    \textbf{5} & 42.3  & 26.2  & 11.61 & 1.16 \\
    \hline
    \textbf{7} & 40.8  & 24.5  & 11.63 & 1.16 \\
    \hline
    \textbf{9} & 41.2  & 24.4  & 11.63 & 1.16 \\
    \hline
    \textcolor{orange}{\textbf{11}} & 40.8  & \textbf{26.3} & 11.61 & 1.16 \\
    \hline
    \textbf{13} & 41    & 24.2  & 11.6  & 1.16 \\
    \hline
    \textbf{15} & 42    & 23.7  & 11.63 & 1.16 \\
    \hline
    \textbf{17} & \textbf{43} & 24.2  & 11.61 & 1.16 \\
    \hline
    \textbf{19} & 42    & 24.3  & 11.61 & 1.16 \\
    \hline
    \end{tabular}%
  }
  \label{tab:TW}\vspace{-1mm}%
\end{table}

\begin{table}[!t]
    \centering
  \caption{ASR$_t$ and imperceptibility of time-freq corrosion with time-masking (TM) w.r.t. $t'$.
  $c$ is fixed to {2}.}\vspace{-1mm}
  \resizebox{0.32\textwidth}{!}{%
    \begin{tabular}{|c|c|c|c|c|}
    \hline
    \textbf{$t'$} & \textbf{ASR$_t$-s} &  \textbf{ASR$_t$-d} & \textbf{SNR} & \textbf{PESQ} \\
    \hline
    \textbf{10} & 40.7  & 24.8  & 11.66 & 1.16 \\
    \hline
    \textbf{20} & 40.7  & 25.3  & 11.65 & 1.16 \\
    \hline
    \textbf{30} & 41.7  & 25.1  & 11.64 & 1.16 \\
    \hline
    \textbf{40} & 41.7  & 25.8  & 11.63 & 1.16 \\
    \hline
    \textbf{50} & 41.3  & 25.3  & 11.62 & 1.16 \\
    \hline
    \textbf{60} & 41.8  & 26.4  & 11.61 & 1.16 \\
    \hline
    \textbf{70} & 43.2  & 26.7  & 11.61 & 1.16 \\
    \hline
    \textbf{80} & 42.7  & 26.1  & 11.6  & 1.16 \\
    \hline
    \textcolor{orange}{\textbf{90}} & 43    & \textbf{27.8} & 11.6  & 1.16 \\
    \hline
    \textbf{100} & \textbf{43.8} & 27    & 11.58 & 1.16 \\
    \hline
    \end{tabular}%
  }\vspace{-1mm}
  \label{tab:TM}%
\end{table}

\begin{table}[!t]
    \centering
    \caption{ASR$_t$ and imperceptibility of time-freq corrosion with frequency-masking (FM) w.r.t. $f'$.
    $c$ is fixed to {2}.}\vspace{-1mm}
  \resizebox{0.33\textwidth}{!}{%
      \begin{tabular}{|c|c|c|c|c|}
      \hline
      \textbf{$f'$} & \textbf{ASR$_t$-s} &  \textbf{ASR$_t$-d} & \textbf{SNR} & \textbf{PESQ} \\
      \hline
      \textbf{1} & 38.8  & 23.8  & 11.68 & 1.16 \\
      \hline
      \textbf{2} & 45    & 28.1  & 11.5  & 1.15 \\
      \hline
      \textbf{3} & 47.7  & 32.6  & 11.41 & 1.15 \\
      \hline
      \textbf{4} & 49.6  & 34.1  & 11.37 & 1.15 \\
      \hline
      \textbf{5} & 50.5  & 35.2  & 11.32 & 1.14 \\
      \hline
      \textbf{6} & 52.2  & 34.2  & 11.28 & 1.14 \\
      \hline
      \textcolor{orange}{\textbf{7}} & 52.7  & \textbf{35.6} & 11.25 & 1.14 \\
      \hline
      \textbf{8} & 53.3  & 32.6  & 11.22 & 1.14 \\
      \hline
      \textbf{9} & \textbf{53.5} & 32.3  & 11.2  & 1.14 \\
      \hline
      \textbf{10} & 52.3  & 30.8  & 11.19 & 1.14 \\
      \hline
      \end{tabular}%
    }\vspace{-1mm}
    \label{tab:FM}%
\end{table}

The results are shown in \tablename~\ref{tab:RD}, \tablename~\ref{tab:NF},
\tablename~\ref{tab:SA}, \tablename~\ref{tab:CD}, \tablename~\ref{tab:FD},
\tablename~\ref{tab:TW}, \tablename~\ref{tab:TM}, and \tablename~\ref{tab:FM}
where the best parameter that yields the highest ASR$_t$
under the differ-enroll setting is highlight in \textcolor{orange}{orange} color.
Recall that differ-enroll is more practical than same-enroll.
Though we tune the parameters in \OSITARGET,
we found that most of the selected best parameters produce promising results in both \OSIUNTARGET and \TDSVTARGET
except for CD where the best parameter found for \OSITARGET worsens the transferability
for \OSIUNTARGET and \TDSVTARGET. Thus, we additionally tune the parameters of CD for \OSIUNTARGET and \TDSVTARGET (cf. the table note of \tablename~\ref{tab:CD}).
In addition, though we tune the parameters on the attack XV-C $\to$ XV-P,
we found that they are also promising on other transfer attacks.

\begin{figure}[!t]
	\centering\subfigcapskip=-4pt
    \subfigure[\OSITARGET (same-enroll)]{\label{fig:loss-tailor-B1-same}
 		\begin{minipage}[b]{0.26\textwidth}\centering
			\includegraphics[width=1\textwidth]{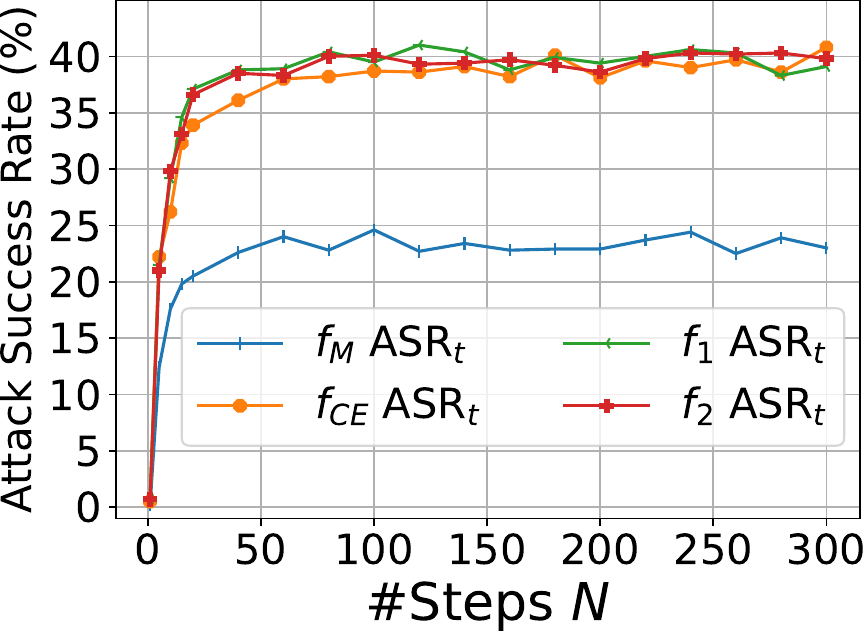}
		\end{minipage}	
    }
    \subfigure[\OSITARGET (differ-enroll)]{\label{fig:loss-tailor-B1-differ}
 		\begin{minipage}[b]{0.26\textwidth}\centering
			\includegraphics[width=1\textwidth]{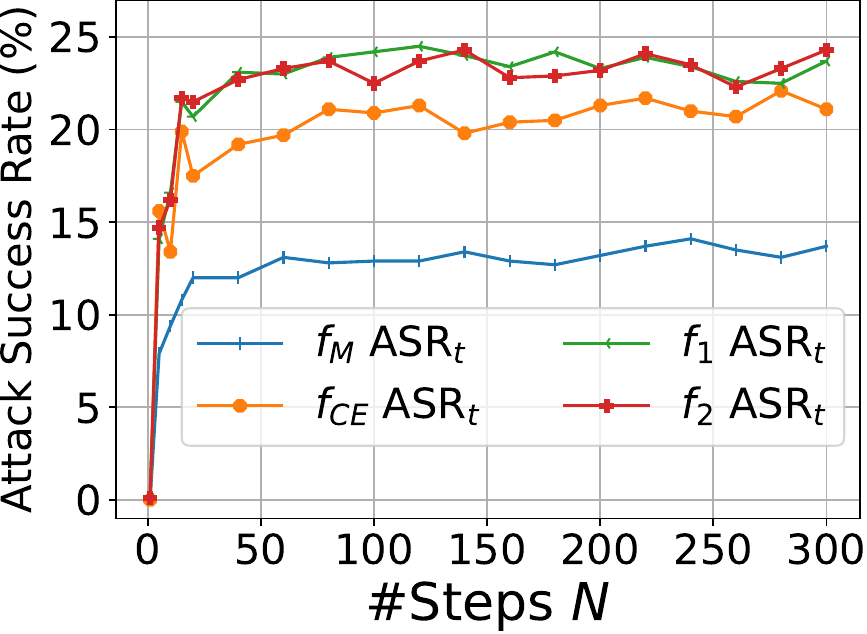}
		\end{minipage}	
    }
    \subfigure[\OSIUNTARGET (same-enroll)]{\label{fig:loss-tailor-B2-same}
 		\begin{minipage}[b]{0.26\textwidth}\centering
			\includegraphics[width=1\textwidth]{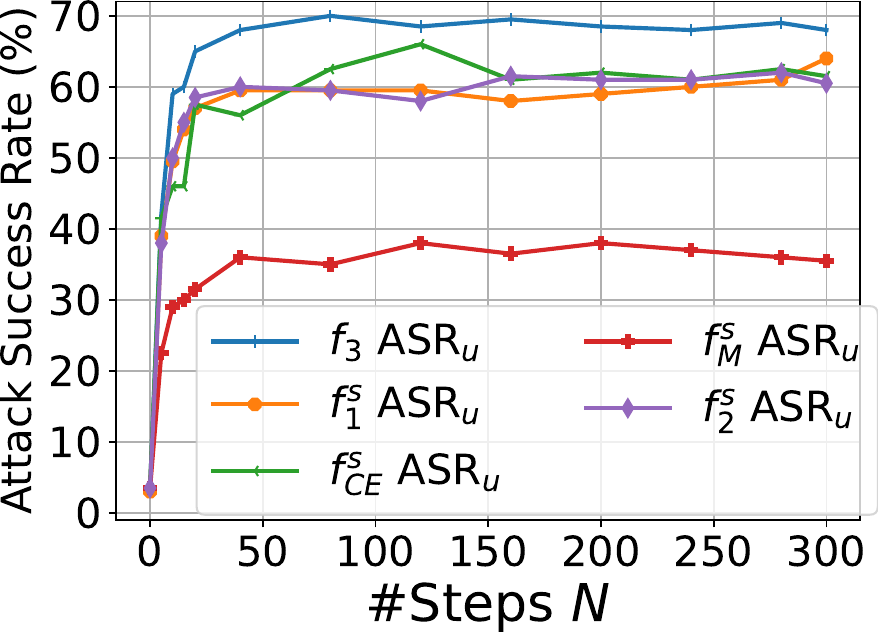}
		\end{minipage}	
    }
    \subfigure[\OSIUNTARGET (differ-enroll)]{\label{fig:loss-tailor-B2-differ}
 		\begin{minipage}[b]{0.26\textwidth}\centering
			\includegraphics[width=1\textwidth]{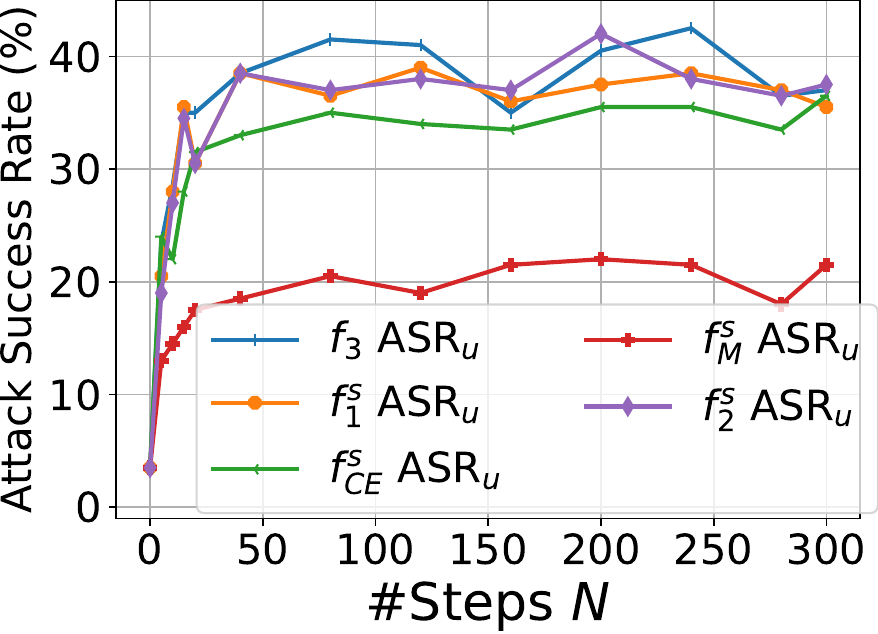}
		\end{minipage}	
    }
    \subfigure[\TDSVTARGET (differ-enroll)]{\label{fig:loss-tailor-B3}
 		\begin{minipage}[b]{0.26\textwidth}\centering
			\includegraphics[width=1\textwidth]{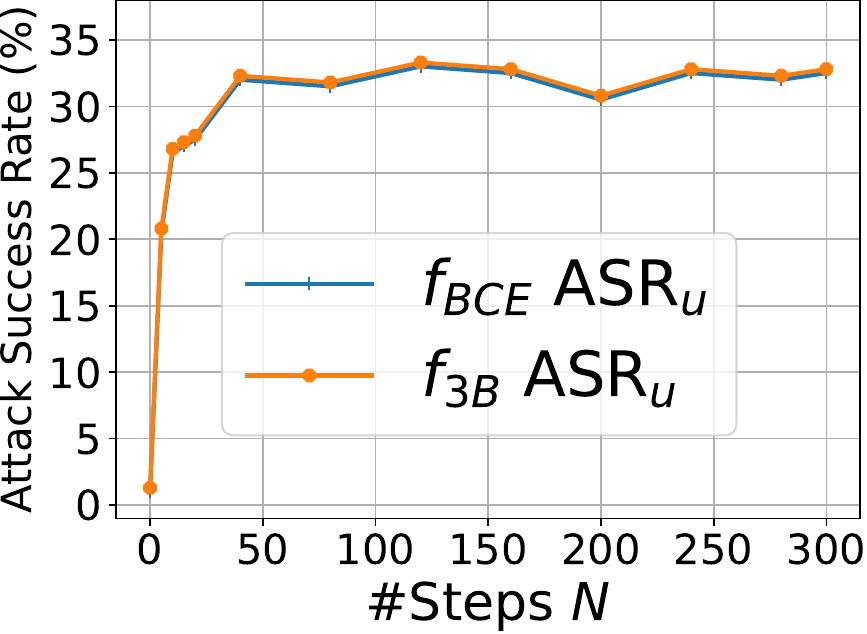}
		\end{minipage}	
    }
    \vspace{-2mm}
	\caption{Transfer attack success rate of different loss functions.}
	\label{fig:loss-tailor}
     \vspace{-2mm}
\end{figure}

\section{Evaluation of Tailored Loss Functions}\label{sec:loss-tailor-exper}
{In this section, we first perform the basic evaluation of tailored loss functions by comparing different losses on one transfer attack with a single surrogate,
then we demonstrate the generalizability of loss functions by showing that the loss comparison results keep consistent across different surrogate and target SRSs,
and between a single surrogate SRS and ensemble of surrogate SRSs even with adapted losses.}

\begin{figure*}
	\centering\subfigcapskip=-6pt
	\begin{minipage}[b]{0.47\textwidth}\centering
    \subfigure[same-enroll]{\label{fig:loss-tailor-srs-ens-all-target-B1-same-enroll}
			\includegraphics[width=.45\textwidth]{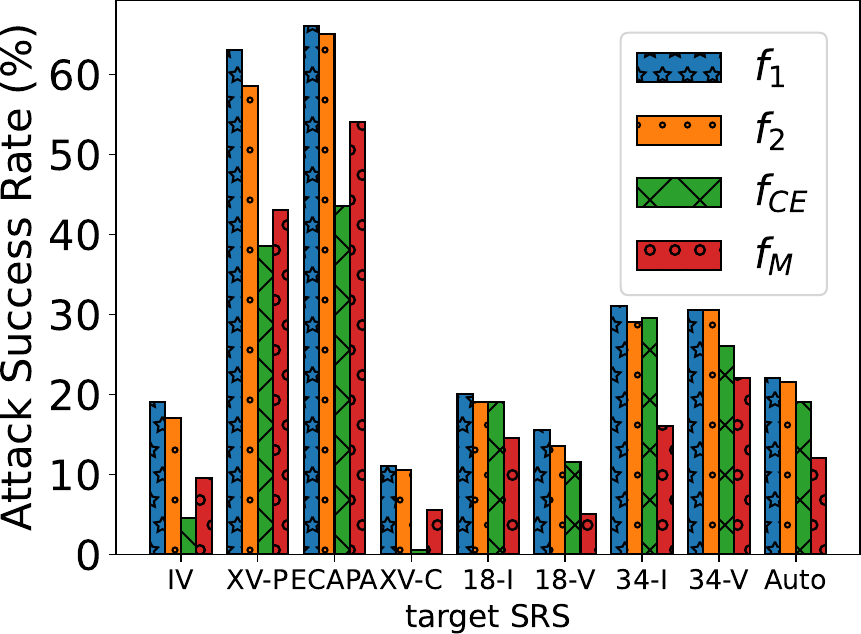}  }
   \quad
    \subfigure[differ-enroll]{\label{fig:loss-tailor-srs-ens-all-target-B1-differ-enroll}
			\includegraphics[width=.45\textwidth]{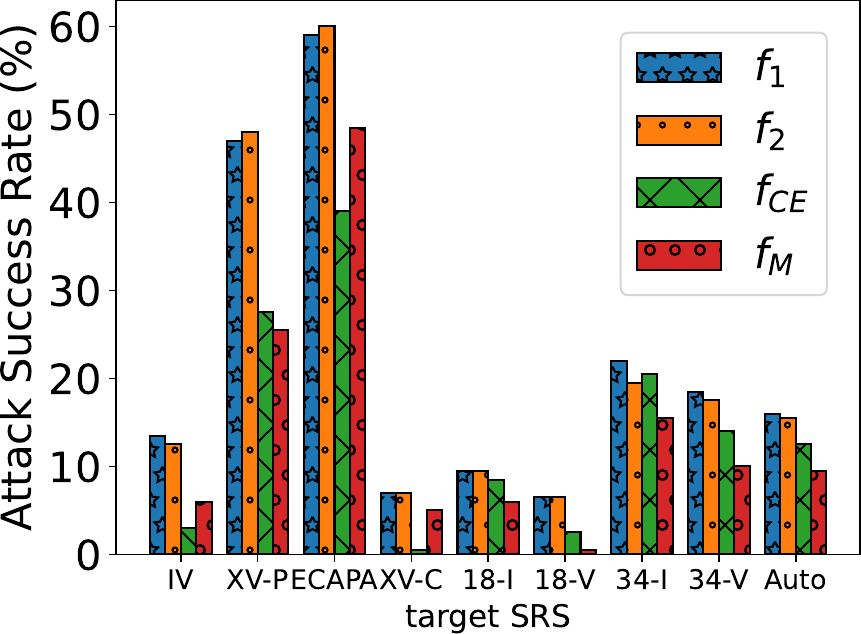} }
  \caption{{Generalizability of loss functions in \OSITARGET.}}
	\label{fig:loss-tailor-srs-ens-all-target-B1}
   \end{minipage}
\quad
	\begin{minipage}[b]{0.47\textwidth}\centering
    \subfigure[same-enroll]{\label{fig:loss-tailor-srs-ens-all-target-B2-same-enroll}
			\includegraphics[width=.45\textwidth]{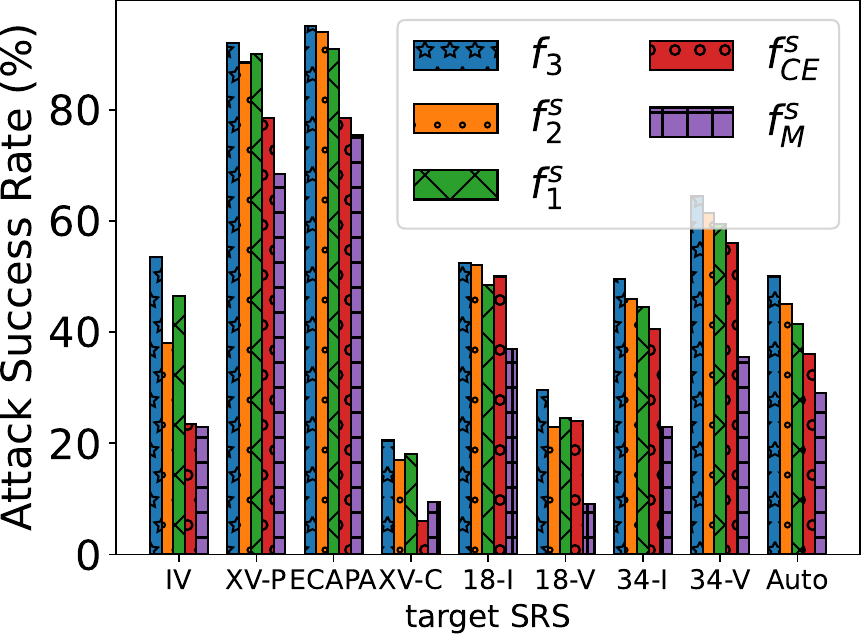} }
   \quad
    \subfigure[differ-enroll]{\label{fig:loss-tailor-srs-ens-all-target-B2-differ-enroll}
			\includegraphics[width=.45\textwidth]{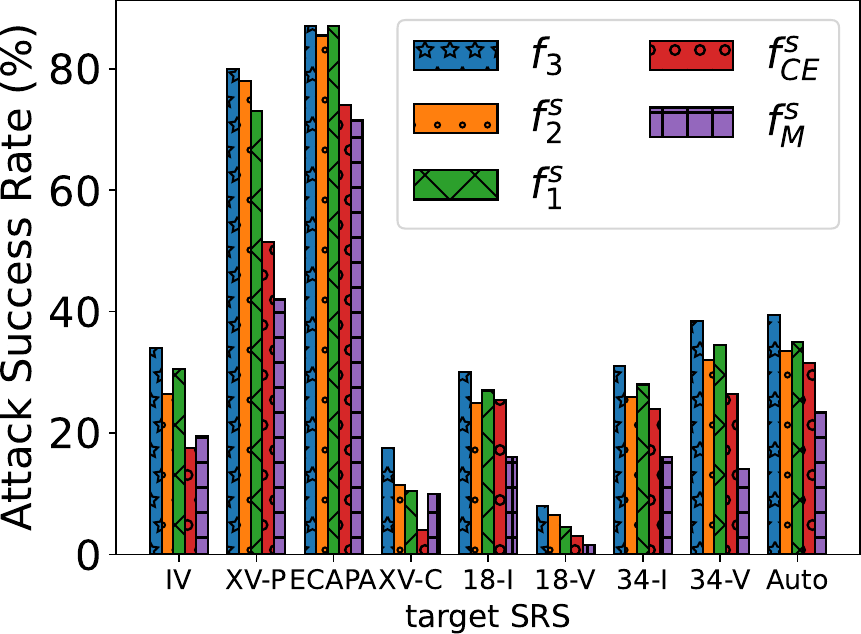}  }
	\caption{{Generalizability of loss functions in \OSIUNTARGET.}}
	\label{fig:loss-tailor-srs-ens-all-target-B2}
\end{minipage}
\end{figure*}

\subsection{Basic Evaluation}\label{sec:loss-tailor-exper-basic}
\noindent {\bf Setting.}
We consider the transfer attack {XV-C} $\to$ {XV-P},
since it produces promising ASR$_t$ without applying our approach
(cf. \tablename~\ref{tab:B1-mutual}).
For the scenario \OSITARGET, we use all the 1000 voices in Spk$_{10}$-imposter as the seed voices,
and randomly select the target enrolled speaker for each seed voice.
For \OSIUNTARGET, we randomly choose 20 voices as the seed voices per imposter in Spk$_{10}$-imposter, leading to $20\times 10=200$ seed voices in total.
For \TDSVTARGET, we use the other four speakers as the imposters
and randomly select one voice with the same text per imposter as the seed voices for each enrolled speaker and each text in Spk$_5$-TD-P$_1$,
leading to $5\times 4\times 10=200$ seed voices.
We vary the number of steps $N$ from 1 to 300 to guarantee that the transferability converges,
in order to fairly compare different loss functions.
We set $L_\infty$ perturbation budget $\varepsilon=0.02$ and step size $\alpha=\frac{\varepsilon}{5}=0.004$.

\noindent {\bf Results of scenario \OSITARGET.}
The results are shown in \figurename~\ref{fig:loss-tailor-B1-same} and \figurename~\ref{fig:loss-tailor-B1-differ}.
Overall, the loss functions $f_1$ and $f_2$ are comparable and result in the best transferability
while $f_\text{M}$ leads the worst transferability. We also observe that
the transfer attack with differ-enroll is more difficult than that with same-enroll, as differ-enroll increases the gap between the surrogate and target SRSs.
The loss functions $f_\text{M}$ and $f_\text{CE}$ perform worse than $f_1$ and $f_2$, because both $f_\text{M}$ and $f_\text{CE}$ are aimed
to simultaneously increase the score of the target speaker and decrease the scores of other enrolled speakers, consequently,
it is possible that they are largely reduced but the score of the target speaker
does not get increased to exceed the threshold $\theta$, resulting in failed transfer attacks.
Note that
\begin{center}
${f_\text{CE}=\log [\Softmax(S(x))]_t=-[S(x)]_t+\log \sum_{i\in G}e^{[S(x)]_i} \text{.}}$
\end{center}
In contrast, $f_1$ only penalizes the score of the target speaker and
$f_2=\theta-[S(x)]_t$ at the beginning since $\max_{i\in G,i\neq t}[S(x)]_i\leq \theta$, which is identical to $f_1$
(Note that the threshold $\theta$ does not impact the gradient).
Thus, reducing $f_1$ or $f_2$ ensures that the score of the target speaker
is increased sufficiently to surpass  $\theta$, leading to successful transfer attacks.
We highlight that while the Cross Entropy Loss $f_\text{CE}$
and the Margin Loss $f_\text{M}$ are widely used to craft adversarial images
for image classification~\cite{goodfellow2014explaining, carlini2017towards},
here we showcase that they exhibit inferior performance for transfer attacks on SRSs,
mainly due to the unique threshold-based decision making of SRSs.

\noindent {\bf Results of scenario \OSIUNTARGET.}
The results are shown in \figurename~\ref{fig:loss-tailor-B2-same} and \figurename~\ref{fig:loss-tailor-B2-differ}.
The comparison of $f_\text{M}^s$, $f_\text{CE}^s$, $f_1^{s}$, and $f_2^s$ is consistent
with that of $f_\text{M}$, $f_\text{CE}$, $f_1$, and $f_2$,
due to the same reason mentioned for \OSITARGET.
In general, the most effective loss function
under both the same-enroll and differ-enroll settings is $f_3$, as $f_3$
dynamically adjusts the target speaker according to the scores of the intermediate voices,
while the others statically determine the target speaker based on the seed voice.
It indicates that the speaker with the largest score often changes during the optimization,
hence $f_3$ achieves better performance.

\noindent {\bf Results of scenario \TDSVTARGET.}
The result of \TDSVTARGET is shown in \figurename~\ref{fig:loss-tailor-B3}.
The loss functions $f_\text{BCE}$ and $f_{3B}$ have identical performance.
It is because our attack optimizes them using the signs of the gradients instead of the gradients,
and the signs of gradients of the two loss functions are the same.

{In summary, {\it \uline{the loss functions $f_1$ and $f_2$ are comparable and outperform the other loss functions for \OSITARGET,
the loss function $f_3$ in general performs better than the other loss functions for \OSIUNTARGET,
and  the loss functions $f_\text{BCE}$ and $f_{3B}$ have the same performance for \TDSVTARGET.}}}

{\subsection{Generalizability of Loss Functions}\label{sec:generalizability-loss}
Here we perform the loss function comparison on more surrogate and target SRSs
to check {{whether the comparison results among loss functions keep consistent across different surrogate and target SRSs.}}
We consider transfer attacks Ens-w/o-Y $\to$ Y,
where Ens-w/o-Y denotes the ensemble of all the surrogate SRSs excluding the target SRS Y.
We use our dynamic weighting for \OSITARGET and \OSIUNTARGET, and additionally
our summation-based global ranking for  \OSIUNTARGET.
Together with Appendix~\ref{sec:loss-tailor-exper-basic}, this also allows us to check {whether the comparison results keep consistent
between a single surrogate SRS and the ensemble of SRSs even with adapted losses.}
\TDSVTARGET is omitted since losses $f_\text{BCE}$ and $f_{3B}$ have the same performance.}

{
The results are depicted in \figurename~\ref{fig:loss-tailor-srs-ens-all-target-B1}
and \figurename~\ref{fig:loss-tailor-srs-ens-all-target-B2}.
For \OSITARGET, the loss functions $f_1$ and $f_2$ still outperform the other two losses $f_\text{CE}$ and $f_\text{M}$
under both the same-enroll and differ-enroll settings, while $f_1$ outperforms $f_2$ under the same-enroll setting
and is comparable with $f_2$ under the differ-enroll setting.
For \OSIUNTARGET, $f_3$ still dominates the other losses
under same-enroll and differ-enroll settings.
{\it \uline{The comparison results among loss functions
keep consistent across different surrogate and target SRSs, and between a single surrogate SRS
and the ensemble of multiple SRSs with adapted losses (i.e., aligning with the results in the above subsection),
demonstrating the generalizability of loss functions.}}}

\begin{table*}
  \centering
  \setlength{\tabcolsep}{8pt}
 \begin{minipage}[t]{0.99\textwidth}
   \centering
  \caption{ASR$_t$ of time-freq corrosion in  \OSITARGET, where Para denotes RD+NF||SA+CD+FD||TW+TM+FM.}
  \setlength{\tabcolsep}{5pt}
  \label{tab:time-freq-corr-result}
  \scalebox{0.88}{
  \begin{tabular}{|c|c|c|c|c|c|c|c|c|c|c|c|c|c|}
    \hline
          \multirow{2}{*}{} & \multirow{2}{*}{\bf Baseline} & \multicolumn{8}{c|}{\bf Single} & \multicolumn{3}{c|}{\bf Serial} & {\bf Parallel} \\ \cline{3-14}
          & & \textbf{RD} & \textbf{NF}
          & \textbf{SA} & \textbf{CD} & \textbf{FD}
          & \textbf{TW} & \textbf{TM} & \textbf{FM} & {\textbf{RD+NF}} & {\textbf{SA+CD+FD}}  & {\textbf{TW+TM+FM}}
          & \textbf{Para}  \\
    \hline
    {\bf Same-enroll} & 39.1  & 52.2  & 59    & 53.9  & 57.7  & 46.8  & 40.8  & 43    & 52.7  & 62    & 72.6  & 57.6  & \textbf{78.4} \\
   \hline
  {\bf Differ-enroll} & 23.7  & 36.3  & 40.6  & 38.1  & 36.1  & 31    & 26.3  & 27.8  & 35.6  & 45.5  & 53.9  & 37.4  & \textbf{64.1} \\
   \hline
    \end{tabular}}%
\end{minipage}

\medskip
 \begin{minipage}[t]{0.99\textwidth}
\centering
  \caption{ASR$_u$ and imperceptibility of time-freq corrosion in the attack scenario \OSIUNTARGET.  }
 \scalebox{0.76}{
    \begin{tabular}{|c|c|c|c|c|c|c|c|c|c|c|c|c|c|}
    \hline
    \multirow{2}[4]{*}{} & \multirow{2}{*}{\textbf{Baseline}} & \multicolumn{8}{c|}{\textbf{Single}}                          & \multicolumn{3}{c|}{\textbf{Serial}} & \multicolumn{1}{c|}{\textbf{Parallel}} \\
\cline{3-14}          &       & \textbf{RD} & \textbf{NF} & \textbf{SA} & \textbf{CD} & \textbf{FD} & \textbf{TW} & \textbf{TM} & \textbf{FM} & {\textbf{RD+NF}} & {\textbf{SA+CD+FD}}  & {\textbf{TW+TM+FM}} & {\bf Para} \\
    \hline
  \textbf{ASR$_u$-s} & 68    & 86.5  & 96    & 71.5  & 76.5  & 80    & 69.5  & 77    & 89.5  & 94.5  & 79    & 93.5  & \textbf{98} \\
    \hline
  \textbf{ASR$_u$-d} & 37    & 60    & 71.5  & 46    & 46.5  & 51    & 40.5  & 45    & 71    & 73.5  & 57    & 83    & \textbf{85} \\
    \hline
    \textbf{SNR (dB)} & 11.37 & 11.48  & 10.95  & 11.31  & 11.38  & 11.38  & 11.33  & 11.31  & 11.37  & 10.88  & \textbf{11.50 } & 11.29  & 11.00  \\
    \hline
    \textbf{PESQ} & 1.19  & 1.21  & 1.20  & 1.19  & 1.20  & 1.18  & 1.19  & 1.19  & 1.18  & \textbf{1.22 } & 1.19  & 1.18  & 1.19  \\
    \hline
    \end{tabular}%
  }
  \label{tab:time-freq-corr-result-B2}%
\end{minipage}
\end{table*}

\begin{table*}[!t]
 \centering
  \caption{ASR$_t$ and imperceptibility of time-freq corrosion in the attack scenario \TDSVTARGET.
  }
 \scalebox{0.78}{ \begin{tabular}{|c|c|c|c|c|c|c|c|c|c|c|c|c|c|}
    \hline
    \multirow{2}[4]{*}{} & \multirow{2}{*}{\textbf{Baseline}} & \multicolumn{8}{c|}{\textbf{Single}}                          & \multicolumn{3}{c|}{\textbf{Serial}} & \textbf{Parallel} \\
\cline{3-14}          &       & \textbf{RD} & \textbf{NF} & \textbf{SA} & \textbf{CD} & \textbf{FD} & \textbf{TW} & \textbf{TM} & \textbf{FM} & {\textbf{RD+NF}} & {\textbf{SA+CD+FD}}  & {\textbf{TW+TM+FM}} & {\bf Para} \\
    \hline
\textbf{ASR$_t$-d} & 29.5  & 33.5  & 42.5  & 30    & 42    & 39    & 33.5  & 42.5  & 42    & 50    & 43    & 46.5  & \textbf{57} \\
    \hline
    \textbf{SNR (dB)} & 12.21 & 12.31  & 11.89  & 12.16  & 12.22  & 12.20  & 12.18  & 12.12  & 12.12  & 11.91  & \textbf{12.38 } & 12.06  & 11.84  \\
    \hline
    \textbf{PESQ} & 1.23  & \textbf{1.24 } & 1.21  & 1.22  & 1.23  & 1.22  & 1.22  & 1.23  & 1.22  & 1.22  & 1.23  & 1.22  & 1.21  \\
    \hline
    \end{tabular}%
  }
  \label{tab:time-freq-corr-result-B3}%
\end{table*}%

\begin{figure}[!t]
	\centering
		\centering
		 \includegraphics[width=.3\textwidth]{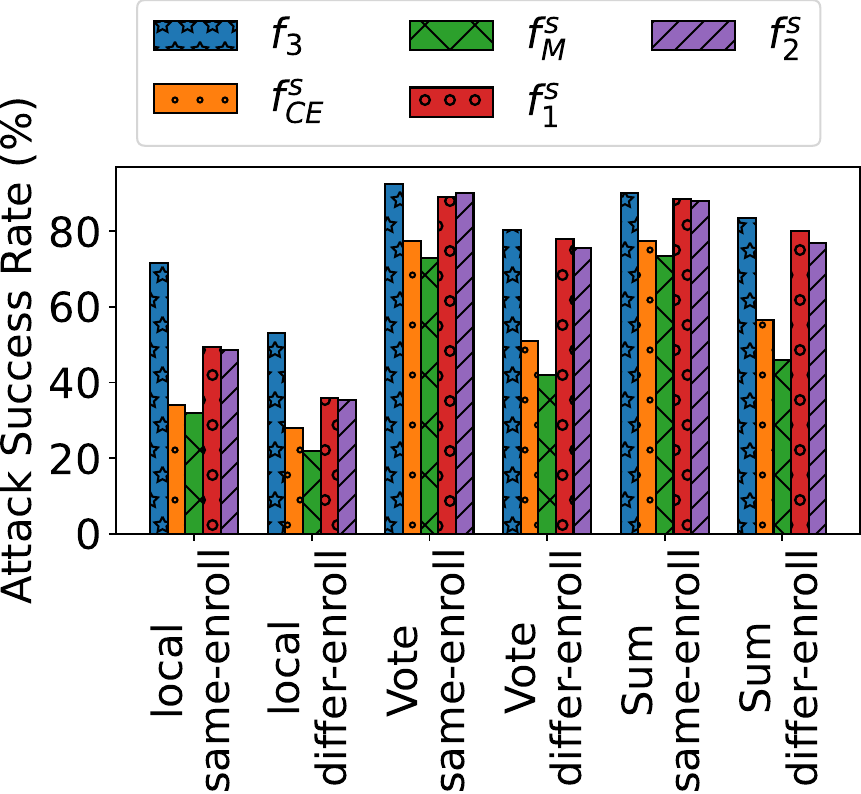}
		 \caption{Comparison of loss functions with different score ranking strategies.}
			\label{fig:loss-tailor-B2-srs-diver-to-xv-plda}
\end{figure}

{We further compare the loss functions of \OSIUNTARGET when SRS ensemble is used with different score ranking strategies
(local-ranking, summation-based global-ranking, and voting-based global-ranking),
using the attack Ens-w/o-XV-P $\to$ XV-P.
Other settings are the same as Appendix~\ref{sec:loss-tailor-exper-basic}.
The results are depicted in \figurename~\ref{fig:loss-tailor-B2-srs-diver-to-xv-plda}.
We can observe  that $f_3$ is still the most effective loss function
under both the same-enroll and different-enrollment settings,
regardless of different ranking strategies.
This is consistent with the above conclusion, further demonstrating the generalizability of
the loss functions.}

\begin{table*}[!t]
    \caption{The effectiveness of SRS ensemble
    for \OSITARGET.
    }
    \label{tab:B1-same-differ-enroll}%
  \centering\setlength{\tabcolsep}{2pt}
  \scalebox{0.68}{ \begin{threeparttable}
    \begin{tabular}{|c|c|c|c|c|c|c|c|c|c|c|c|c|c|c|c|c|c|c|}
    \hline
    \multirow{2}{*}{\diagbox{\bf S}{\bf T}} & \multicolumn{2}{c|}{\textbf{IV}} & \multicolumn{2}{c|}{\textbf{ECAPA}} & \multicolumn{2}{c|}{\textbf{XV-P}} & \multicolumn{2}{c|}{\textbf{XV-C}} & \multicolumn{2}{c|}{\textbf{Res18-I}} & \multicolumn{2}{c|}{\textbf{Res18-V}} & \multicolumn{2}{c|}{\textbf{Res34-I}} & \multicolumn{2}{c|}{\textbf{Res34-V}} & \multicolumn{2}{c|}{\textbf{Auto}} \\
\cline{2-19}          & \textbf{ASR$_t$-s} & \textbf{ASR$_t$-d} & \textbf{ASR$_t$-s} & \textbf{ASR$_t$-d} & \textbf{ASR$_t$-s} & \textbf{ASR$_t$-d} & \textbf{ASR$_t$-s} & \textbf{ASR$_t$-d} & \textbf{ASR$_t$-s} & \textbf{ASR$_t$-d} & \textbf{ASR$_t$-s} & \textbf{ASR$_t$-d} & \textbf{ASR$_t$-s} & \textbf{ASR$_t$-d} & \textbf{ASR$_t$-s} & \textbf{ASR$_t$-d} & \textbf{ASR$_t$-s} & \textbf{ASR$_t$-d} \\
    \hline
    \textbf{Best-single} & 11.9  & 6.7   & 47.1  & 39.8  & 39.1  & 23.7  & 5.8   & 3.4   & 4.8   & 1.2   & 0.6   & 0.5   & 3.9   & 1.5   & 2.2   & 0.5   & 2.2   & 3.8 \\
    \hline
    \textbf{Uniform-Ens (w/o T)} &  {\bf 21.7}  & {\bf 15}    & 58    & 52.7   & 47.5  & 27.2   & {\bf 13.4}  & {\bf 8.1}   & 3.3   & 0  &  0     & 0    & 7.8   & 4.6   & 6.7   & 3.2     & 6.5   & 4.6  \\
    \hline
    \textbf{Dynamic-Ens (w/o T)} & 19.7  & 14    & {\bf 66.6}  & {\bf 60}    & {\bf 64.6}  & {\bf 49.4}  & 12.3  & 6.8   & {\bf 24.3}  & {\bf 11.5}  & {\bf 13.8}  & {\bf 6.5}   & {\bf 30.2}  & {\bf 22.1}  & {\bf 34.5}  & {\bf 21.3}  & {\bf 23.9}  & {\bf 18.1} \\
    \hline
    \end{tabular}%

    \begin{tablenotes}
      \item Note: (1) S and T denote the surrogate and target SRSs, respectively.
      (2) Best single denotes the surrogate SRS that leads to the largest ASR$_t$, which varies with the target.
      (3) ``W/o T'' means that all the SRSs except the target are used as surrogate.
    \end{tablenotes}
  \end{threeparttable}
  }
  \label{tab:addlabel}%
\end{table*}%

\section{Evaluation of Time-Freq Corrosion}\label{sec:time-freq-curr-exper}

\noindent {\bf Setting.}
To keep consistent with Appendix~\ref{sec:loss-tailor-exper},
we consider the transfer attack XV-C $\to$ XV-P with the same seed voices, $L_\infty$ perturbation budget $\varepsilon$ and  step size $\alpha$,
while the number of steps $N$ and sampling size $\beta$ are set to 300 and 5.
We adopt the optimal loss function found in Appendix~\ref{sec:loss-tailor-exper} for each attack scenario.
The parameters of each modification function is set to the one selected in Appendix~\ref{sec:tuning-para}.
The frequency-domain modification functions are applied on the Spectrogram
since Fbank and MFCC are extracted from Spectrogram.
The ASR$_t$ for \OSITARGET are reported in \tablename~\ref{tab:time-freq-corr-result}.
We mainly analyze the results of attacks with differ-enroll.
The results for \OSIUNTARGET and \TDSVTARGET are reported in \tablename~\ref{tab:time-freq-corr-result-B2}
and \tablename~\ref{tab:time-freq-corr-result-B3}, respectively,
where similar conclusions can be drawn.

\noindent {\bf Results of individual modification functions.}
Under the differ-enroll setting, all individual modification functions can improve ASR$_t$ compared to the baseline
(i.e., transfer attack without time-freq corrosion) by at least 1.7\%,
where Noise-flooding (NF) improves the transferability by nearly 17\%.

\noindent {\bf Results of serial combinations.}
The serial combinations further boost the transferability.
SA+CD+FD achieves improves the ASR$_t$ by over 15\% compared to the best one among SA, CD, and FD.
RD+NF (resp. TW+TM+FM) achieves 4.9\% (resp. 1.8\%) higher ASR$_t$
than the best one between RD and NF (resp. among TW, TM, and FM).

\noindent {\bf Results of parallel combination.}
Under the differ-enroll setting,  the parallel combination of RD+NF, SA+CD+FD, and TW+TM+FM
yields the best transferability, specifically, 10\% higher than
the best one among RD+NF, SA+CD+FD, and TW+TM+FM, and 40.4\% higher than the baseline.

The above results show the effectiveness of time-freq corrosion.
In the rest of this section, we will utilize the parallel combination of RD+NF, SA+CD+FD, and TW+TM+FM, denoted by Para,
as the default modification function for time-freq corrosion.

\begin{table}[t]
  \centering
  \caption{ASR$_u$ between pairs of surrogate and target SRSs
  in the attack scenarios \OSIUNTARGET.}
  \scalebox{0.82}{  \begin{tabular}{|c|c|c|c|c|}
    \hline
    \multirow{2}{*}{\diagbox{\bf S}{\bf T}} & \multicolumn{2}{c|}{\textbf{XV-P}} & \multicolumn{2}{c|}{\textbf{Res18-V}} \\
\cline{2-5}          & \textbf{ASR$_u$-s}  & \textbf{ASR$_u$-d}  &  \textbf{ASR$_u$-s}  & \textbf{ASR$_u$-d}  \\
    \hline
    \textbf{IV} & 44.5  & 20    & 0     & 0 \\
    \hline
    \textbf{ECAPA} & 60.5  & 36    & 0     & 0 \\
    \hline
    \textbf{XV-P} & -   & -   & 0     & 0 \\
    \hline
    \textbf{XV-C} & 68    & 37    & 0     & 0 \\
    \hline
    \textbf{Res18-I} & 4.5   & 1     & 0.5   & 0.5 \\
    \hline
    \textbf{Res18-V} & 3.5   & 0     & -   & - \\
    \hline
    \textbf{Res34-I} & 4     & 0.5   & 0     & 0 \\
    \hline
    \textbf{Res34-V} & 4     & 0     & 0.5   & 0 \\
    \hline
    \textbf{Auto} & 4     & 1.5   & 2.5   & 0 \\
    \hline
    \end{tabular}%
  }
  \label{tab:B2-mutual}%
\end{table}%

\begin{table}[t]
  \centering
  \caption{ASR$_t$ between pairs of surrogate and target SRSs
  in the attack scenarios \TDSVTARGET.}
  \scalebox{0.82}{
    \begin{tabular}{|c|c|c|}
    \hline
    \multirow{2}{*}{\diagbox{\bf S}{\bf T}} & \textbf{XV-P} & \textbf{Res18-V} \\
\cline{2-3}          &  \textbf{ASR$_t$-d}& \textbf{ASR$_t$-d} \\
    \hline
    \textbf{IV} & 16.5  & 0 \\
    \hline
    \textbf{ECAPA} & 19.5  & 0 \\
    \hline
    \textbf{XV-P} & -   & 0 \\
    \hline
    \textbf{XV-C} & 29.5  & 0 \\
    \hline
    \textbf{Res18-I} & 0     & 1.5 \\
    \hline
    \textbf{Res18-V} & 0     & - \\
    \hline
    \textbf{Res34-I} & 0     & 0.5 \\
    \hline
    \textbf{Res34-V} & 0     & 0 \\
    \hline
    \textbf{Auto} & 0.5   & 1.5 \\
    \hline
    \end{tabular}
  }
  \label{tab:B3-mutual}%
\end{table}

\begin{table*}[t]
  \centering\setlength{\tabcolsep}{4pt}
  \caption{ASR$_t$ between pairs of surrogate and target SRSs
  in the attack scenarios \OSITARGET.
  }
  \resizebox{1\textwidth}{!}{
    \begin{tabular}{|c|c|c|c|c|c|c|c|c|c|c|c|c|c|c|c|c|c|c|}
    \hline
    \multirow{2}{*}{\diagbox{\bf S}{\bf T}} & \multicolumn{2}{c|}{\textbf{IV}} & \multicolumn{2}{c|}{\textbf{ECAPA}} & \multicolumn{2}{c|}{\textbf{XV-P}} & \multicolumn{2}{c|}{\textbf{XV-C}} & \multicolumn{2}{c|}{\textbf{Res18-I}} & \multicolumn{2}{c|}{\textbf{Res18-V}} & \multicolumn{2}{c|}{\textbf{Res34-I}} & \multicolumn{2}{c|}{\textbf{Res34-V}} & \multicolumn{2}{c|}{\textbf{Auto}} \\
\cline{2-19}          & \textbf{ASR$_t$-s} & \textbf{ASR$_t$-d}  & \textbf{ASR$_t$-s} & \textbf{ASR$_t$-d}  & \textbf{ASR$_t$-s} & \textbf{ASR$_t$-d}  & \textbf{ASR$_t$-s} & \textbf{ASR$_t$-d}  & \textbf{ASR$_t$-s} & \textbf{ASR$_t$-d}  & \textbf{ASR$_t$-s} & \textbf{ASR$_t$-d}  & \textbf{ASR$_t$-s} & \textbf{ASR$_t$-d}  & \textbf{ASR$_t$-s} & \textbf{ASR$_t$-d}  & \textbf{ASR$_t$-s} & \textbf{ASR$_t$-d}  \\
    \hline
    \textbf{IV} & -   & -   & 14.7  & 12.1  & 15    & 5.4   & 1.9   & 1.8   & 0     & 0     & 0     & 0     & 0     & 0     & 0     & 0     & 0     & 0     \\
    \hline
    \textbf{ECAPA} & 4.2   & 2.6   & -   & -   & 26.6  & 16.2  & 4.2   & 2.2   & 0     & 0     & 0     & 0     & 0     & 0     & 0     & 0     & 0.1   & 0.4    \\
    \hline
    \textbf{XV-P} & 11.9  & 4.7   & 47.1  & 39.8  & -   & -   & 5.8   & 3.4   & 0     & 0     & 0     & 0     & 0     & 0     & 0     & 0     & 0     & 0.1    \\
    \hline
    \textbf{XV-C} & 9.9   & 6.7   & 43.8  & 36.5  & 39.1  & 23.7  & -   & -   & 0     & 0     & 0     & 0     & 0     & 0     & 0     & 0     & 0.5   & 0.8    \\
    \hline
    \textbf{Res18-I} & 0     & 0     & 1     & 0.8   & 0.5   & 0     & 0     & 0     & -   & -   & 0.3   & 0.5   & 1.6   & 1.5   & 0.9   & 0.2   & 0.6   & 1.8   \\
    \hline
    \textbf{Res18-V} & 0     & 0     & 0.5   & 0.5   & 0.4   & 0     & 0     & 0     & 4.8   & 1.2   & -   & -   & 3.9   & 1.5   & 2.2   & 0.5   & 2.2   & 3.8   \\
    \hline
    \textbf{Res34-I} & 0     & 0     & 0.7   & 0.5   & 0.3   & 0     & 0     & 0     & 0.6   & 0     & 0     & 0     & -   & -   & 0.9   & 0.3   & 0.2   & 1.2    \\
    \hline
    \textbf{Res34-V} & 0     & 0     & 0.5   & 0.8   & 1.1   & 0.3   & 0     & 0     & 0.9   & 0.1   & 0.2   & 0     & 1.8   & 0.2   & -   & -   & 1.1   & 2.3    \\
    \hline
    \textbf{Auto} & 0     & 0     & 3.3   & 2.8   & 1.2   & 0.1   & 0     & 0     & 1.2   & 0.1   & 0.6   & 0     & 2.5   & 1.1   & 0.6   & 0.1   & -   & -   \\
    \hline
\end{tabular}
  }
  \label{tab:B1-mutual}
\end{table*}

\section{Evaluation of SRS Ensemble}\label{sec:srs-diverse-exper}

\noindent {\bf Setting.}
The seed voices, $L_\infty$ perturbation budget $\varepsilon$,  step size $\alpha$
and the number of steps $N$ are the same as Appendix~\ref{sec:time-freq-curr-exper}.
We set the sampling size $\beta=1$. 
For the surrogate and target SRSs, we adopt the ``leave-one-out'' strategy,
i.e., for each of 9 open-source SRSs (cf. \cref{sec:setting}),
we regard it as the black-box target SRS, and the others SRSs are used as surrogate SRSs.
For comparison, the transfer attack with a single surrogate SRS is used as a baseline
and we only compare with the best baseline for better readability (results of baselines refer to \tablename~\ref{tab:B2-mutual}, \tablename~\ref{tab:B3-mutual}, and \tablename~\ref{tab:B1-mutual}).
We mainly analyze the results of attacks for \OSITARGET whose ASR$_t$ are reported in \tablename~\ref{tab:B1-same-differ-enroll},
where ASR$_t$-s and ASR$_t$-d denote ASR$_t$ under the same-enroll and differ-enroll settings, respectively.
The results for \OSIUNTARGET and \TDSVTARGET are shown in Appendix~\ref{sec:evaluate-qfa2sr-open}  where similar
conclusions can be drawn.

\noindent {\bf Uniform weight vs. dynamic weight.}
We compare our SRS ensemble using dynamic weight with the uniform weight used in the image domain~\cite{LCLS17}.
Our dynamic weight achieves higher ASR$_t$ on all the target SRSs (except for IV and XV-C)
in both the same-enroll and differ-enroll settings, with improvement from 6.5\% to 27.8\%.
We note that uniform weight is slightly better than ours (no more than 2\%) on IV and XV-C.
It is because XV-P is best surrogate SRS against IV and XV-C (cf. \tablename~\ref{tab:B1-mutual})
and has larger score scale than most of the other surrogate SRSs due to its PLDA-based scoring backend.
Uniform weight will make the loss of XV-P dominant while our dynamic weight will tend to balance different losses and reduce the impact of XV-P on the transferability,
hence uniform weight achieves slightly better transferability.

Due to the overall advantage,
we will use our dynamic weight for SRS ensemble in the rest of this section.

\noindent {\bf Results of SRS ensemble.}
SRS ensemble improves ASR$_t$ by at least 6.5\% and 3.4\%
under the same-enroll and differ-enroll settings, respectively.
In particular, against {Res34-V} (resp. {XV-P}) under the same-enroll (resp. differ-enroll) setting,
SRS ensemble improves the transferability from 2.2\% (resp. 23.7\%) to 34.5\% (resp. 49.4\%),
compared with the best baseline.

\begin{figure}[t]
  \centering
		\includegraphics[width=.3\textwidth]{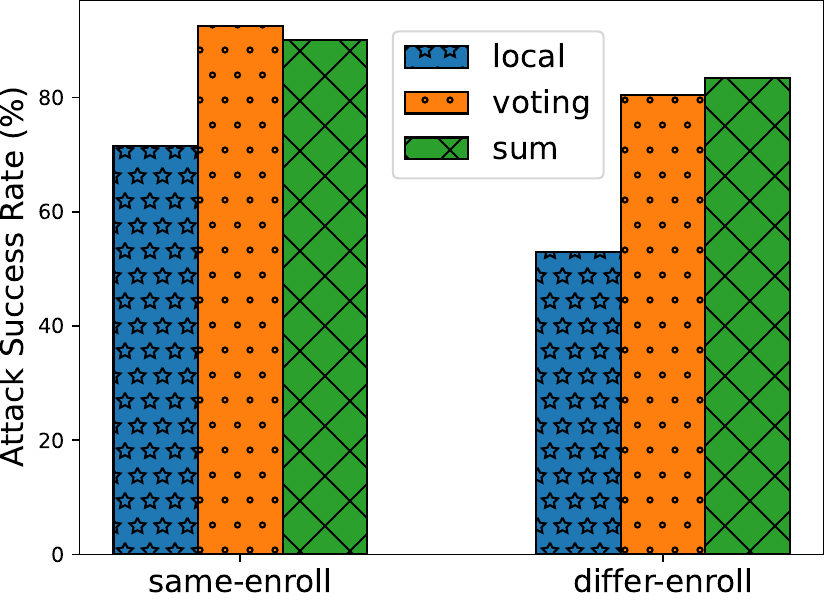}\vspace{-2mm}
		\caption{Comparison of score ranking strategies for SRS ensemble in \OSIUNTARGET.}
       \label{fig:B2-local-global}\vspace{-2mm}
\end{figure}

\noindent {\bf Score ranking strategies for \OSIUNTARGET.}
We compare score ranking strategies, i.e., local-ranking, summation-based global-ranking (\SUMGLOBAL),
and voting-based global-ranking (\VOTEGLOBAL) used in defining the loss functions of SRS ensemble
for \OSIUNTARGET. The results are shown in \figurename~\ref{fig:B2-local-global}.
Both \SUMGLOBAL and \VOTEGLOBAL achieve much higher ASR$_u$
than the local ranking,
indicating the necessity of using global ranking.
This conclusion also holds for other loss functions (cf. Appendix~\ref{sec:generalizability-loss}).
Since \SUMGLOBAL and \VOTEGLOBAL are generally comparable,
we will simply use \SUMGLOBAL in the rest of this section.

\section{Details of the Compared Attacks}\label{sec:detail-compared-attack}
\noindent {\bf Adversarial attacks.} The $L_\infty$ perturbation budget (except for Kenansville)
and the seed voices of all the adversarial attacks are the same as in Appendix~\ref{sec:srs-diverse-exper}
except that we discard those seed voices that are falsely accepted by the target commercial SRSs.
The step size, number of steps of \toolname are also the same as in Appendix~\ref{sec:srs-diverse-exper},
and sampling size $\beta$ is set to 5.
BIM is implemented as a special case of \toolname
with only one surrogate SRS but without time-freq corrosion.
For FakeBob (resp. SirenAttack),
we set the number of iterations (resp. maximum number of epochs) to 1500 (resp. 100),
which is sufficient for the attacks to converge according to our experiments,
while other parameters are the same as the original work\cite{chen2019real,du2020sirenattack}.
Additionally, we set the confidence value $\kappa=5\times \theta$ in FakeBob and SirenAttack
where $\theta$ is the threshold of the surrogate SRS.
This enables the attacks to continue searching for high-confident adversarial voices
instead of early-stopping, which may benefit the transferability~\cite{carlini2017towards, chen2019real}.
For Kenansville, we use the Fast Fourier Transform (FFT) method to perturb a voice
with 15 maximal number of iterations (the same as the original work~\cite{Kenansville}),
while the Singular Spectrum Analysis method is not considered
since it is comparable with FFT method regarding the transferability,
but is much less efficient.
{PGD is the same as the BIM attack except that it starts from a randomly perturbed example,
which may help the attack find a better local optimum.
We run the random start 10 times and select the adversarial voice with the minimal loss that is more likely to transfer,
and other settings are the same as BIM.
For CW attack, we adopt its $L_\infty$ version, set the confidence value $\kappa=5\times \theta$,
and adopt the efficient implementation in \cite{madry2017towards}.}

\noindent {\bf Hidden voice attack.} We exploit Time Domain Inversion (TDI) to perturb a voice
since it is one of the most effective method~\cite{AbdullahGPTBW19}.
TDI features the parameter window size $w$, where
the smaller $w$, the less comprehensible the voices for human
and the harder the voices to be correctly recognized by the SRS.
To produce the least understandable voices for human
when ensuring the correct recognition of the SRS,
we start from $w=1$ milliseconds (ms), gradually increase to $10$ ms with step of $0.5$ ms,
and stop the attack once the SRS correctly recognizes the perturbed voice.

\begin{table*}[!t]\centering
  \caption{The effectiveness and imperceptibility of \toolname
    in the attack scenario \OSITARGET.
    }\vspace{-2mm}
    \label{tab:B1-same-differ-enroll-appendix}%
  \begin{minipage}{1.0\textwidth}
    \centering\setlength\tabcolsep{3pt}
    \resizebox{1\textwidth}{!}{
      \begin{tabular}{|c|c|c|c|c|c|c|c|c|c|c|c|c|c|c|c|c|c|c|c|c|}
        \hline
        \multirow{2}[3]{*}{} & \multicolumn{4}{c|}{\textbf{IV}} & \multicolumn{4}{c|}{\textbf{ECAPA}} & \multicolumn{4}{c|}{\textbf{XV-P}} & \multicolumn{4}{c|}{\textbf{XV-C}} & \multicolumn{4}{c|}{\textbf{Res18-I}} \\
    \cline{2-21}          &
    \textbf{ASR$_t$-s} &  \textbf{ASR$_t$-d} & {\bf SNR}  & \textbf{PESQ} &
    \textbf{ASR$_t$-s} &  \textbf{ASR$_t$-d} & {\bf SNR}  & \textbf{PESQ} &
    \textbf{ASR$_t$-s} &  \textbf{ASR$_t$-d} & {\bf SNR}  & \textbf{PESQ} &
    \textbf{ASR$_t$-s} &  \textbf{ASR$_t$-d} & {\bf SNR}  & \textbf{PESQ} &
    \textbf{ASR$_t$-s} &  \textbf{ASR$_t$-d} & {\bf SNR}  & \textbf{PESQ} \\
        \hline
        \begin{tabular}[c]{@{}c@{}} {\bf BIM} \end{tabular} & 8.1  &  5.2  & \begin{tabular}[c]{@{}c@{}}{\bf 11.50} \\ (11.29)\end{tabular} & \begin{tabular}[c]{@{}c@{}} 1.19 \\ (1.18) \end{tabular} & 44.3  &  34.6 & \textbf{11.51} & 1.18  &  39.1 &  21.2 & 11.25  & 1.18  &  3.9  &  1.9  & {11.47} & 1.17  &  4.8  &  1.0  & {11.73} & \textbf{1.21} \\
        \hline
        \begin{tabular}[c]{@{}c@{}} {\bf BIM} +  {\small {\Circled[outer color=green]{1}}} \end{tabular} & 11.9  & 6.7   & \begin{tabular}[c]{@{}c@{}}{\bf 11.50} \\ (11.28)\end{tabular} & \begin{tabular}[c]{@{}c@{}} 1.18 \\ (1.19) \end{tabular} & 47.1  & 39.8  & {11.50} & 1.18  & 39.1  & 23.7  & 11.28  & 1.19  & 5.8   & 3.4   & \textbf{11.50} & 1.18  & 4.8   & 1.2   & \textbf{11.74} & \textbf{1.21} \\
        \hline
        \begin{tabular}[c]{@{}c@{}} {\bf BIM} +  {\small {\Circled[outer color=green]{1}}}  {\small {\Circled[outer color=blue]{2}}} \end{tabular} & 19.7  & 14    & 11.43  & 1.20  & 66.6  & 60    & 11.43  & 1.20  & 64.6  & 49.4  & \textbf{11.44} & 1.20  & 12.3  & 6.8   & 11.43  & 1.20  & 24.3  & 11.5  & 11.41  & 1.20  \\
        \hline
        \makecell[c]{\textbf{BIM} + {\small {\Circled[outer color=green]{1}}}  {\small {\Circled[outer color=blue]{2}}}  {\small {\Circled[outer color=red]{3}}} \\ (\textbf{\toolname})} & \textbf{56.4} & \textbf{49.6} & 11.11  & \textbf{1.21} & \textbf{93.5} & \textbf{89.4} & 11.13  & \textbf{1.21} & \textbf{95.1} & \textbf{89.6} & 11.13  & \textbf{1.21} & \textbf{28.9} & \textbf{16.5} & 11.10  & \textbf{1.21} & \textbf{57.5} & \textbf{32.1} & 11.12  & \textbf{1.21} \\
        \hline
    \end{tabular}%
}
  \end{minipage}
  \begin{minipage}{1.0\textwidth}
     \centering\setlength\tabcolsep{7pt}
    \centering
    \resizebox{1\textwidth}{!}{\begin{threeparttable}
      \begin{tabular}{|c|c|c|c|c|c|c|c|c|c|c|c|c|c|c|c|c|}
        \hline
        \multirow{2}[3]{*}{} & \multicolumn{4}{c|}{\textbf{Res18-V}} & \multicolumn{4}{c|}{\textbf{Res34-I}} & \multicolumn{4}{c|}{\textbf{Res34-V}} & \multicolumn{4}{c|}{\textbf{Auto}} \\
    \cline{2-17}          &
        \textbf{ASR$_t$-s} &  \textbf{ASR$_t$-d} & {\bf SNR}  & \textbf{PESQ} &
    \textbf{ASR$_t$-s} &  \textbf{ASR$_t$-d} & {\bf SNR}  & \textbf{PESQ} &
    \textbf{ASR$_t$-s} &  \textbf{ASR$_t$-d} & {\bf SNR}  & \textbf{PESQ} &
    \textbf{ASR$_t$-s} &  \textbf{ASR$_t$-d} & {\bf SNR}  & \textbf{PESQ} \\
        \hline
        \begin{tabular}[c]{@{}c@{}} {\bf BIM} \end{tabular} & 0.3   & 0.4   &  \begin{tabular}[c]{@{}c@{}}11.63 \\ {\bf (11.81)} \end{tabular} &  \begin{tabular}[c]{@{}c@{}}1.19 \\ {(1.20)} \end{tabular} &  3.8  & 1.2   & {11.72} & \textbf{1.21} & 2.1   &  0.4 & \textbf{11.74} & {1.20} &  1.9 &  3.3  & \textbf{11.75} & \textbf{1.21} \\
        \hline
        \begin{tabular}[c]{@{}c@{}} {\bf BIM} +  {\small {\Circled[outer color=green]{1}}} \end{tabular} & 0.6   & 0.5   &  \begin{tabular}[c]{@{}c@{}}11.62 \\ {(11.79)} \end{tabular} &  \begin{tabular}[c]{@{}c@{}}1.20 \\ {\bf (1.21)} \end{tabular} & 3.9   & 1.5   & \textbf{11.74} & \textbf{1.21} & 2.2   & 0.5   & \textbf{11.74} & \textbf{1.21} & 2.2   & 3.8   & {11.74} & \textbf{1.21} \\
        \hline
        \begin{tabular}[c]{@{}c@{}} {\bf BIM} +  {\small {\Circled[outer color=green]{1}}} {\small {\Circled[outer color=blue]{2}}} \end{tabular} & 13.8  & 6.5   & 11.41  & 1.20  & 30.2  & 22.1  & 11.37  & 1.20  & 34.5  & 21.3  & 11.39  & 1.20  & 23.9  & 18.1  & 11.42  & 1.20   \\
        \hline
        \makecell[c]{\textbf{BIM} + {\small {\Circled[outer color=green]{1}}} {\small {\Circled[outer color=blue]{2}}}  {\small {\Circled[outer color=red]{3}}} \\ (\textbf{\toolname})} & \textbf{41} & \textbf{14.7} & 11.12  & \textbf{1.21} & \textbf{53.4} & \textbf{35.5} & 11.06  & \textbf{1.21} & \textbf{66} & \textbf{36.6} & 11.09  & \textbf{1.21} & \textbf{74} & \textbf{53.3} & 11.11  & \textbf{1.21}  \\
        \hline
        \end{tabular}%

        \begin{tablenotes}
          \item Note: For BIM, we report the best transferability among all open-source SRSs excluding the target SRS.
          In case that the best surrogate SRS differs with the enrollment settings, the SNR and PESQ of different-enrollment are included in the bracket.
          {\small {\Circled[outer color=green]{1}}}, {\small {\Circled[outer color=blue]{2}}}, {\small {\Circled[outer color=red]{3}}}
        denote Tailored Loss Functions, SRS Ensemble, and Time-Freq Corrosion, respectively.
        \end{tablenotes}
      \end{threeparttable}
}
  \end{minipage}
\vspace{-2mm}\end{table*}

\section{Ablation Study of Contributions on Open-source SRSs}\label{sec:evaluate-qfa2sr-open}
Following the ablation study on commercial APIs in \cref{sec:commercial-APIs},
we repeat the ablation study on open-source SRSs to evaluate the contributions of individual methods of \toolname.
For \OSITARGET, we adopt all the nine open-source SRSs as target SRSs,
while for \OSIUNTARGET and \TDSVTARGET, we choose two target SRSs, {XV-P} and {Res18-V},
which represent the easily and hardly transferable SRSs,
according to the results in \tablename~\ref{tab:B1-mutual}.
The results shown in \tablename~\ref{tab:B1-same-differ-enroll-appendix}, \tablename~\ref{tab:B2-same-differ-enroll} and \tablename~\ref{tab:B3-differ-enroll}
demonstrate that all of tailored loss functions, SRS ensemble, and time-freq corrosion boost the transferability,
consistent with the conclusion drawn in \cref{sec:commercial-APIs}.

\begin{table}[t]
  \centering
  \caption{ASR$_u$ and imperceptibility of \toolname
    for \OSIUNTARGET.}\vspace{-2mm}\setlength{\tabcolsep}{2pt}
  \resizebox{0.47\textwidth}{!}{%
    \begin{tabular}{|c|c|c|c|c|c|c|c|c|}
    \hline
    \multirow{2}[3]{*}{} & \multicolumn{4}{c|}{\textbf{XV-P}} & \multicolumn{4}{c|}{\textbf{Res18-V}} \\
\cline{2-9}          &
\textbf{ASR$_u$-s} &  \textbf{ASR$_u$-d} & {\bf SNR}  & \textbf{PESQ} &
\textbf{ASR$_u$-s} &  \textbf{ASR$_u$-d} & {\bf SNR}  & \textbf{PESQ} \\
    \hline
    \textbf{BIM} & 61.5 & 36.5 & 11.37 & 1.20 & 1.0 & 0 & \begin{tabular}[c]{@{}c@{}} 11.68 \\ {(11.80)} \end{tabular} & \begin{tabular}[c]{@{}c@{}} 1.19 \\ {(1.20)} \end{tabular} \\
    \hline
    \begin{tabular}[c]{@{}c@{}} {\bf BIM} +  {\small {\Circled[outer color=green]{1}}} \end{tabular} & 68    & 37    & 11.37  & 1.19  & 2.5   & 0.5   & \begin{tabular}[c]{@{}c@{}} 11.67 \\ {\bf (11.83)} \end{tabular} & \begin{tabular}[c]{@{}c@{}} 1.20 \\ {(1.21)} \end{tabular} \\
    \hline
    \begin{tabular}[c]{@{}c@{}} {\bf BIM} +  {\small {\Circled[outer color=green]{1}}}  {\small {\Circled[outer color=blue]{2}}} \end{tabular} & 92.5  & 83.5    & \textbf{11.55} & 1.21  & 31.5    & 10.5     & 11.51  & 1.20  \\
    \hline
    \begin{tabular}[c]{@{}c@{}} {\bf BIM} +  {\small {\Circled[outer color=green]{1}}}   {\small {\Circled[outer color=blue]{2}}} {\small {\Circled[outer color=red]{3}}} \\ (\textbf{\toolname}) \end{tabular} & \textbf{98.5} & \textbf{95.5} & 11.35  & \textbf{1.23} & \textbf{59} & \textbf{18.5} & 11.33  & \textbf{1.22} \\
    \hline
    \end{tabular}%
}\vspace{-1mm}
  \label{tab:B2-same-differ-enroll}%
\end{table}%

\begin{table}[t]
  \centering
  \caption{ASR$_t$ and imperceptibility of \toolname
    for \TDSVTARGET.}\vspace{-2mm}
 \scalebox{0.75}{
    \begin{tabular}{|c|c|c|c|c|c|c|}
    \hline
    \multirow{2}[3]{*}{} & \multicolumn{3}{c|}{\textbf{XV-P}} & \multicolumn{3}{c|}{\textbf{Res18-V}} \\
\cline{2-7}          & \textbf{ASR$_t$-d} &  {\bf SNR} & \textbf{PESQ} & \textbf{ASR$_t$-d} &  {\bf SNR}& \textbf{PESQ} \\
    \hline
    \textbf{BIM} & 29.5  & 12.21  & 1.23  & 1.5   & \textbf{12.54} & 1.25 \\
    \hline
    \begin{tabular}[c]{@{}c@{}} {\bf BIM} +  {\small {\Circled[outer color=green]{1}}} \end{tabular} & 29.5  & 12.21  & 1.23  & 1.5   & \textbf{12.54} & 1.25\\
    \hline
    \begin{tabular}[c]{@{}c@{}} {\bf BIM} +  {\small {\Circled[outer color=green]{1}}}  {\small {\Circled[outer color=blue]{2}}} \end{tabular} & 50    & \textbf{12.33} & 1.25  & 8     & 12.28  & 1.24  \\
    \hline
    \begin{tabular}[c]{@{}c@{}} {\bf BIM} +  {\small {\Circled[outer color=green]{1}}}  {\small {\Circled[outer color=blue]{2}}}  {\small {\Circled[outer color=red]{3}}} \\ (\textbf{\toolname}) \end{tabular} & \textbf{78.5} & 12.01  & \textbf{1.26} & \textbf{37} & 11.99  & \textbf{1.26} \\
   \hline
    \end{tabular}%
 }\vspace{-1mm}
  \label{tab:B3-differ-enroll}%
\end{table}%

\section{More Details of Experimental Setting on Attacking Voice Assistants}\label{sec:voice-assistant-dataset}
\noindent {\bf Datasets.} The activation phrase as well as the recording number
is shown in \tablename~\ref{tab:Activation-Phrase}.
For Google Assistant and Apple Siri, these activation phrases
are used for both the enrollment voices and the seed voices for the attack.
For TMall Genie, ``TMall Genie'' is used for enrolling and ``TMall Genie, who am I''
is used as the attack seed voices.
The reason is that the activation of TMall Genie by ``TMall Genie'' is speaker-independent,
and we have to ask the TMall Genie ``who am I'' to determine the identity of the speaker.

\noindent {\bf Attack success rate.}
For Google Assistant and Apple Siri, we count a successful attack
only when the voice assistants are activated within the number of allowed queries to the target SRS.
For TMall Genie, there are three kinds of response to ``TMall Genie, who am I'',
each reflecting the confidence that TMall Genie recognizes the voice as from the speaker {\tt SPK\_ID},
namely, ``Hello, {\tt SPK\_ID}, happy to serve you.'' (high-confidence),
``I think you are {\tt SPK\_ID}, am I right?'' (medium confidence),
and ``I am unfamiliar with your voice.'' (low confidence).
We regard an attack as a successful attack when one of the following conditions holds:
(1) The seed voice receives the low-confidence response or the other two responses
where {\tt SPK\_ID} is different from the target speaker,
and the adversarial voice receives the medium or high confidence response
where {\tt SPK\_ID} is identical to the target speaker.
(2) The seed voice receives the medium confidence response,
the adversarial voice receives the high confidence response,
and both of their {\tt SPK\_ID} are identical to the target speaker.

\begin{table}[t]
  \centering
  \caption{The datasets used for enrolling and attacking voice assistants. }\vspace{-2mm}
   \scalebox{0.8}{
      \begin{tabular}{|c|c|c|}
    \hline
    \textbf{Voice Assitant} & \textbf{Activation Phrase} & \textbf{Number} \\
    \hline
    \textbf{Google  Assitant} & Ok Google & 5 \\
    \hline
    \multirow{5}{*}{\textbf{Apple Siri}} & Hey Siri & 1 \\
\cline{2-3}          & Hey Siri, send a message & 1 \\
\cline{2-3}          & Hey Siri, how's the weather today & 1 \\
\cline{2-3}          & Hey Siri, set a timer for three minutes & 1 \\
\cline{2-3}          & Hey Siri, play some music & 1 \\
    \hline
    \multirow{2}{*}{\textbf{TMall Genie}} & TMall Genie (Chinese) & 3 \\
\cline{2-3}          & TMall Genie, who am I (Chinese) & 5 \\
    \hline
    \end{tabular}%
  }\vspace{-2mm}
  \label{tab:Activation-Phrase}%
\end{table}%

\end{document}